\documentclass[graybox]{raa_gam}

\usepackage{helvet}         
\usepackage{courier}        
\usepackage{times}
\usepackage{type1cm}        
\usepackage{makeidx}         
\usepackage{graphicx}        
\usepackage{multicol}        
\usepackage{journals}
\usepackage{natbib}
\usepackage{color}
\usepackage{epstopdf}
\usepackage[bottom]{footmisc}

\definecolor{darkgreen}{rgb}{0,0.7,0}
\definecolor{grey}{rgb}{0.5,0.6,0.7}

\makeindex             

\def\simlt{\lower.5ex\hbox{$\; \buildrel < \over \sim \;$}}
\def\simgt{\lower.5ex\hbox{$\; \buildrel > \over \sim \;$}}

\begin{document}
\volnopage{{\bf 2012} Vol {\bf 12} No. {\bf 8}, 917--946}

\title{The Current Status of Galaxy Formation}

\author{Joseph Silk$^{1,2,3}$, Gary A. Mamon$^1$}

\institute{
$^1$ Institut d'Astrophysique de Paris (UMR 7095: CNRS \& UPMC), 98 bis
  Boulevard Arago, Paris 75014, France\\ 
$^2$ Department of Physics and Astronomy, 3701 San Martin Drive, The Johns Hopkins University, Baltimore MD 21218, USA\\
$^3$ Beecroft Institute of Particle Astrophysics and Cosmology, 1 Keble Road,
  University of Oxford, Oxford OX1 3RH UK \\
\vspace{\baselineskip}
{\small Received: 2012 July 12; accepted: 2012 July 17}
}

%
%


\abstract{
Understanding galaxy formation is one of the most pressing issues  in cosmology.  We review the current status of galaxy formation  from both an observational and a theoretical perspective,
 and summarize the prospects for future advances. }
\maketitle
\section{Introduction}
\label{}
Numerical simulations  of  large-scale structure have met with great success. However these  same simulations fail 
to account for several of the observed properties of galaxies.  On large
scales, $\sim 0.01-100 \rm\, Mpc$, the ansatz of cold, weakly interacting
dark matter has led to realistic maps of the galaxy distribution, under the
assumptions that light traces mass and that the initial conditions are
provided by the observed temperature fluctuations in the cosmic microwave
background. On smaller scales, light no longer traces mass because of the
complexity of galaxy and star formation. Baryon physics must be added to the
simulations in order to produce realistic galaxies. It is here that the
modelling is still inadequate.

In this review, we will begin with the standard phenomenology of galaxy
formation, then discuss methods and present the recent observational and
modeling advances, finishing with a summary of the numerous outstanding issues in galaxy
formation theory.

\section{Phenomenology}

\subsection{The luminosity function}
\begin{figure}[ht]
\centering
\includegraphics[width=6.3cm,bb=10 175 430 700,angle=-90]{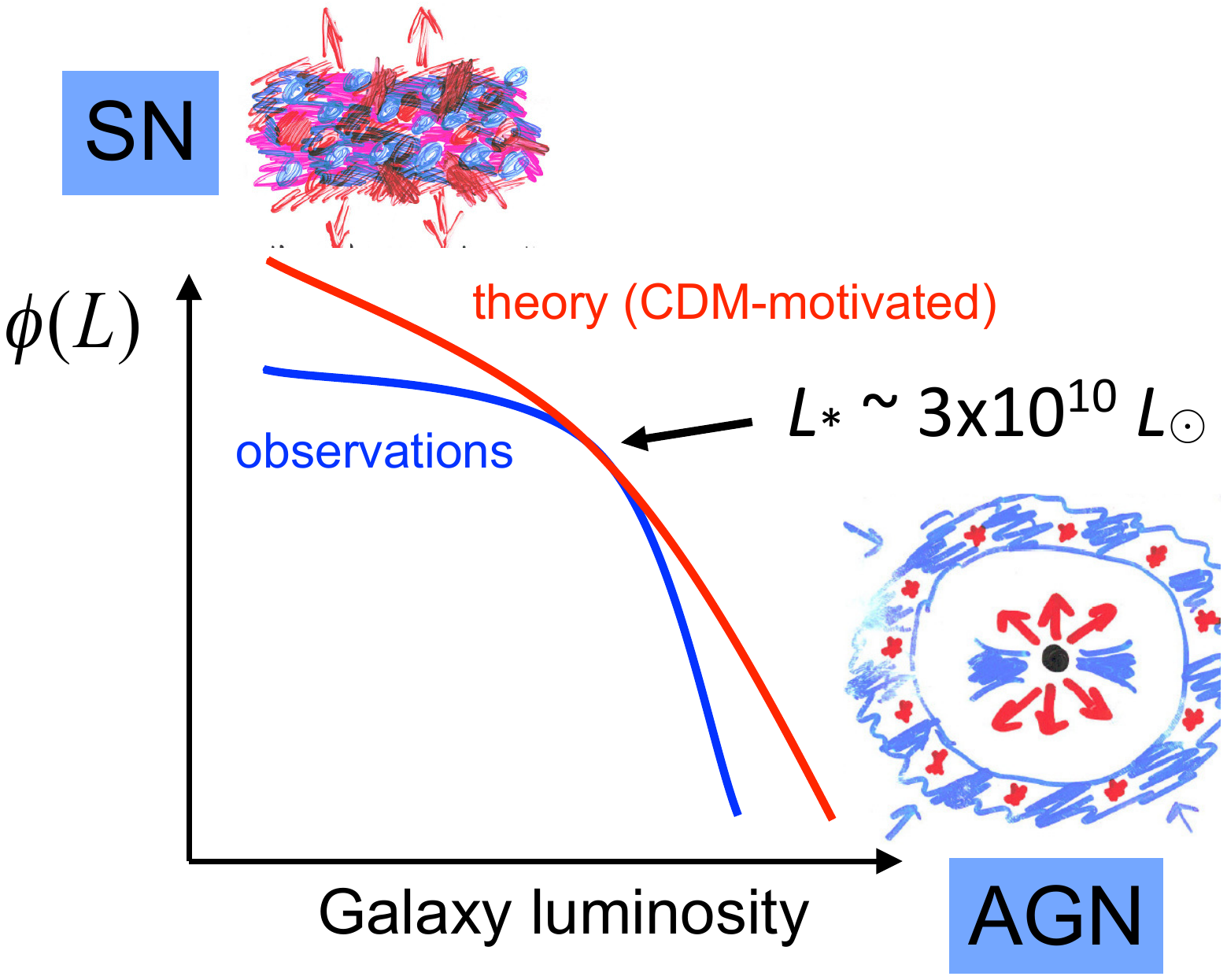}
\caption{Role of feedback in modifying the galaxy luminosity function}
\label{lfun}
\end{figure}

Theory provides the mass function of dark halos. Observation yields the
luminosity function of galaxies, usually fit by a \cite{Schechter76} 
function. Comparison of the two is at first sight disconcerting.  One can calculate the $M/L$ ratio for the two functions to overlap at one point,  for a mass $M^\ast$ corresponding to $L_\ast.$
Define $t_{\rm cool}={3/ 2}nkT/ [n^2 \Lambda(T)]$ and 
$ t_{\rm dyn}= {3}/( \sqrt {32\pi G \rho})$.  For star formation to occur, cooling is essential, and the condition 
$t_{\rm cool}<t_{\rm dyn}$ guarantees cooling in an inhomogeneous galactic halo where gas clouds collide at the virial velocity.
One finds that 
$$
M_{\rm cool}^\ast={\alpha^{3}\over \alpha_g^2}\,{m_p\over m_e}\,{t_{\rm cool}\over
  t_{\rm dyn}}\, T^{1+2\beta} \ ,$$
where $\alpha=e^2/(\hbar c)$ and $\alpha_g=G m_p^2 /e^2$ are the electromagnetic and gravitational fine structure constants.
For a cooling function $\Lambda(T)\propto T^\beta,$ over the relevant temperature range ($10^5-10^7$ K), one can take 
$\beta\approx -1/2$ for a low metallicity plasma \citep{GS07}. The result is
that one finds a characteristic galactic halo mass, in terms of fundamental
constants, to be of order $10^{12} \rm M_\odot$ \citep{Silk77}. The inferred value of the mass-to-light ratio $M/L$ is similar to that observed for $L_\ast$ galaxies. This is a success for theory: dissipation provides a key ingredient in understanding the stellar masses of galaxies, at least for the ``typical'' galaxy.  
The characteristic galactic mass is understood by the requirement that
cooling within a dynamical time is a necessary condition for efficient star
formation (Fig.~\ref{lfun}).

However,
the na\"{\i}ve assumption that stellar mass follows halo mass,
leads to too many small galaxies, too many big galaxies in the nearby
universe, too few massive galaxies at high redshift, and too many baryons
within the galaxy halos. In addition there are structural problems: for
example, massive galaxies with thin disks and/or without bulges are missing,
and the concentration and cuspiness of cold dark matter is found to be
excessive in barred galaxies and in dwarfs.  The resolution to all of these
difficulties must lie in feedback. There are various flavors of feedback
that span the range of processes including reionization at very high
redshift, supernova (SN) explosions, tidal stripping and input from active
galactic nuclei (AGN). All of these effects no doubt have a role, but we shall see
that what is missing is a robust theory of star formation as well as adequate
numerical resolution to properly model the interactions between baryons,
dynamics and dark matter.

\subsection{Star formation rate and efficiency}

In addressing star-forming galaxies, the problem reduces to our fundamental
ignorance of star formation. Phenomenology is used to address this gap in our
knowledge. Massive star feedback in giant molecular clouds, the seat of most
galactic star formation, implies a star formation efficiency (SFE), defined
as star formation rate (SFR) divided by the ratio of gas mass to dynamical or disk
rotation time, of around 2\%. This is also found to be true globally in the
Milky Way (MW) disk.

Remarkably, a similar SFE is found in nearby star-forming disk galaxies. Indeed,  SFRs per unit area in disk galaxies, both near and far, can be described by a simple law,
with SFE being the controlling parameter \citep{Silk97,Elmegreen97}:
\begin{equation}
\rm SFE ={SFR  \times  DYNAMICAL \, TIME  \over  GAS \, MASS} \approx  0.02.
\label{sfeeq}
\end{equation}
The motivation comes from the gravitational instability of cold gas-rich
disks, which provides the scaling, although the normalization depends on
feedback physics. For the global law, in terms of SFR and gas
mass per unit area, SN regulation provides the observed efficiency of about
2\% which fits essentially all local star--forming galaxies.  One finds from
simple momentum conservation that ${\rm SFE} = {\sigma_{\rm gas} v_{\rm cool}
  m*_{\rm SN}} / {E_{\rm SN}^{\rm initial}} \approx 0.02$. Here, $v_{\rm
  cool}$ is the 
SN-driven swept-up shell velocity at which approximate momentum conservation sets in and 
$m*_{\rm SN}\approx 150 \rm M_\odot$ is the mass formed in stars per SNII, in
this case for a \cite{Chabrier03} initial mass function (IMF).
This is a crude
estimator of the efficiency of SN momentum input into the interstellar
medium, but it reproduces the observed global normalization of the star
formation law.

The fit applies not only globally but to star formation complexes in
individual galaxies such as M51 and also to starburst galaxies. The star
formation law is known as the Schmidt-Kennicutt law \citep{Kennicutt+07}, and
its application reveals that molecular gas is the controlling gas ingredient.
In the outer parts of galaxies, where the molecular fraction is reduced due
to the ambient UV radiation field and lower surface density, the SFR 
per unit gas mass also declines \citep{BLW11}.

\begin{figure}[ht]
\centering
\includegraphics[width=0.77\hsize]{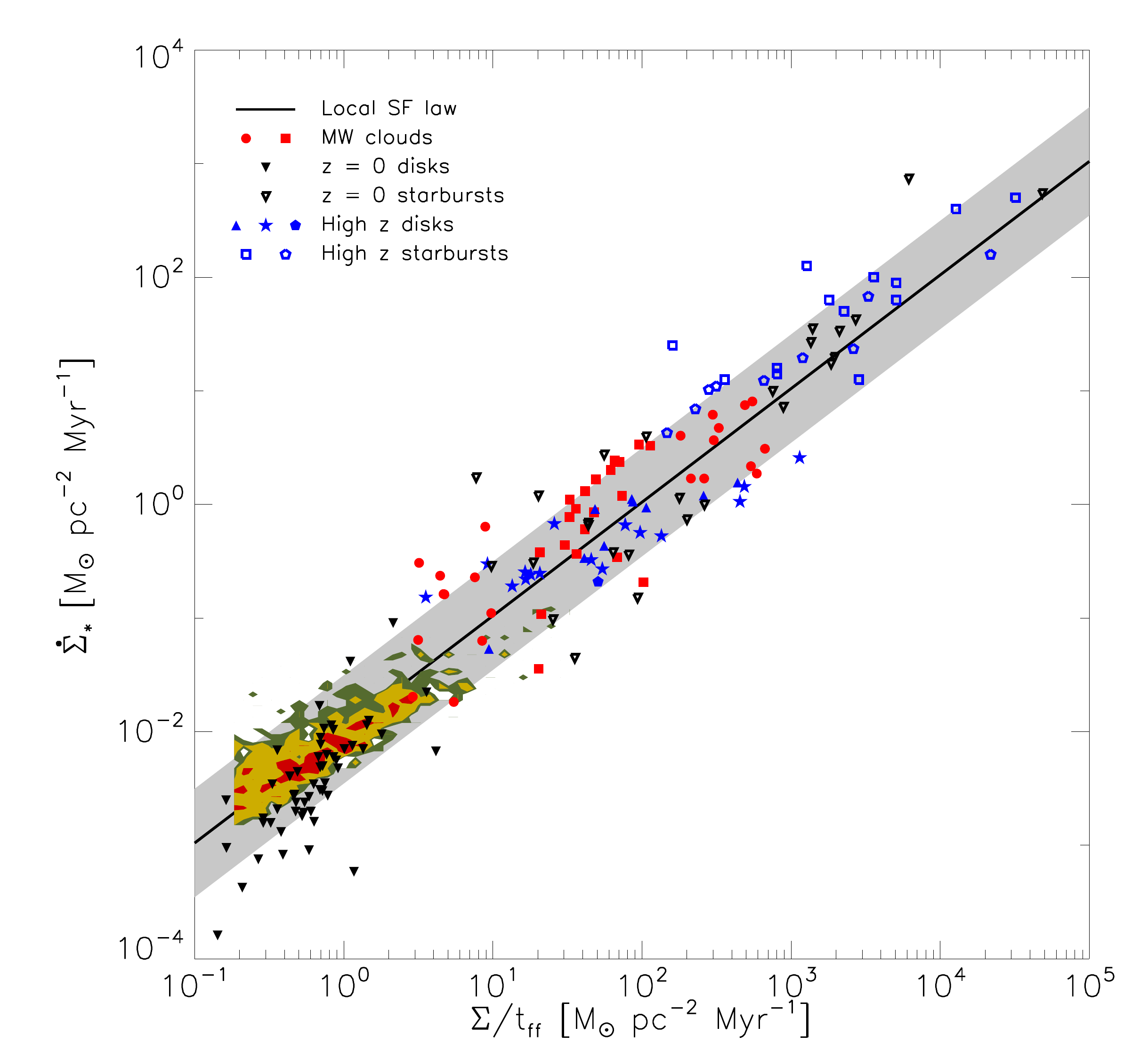}
\caption{Schmidt-Kennicutt laws on nearby (including Local Group galaxies as
  \emph{shaded regions}) and distant galaxies, as well as
  Milky Way Giant Molecular Clouds \citep{KDM12}. The solid line is similar
  to equation~(\ref{sfeeq}). }
\label{SKlaw}
\end{figure}
For disk instabilities to result in cloud formation, followed by cloud
agglomeration and consequent star formation, one also needs to maintain a
cold disk by accretion of cold gas. There is ample evidence of a supply of
cold gas, for example in the M33 group.
Other spiral galaxies show extensive reservoirs of HI in their outer regions,
for example  NGC 6946 \citep{Boomsma+08} and UGC 2082  \citep{Heald+11a}.  
Recent data extends the Schmidt-Kennicutt law to $z\sim 2,$ with
 a tendency for ultraluminous starbursts at $z\sim 2$ to have somewhat higher
 SFE (\citealp{Genzel+10}, see Fig.~\ref{SKlaw}).

A more refined theoretical model needs to take account of star formation in a
multi-phase interstellar medium.
One expects self-regulation to play a role. If the porosity in the form of SN
remnant-driven bubbles is low, there is no venting and the pressure is
enhanced, clouds are squeezed, and SN explosions are triggered by massive
star formation. This is followed by high porosity and blow-out,
and the turbulent pressure drops. Eventually halo infall replenishes the cold
gas, the porosity is lowered and the cycle recommences. Some of this
complexity can be seen in numerical simulations \citep{ATM11}. SNe provide
recirculation and venting of gas into fountains, thereby reducing the SFE and
prolonging the duration of star formation in normal disk galaxies.


%
\begin{figure}[ht]
\centering
\includegraphics[width=8cm]{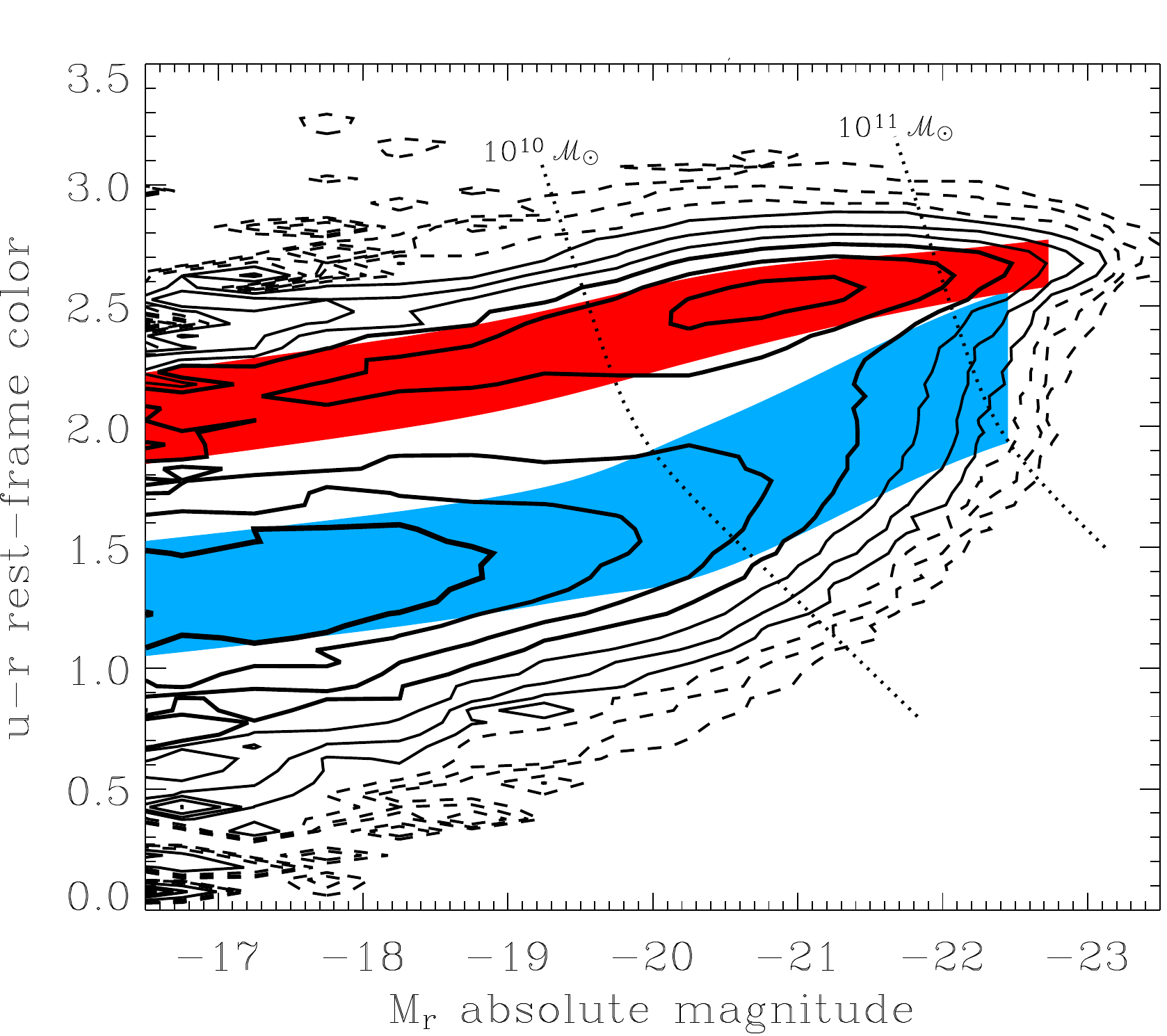} 
\caption{Illustration of galaxy bimodality. The contours are the density of SDSS
  galaxies in color-luminosity space, after correction for selection effects
  \citep{Baldry+04}.}
\label{bimod}
\end{figure}
In fact, galaxy colors illustrate the \emph{bimodality} of SFRs.
\emph{Elliptical and lenticular galaxies are red, spirals are blue.} 
This lyric does not hide a continuity in galaxy properties:
most galaxies lie in either the \emph{Red Sequence} or the
\emph{Blue Cloud} (see Fig.~\ref{bimod}).
This suggests that star formation in galaxies is either ongoing or was quenched
several Gyr ago. The small fraction of intermediate population, \emph{Green
  Valley} galaxies suggests that some galaxies have experienced a recent quenching of
their star formation.
\begin{figure}[ht]
\centering
\includegraphics[width=7.5cm,angle=90]{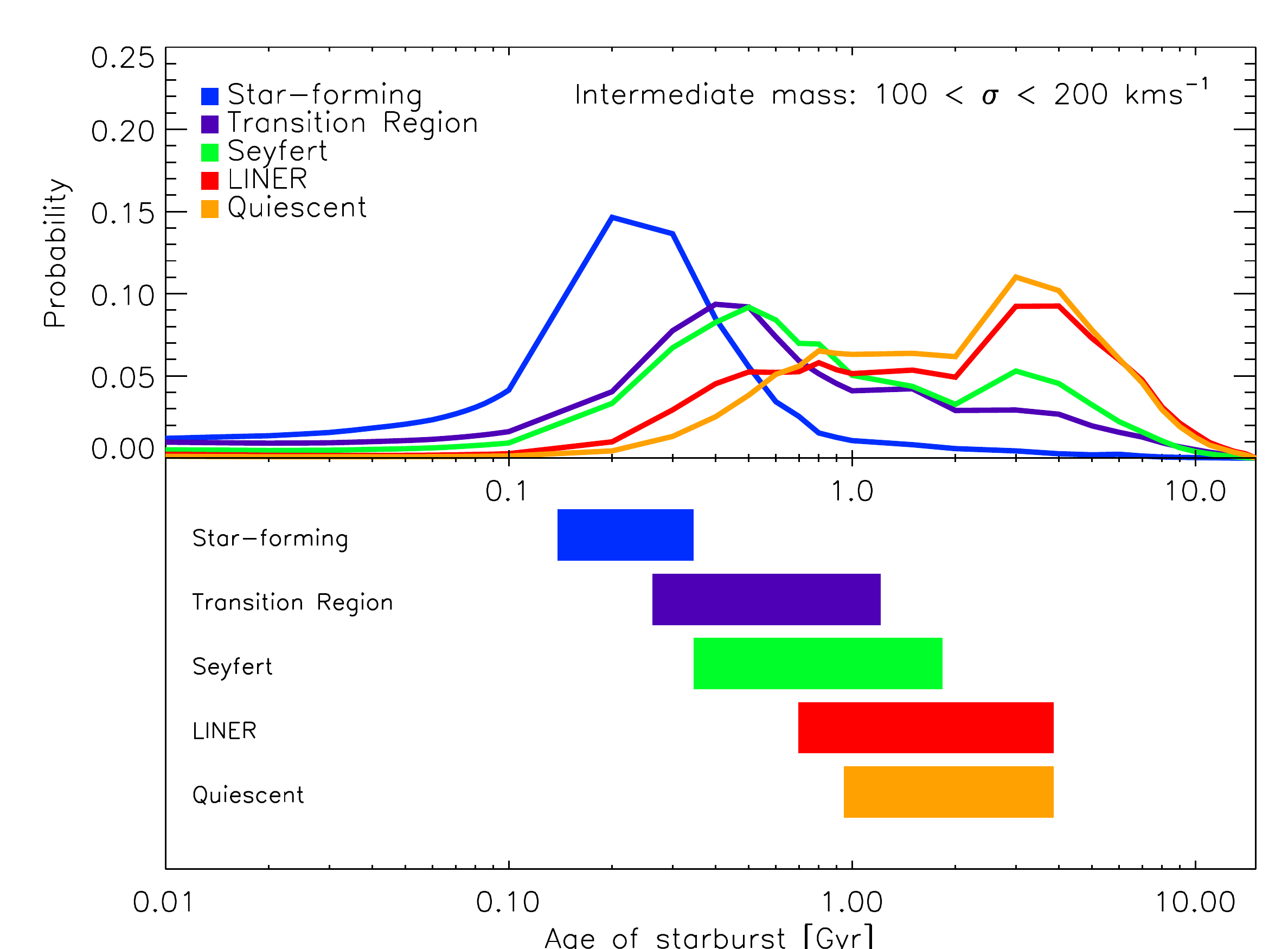}
\caption{Ages of galaxies of different activity \citep{Schawinski+07}}
\label{ages}
\end{figure}
Seyfert galaxies have intermediate age stellar populations (\citealp{Schawinski+07}, see
fig.~\ref{ages}) and mostly lie in the  Green Valley \citep{Schawinski12}.
This suggests that star formation is quenched by nuclear activity.

\subsection{Scaling relations}

The global properties of early-type galaxies are known to correlate: early
work focussed on $L\sim \sigma_v^4$ \citep{FJ76}.
The early work found a slope of 4 because of the inclusion of bright and
faint galaxies. The modern work finds a slope of 5 for luminous galaxies
($M_B\la-20.5$, core-S\'ersic galaxies) and a slope of 2 for the less
luminous spheroids, and has been distilled into the Fundamental Plane linking
mass, mass-to-light ratio, and mean 
surface brightness at the effective radius \citep{BBF92}.
\citep{BBF92}. 
\begin{figure}[ht]
\centering
\includegraphics[width=9.5cm]{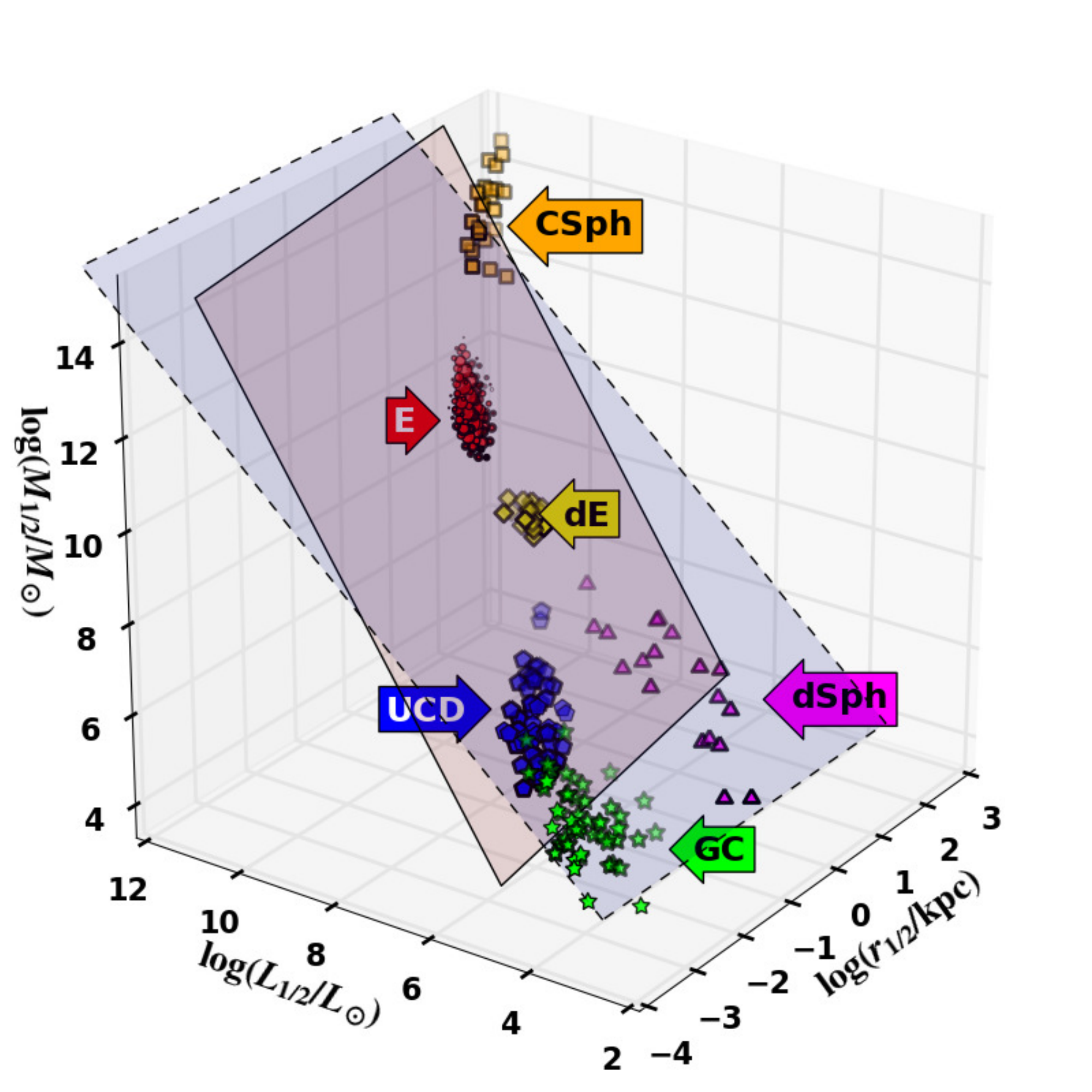}
\caption{3D view of scaling relations  of spheroidal systems from globular
  clusters (GC) to clusters of galaxies (CSph), via ultra-compact dwarfs
  (UCD), dwarf spheroidals (dSph), dwarf ellipticals (dE) and giant
  ellipticals (E), where the axes are
  half-luminosity, half-luminosity radius and total mass within
  half-luminosity radius
  \citep{TBGW11}. 
The \emph{red} and \emph{blue planes} respectively represent the Fundamental Plane and
the ``virial plane'' of constant $M/L$.}
\label{3dscale}
\end{figure}

Figure~\ref{3dscale} shows a more modern version of the
properties of early-type galaxies, to which are added globular clusters and
clusters of galaxies. It is not yet understood what makes the continuity of
the global properties of massive systems fragment into two branches with
ultra-compact dwarfs and globular clusters on one side and dwarf spheroidals
on the other.

\subsection{Evolution of low mass galaxies}

\begin{figure}[ht]
\centering
\includegraphics[width=8.5cm]{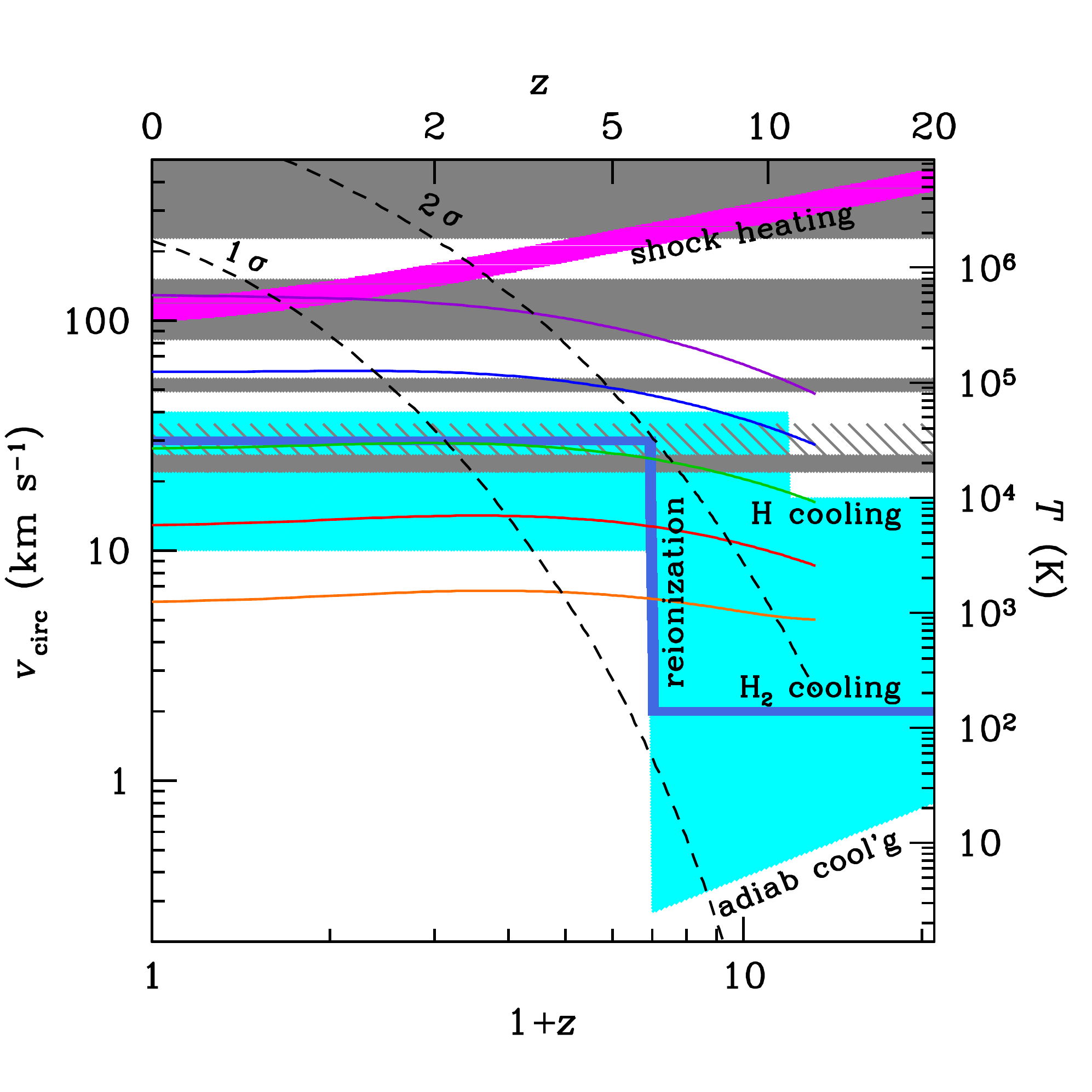} 
\caption{Evolution of circular velocities (at virial radius) for halos with
  efficient star formation \citep{MTTC12}. The \emph{smooth curves} indicate the
  mean evolution of halos (with final masses $\log h M = 8$ to 12 going
  upwards).
The \emph{blue broken line} is a model for the
  evolution of the minimum mass for galaxy formation (set by entropy feedback
  and related to the temperature of the IGM) and the \emph{cyan shaded
    region} represents our ignorance of this parameter. The \emph{magenta curve}
is the maximum circular velocity for efficient gas infall. 
The \emph{grey shaded
    bands} represent regions of thermal instability. The \emph{dashed curves}
shows 1 and $2\,\sigma$ fluctuations from the $\Lambda$CDM
primordial density fluctuation spectrum.}
\label{vminvsz}
\end{figure}

The accepted solution for gas disruption and dispersal in intermediate mass
and massive dwarfs (halo mass $\sim 10^8 -10^{10}\, \rm M_\odot$) is by SN
feedback.  SNe expel the remaining baryons in systems of halo mass up to
$\sim 10^8\rm\, M_\odot$,  leaving behind dim remnants of dwarf galaxies
\citep{DS86}. Presumably the luminous dwarfs accrete gas at later
epochs. Most gas is ejected by the first generations of SNe for
systems with escape velocity $\simlt 50 \rm\, km/s,$ leaving dim stellar
remnants behind.

In very
low-mass halos gas cannot even fall in, because its specific entropy is too
high \citep{Rees86}. This \emph{entropy barrier} amounts to a temperature barrier
since the gas density, which to first order is proportional to the total mass
density, is the same in different halos at a given epoch.
Only halos of mass $\simgt 10^5\rm \,M_\odot$ trap baryons that are able to
undergo early $\rm H_2$ cooling and eventually form stars. 
Hydrodynamical simulations indicate that this lower limit is sharp \citep{Gnedin00,OGT08}.
Reionization
reinforces this limit by heating the intergalactic gas to high entropy,
hence suppressing subsequent
star formation (see Fig.~\ref{vminvsz}).  
The abrupt increase of the sound speed to $\sim 10-20 \rm\, km/s$
at $z\sim 10$ means that dwarfs of halo mass $\sim 10^6-10^7\rm \,M_\odot,$ which
have not yet collapsed and fragmented into stars, will be disrupted. However
massive dwarfs are unaffected, as are the high $\sigma$ peaks that develop
into early collapsing, but rare, low mass dwarfs.

\subsection{Specific SFR}
 Other  serious, not unrelated,  problems  arise with low mass galaxies. In the hierarchical
 approach, these generically form early. Theoretical models, both SAMs and
 hydrodynamical, appear to fail to account for the observed specific star
 formation rates (SFR per unit stellar mass or SSFR, \citealp{Weinmann+12}), producing too little star formation at
 late times. Metallicity-dependent star formation alleviates the high
 redshift problem, reducing the stellar mass that is in place early and
 enhancing the SSFR as needed \citep{KD12}. However, it leads to inconsistency 
at low redshift, because the change in metallicity 
and the gas fraction anti-correlate, hence
 leading to too little evolution in the SSFR.

\begin{figure}[ht]
\centering
\includegraphics[width=0.45\hsize]{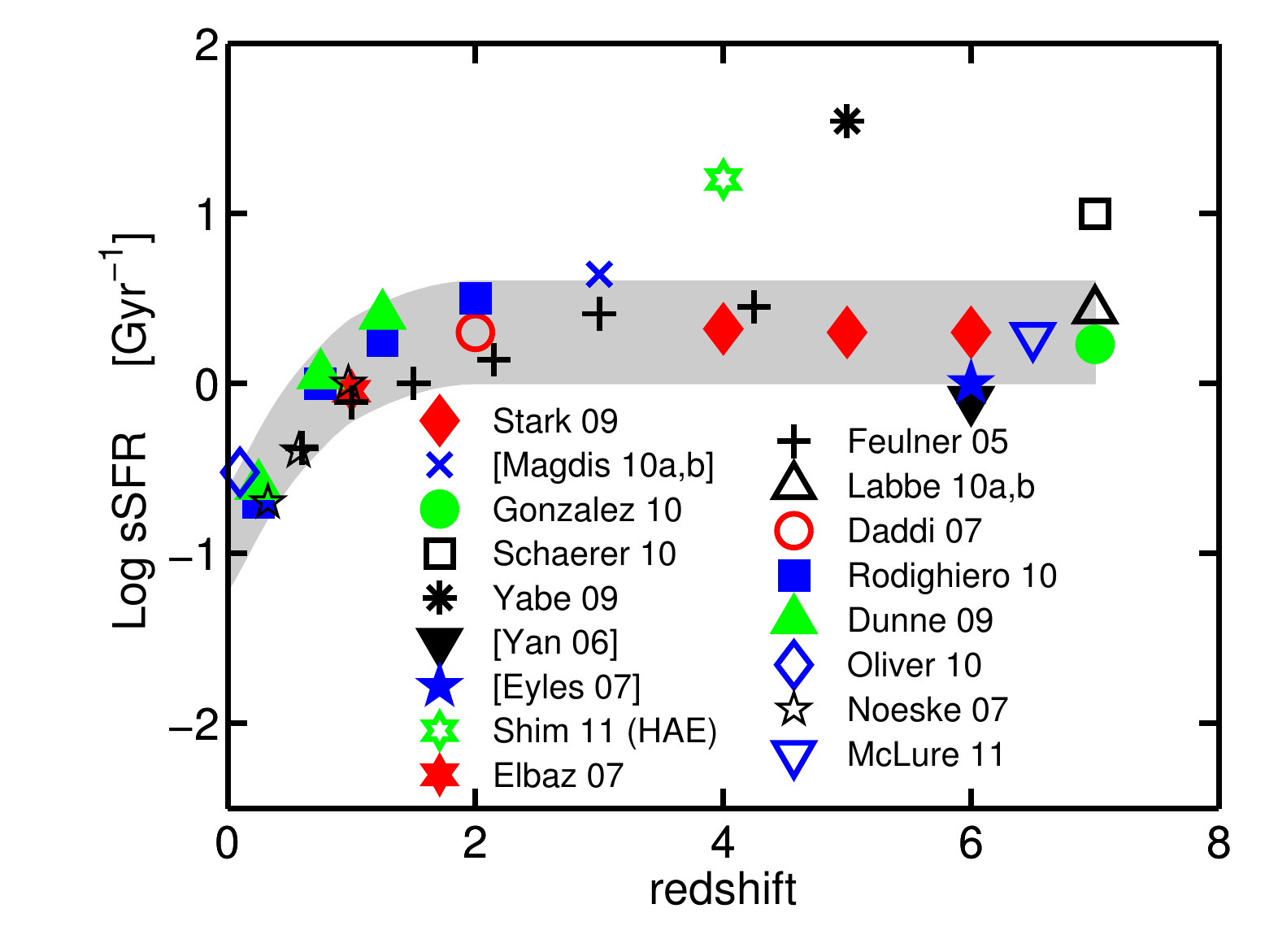} \\
\includegraphics[width=0.8\hsize]{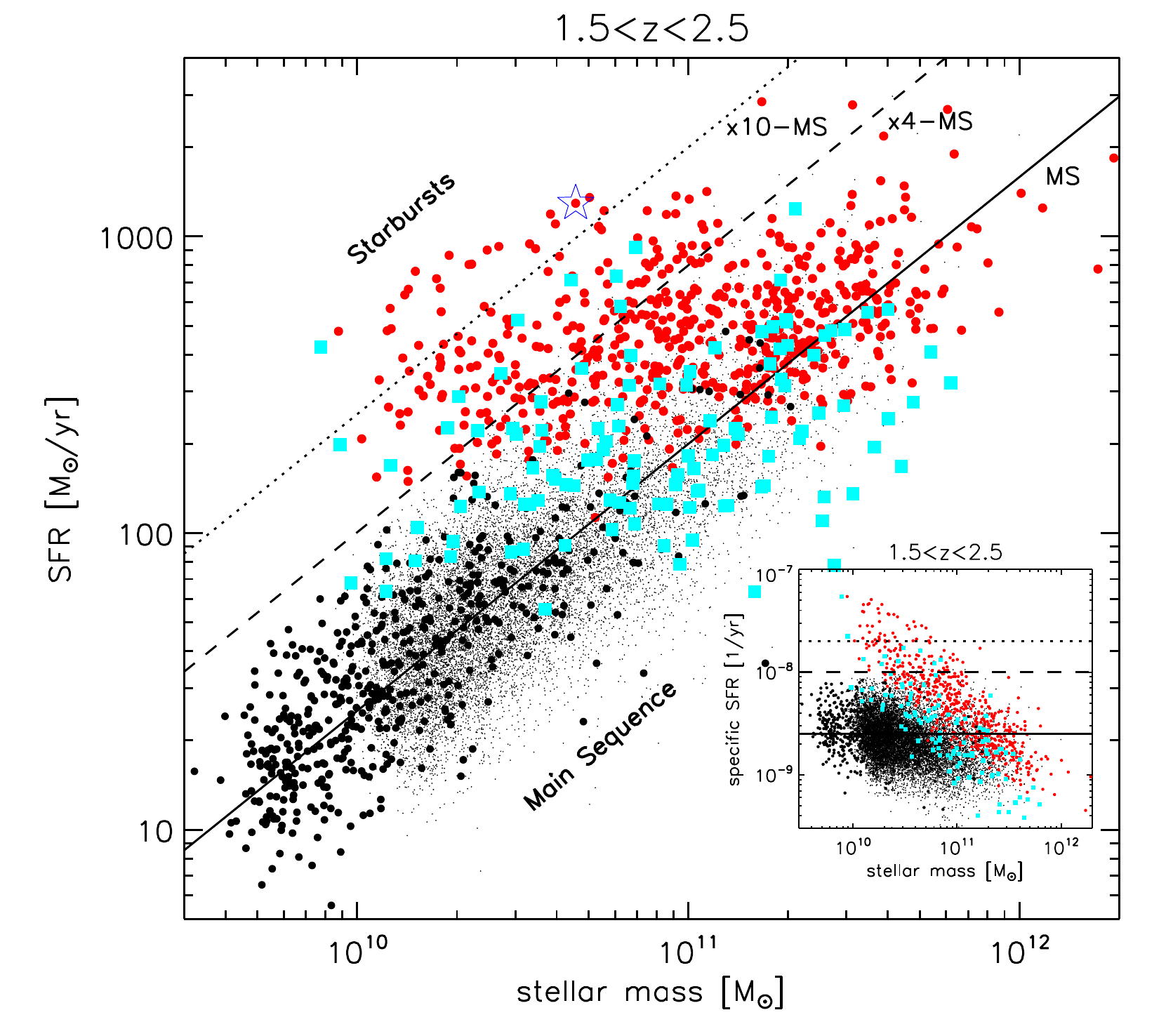} 
\caption{Evolution of the specific SFR  (SSFR) of galaxies of
  stellar mass $0.2-1\times 10^{10} \rm \,M_\odot$ (\citealp{WND11},
\emph{top}); SFR  for galaxies from different samples (different color
symbols), highlighting the ``Main Sequence'' and a population of 
starbursts, with
SSFR mass-dependence inset (\citealp{Rodighiero+11}, \emph{bottom}). 
}
\label{ssfr}
\end{figure}

As shown in Fig.~\ref{ssfr}, 
the star formation time-scale (or 1/SSFR) goes from the MW
value of $\sim 10\,\rm Gyr$ at low redshift to $\sim 0.5\,\rm Gyr$ at $z\simgt 2$. This
result suggests two distinct feedback-regulated
modes of star formation: at low redshift via SNe and without  AGN, and
at high redshift  with, most plausibly,  quenching and possibly  triggering
by AGN playing a central role. One would expect a transition between these
two modes as the AGN duty cycle becomes shorter beyond $z\sim 1.$ 

A related triggering mechanism appeals to enhanced rate of merging at high $z$
\citep{KS11}. Alternatively, it has been argued that intensified halo cold
gas accretion at early epochs may account for all but the most the extreme
SFRs at high $z$, although this may require an implausibly
high SFE \citep{Dekel+09}.

\subsection{Spheroidal galaxies}
The baryon fraction is far from its primordial value in all systems other than massive galaxy clusters. SNe cannot eject significant amounts of gas from massive galaxies. 
Baryons continue to be accreted over a Hubble time and the stellar mass grows. One consequence is  that massive galaxies are overproduced in the models, and that the massive galaxies are also too blue. 

Galaxies like the MW have peanut-shaped  pseudobulges, in contrast with 
the classical  bulges of more massive spirals.  If formed by secular gas-rich disk instabilities, they
should have an age distribution similar to that of the old disk. However the
formation time would be at least $\sim 1 \rm\, Gyr.$ The elevated $\alpha/[\rm Fe]$ 
ratio of our bulge favors a shorter formation time. This would be more
consistent with an early disk instability phase reminiscent of that
associated with clumpy gas-rich galaxies observed at $z\sim 2.$ Massive clump
merging provides a possible solution for forming bulges at a relatively late
epoch \citep{CDB10}. 

However the time-scale (several Gyr) is too long to
result in the enhanced $\alpha/[\rm Fe] $ ratios characteristic of massive
spheroids, or even the less extreme enhancement in the MW bulge.
The shorter time-scales required arise in more plausible cosmological
 initial conditions that result in a redshift $z>2$ for pseudobulge formation
 \citep{Okamoto12}.

\subsection{The role of AGN }

SNe have little impact on the formation of massive galaxies.  Feedback
from SN explosions fails to stop the streaming of cold flows towards
the centre~\citep{PSD11}. The SN ejecta tends to be driven out with
only modest interaction with, and entrainment of, cold infalling gas. A more
coherent and effective interaction is provided by AGN feedback from
supermassive black holes (SMBH).
\begin{figure}[ht]
\centering
\includegraphics[width=0.5\hsize]{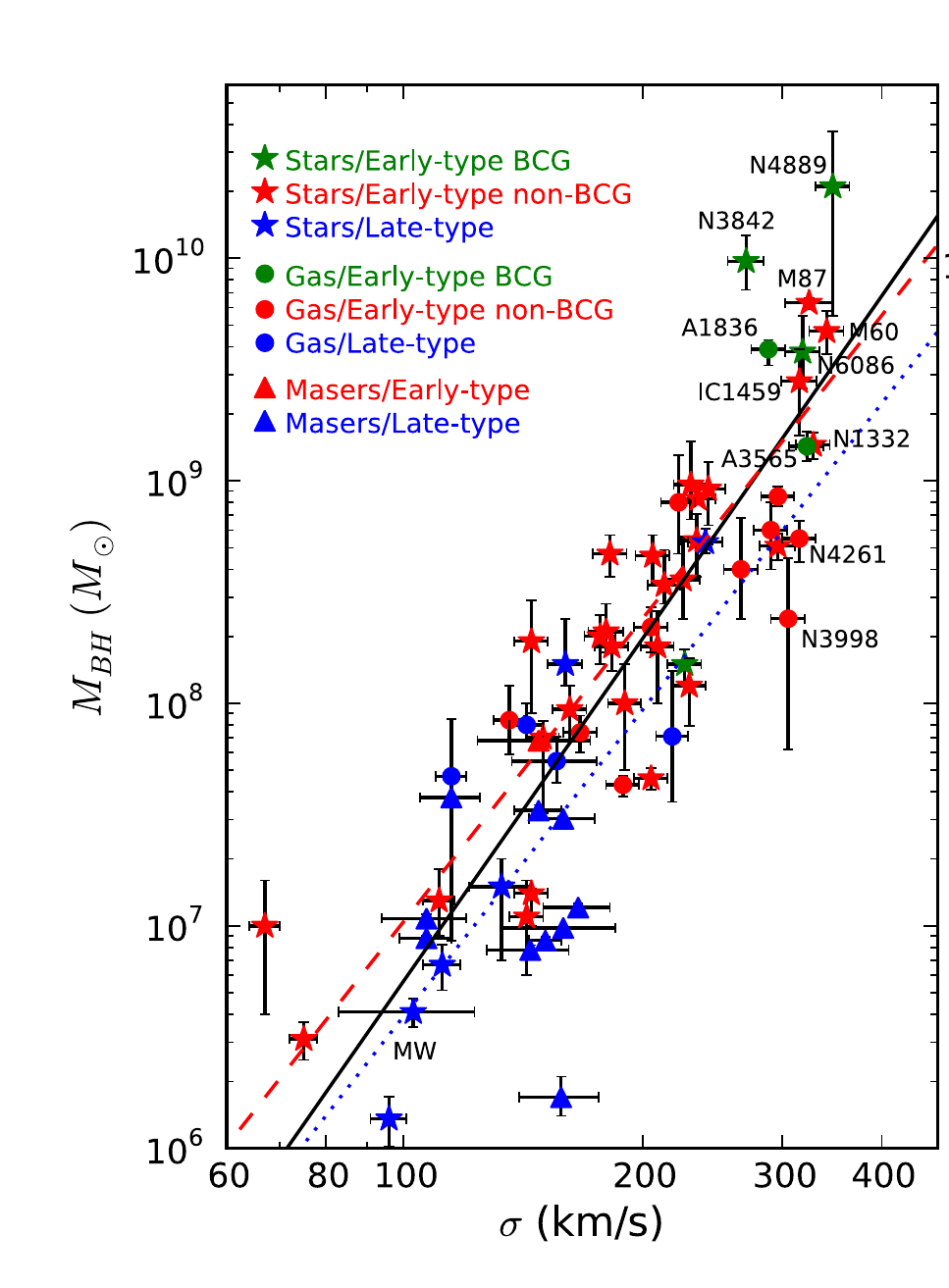} 
\caption{Black hole mass versus spheroid velocity dispersion
  (luminosity-weighted within one effective radius), from
  \cite{McConnell+11}}
\label{smbhscale}
\end{figure}
A clue towards a solution for these dilemmas comes from the accepted
explanation of the \citeauthor{Magorrian+98} relation, which relates SMBH 
mass to spheroid mass \citep{Magorrian+98} and velocity dispersion
(\citealp{FM00}, see Fig.~\ref{smbhscale}).
  
This requires collusion
between black hole growth and the initial gas content of the galaxy when the
old stellar spheroid formed. One conventionally appeals to outflows from the
central black hole that deliver momentum to the protogalactic gas. When the
black hole is sufficiently massive, the Eddington luminosity is high enough
that residual gas is ejected.  An estimate of the available momentum supply
come from equating the Eddington momentum with self-gravity on circumgalactic
gas shells, $L_{\rm Edd}/c=4\pi G M/\kappa = GMM_{\rm gas}/r^2$, where
$\kappa$ us the opacity. Blowout occurs and star
formation terminates when the SMBH--$\sigma_v$ relation saturates. This
occurs for $M_{\rm BH}\propto\sigma_v^{4}$, close to the observed slope of
$\ga 5$
\citep{GOAC11}, and gives the correct
normalization of the relation, at least in order of magnitude. This is the early feedback quasar mode.

There is also a role for AGN feedback at late epochs, when the AGN radio mode
drives jets and cocoons that heat halo gas, inhibit cooling, resolve the
galaxy luminosity function bright end problem and account for the red colors
of massive early-type galaxies. AGN feedback in the radio mode may also
account for the suppression in numbers of intermediate mass and satellite
galaxies (e.g., \citealp{Cattaneo+09} and references therein).  
Feedback from AGN in the host galaxies also preheats the halo gas
that otherwise would be captured by satellites.

\subsection{Galaxies downsize} 
\begin{figure}[ht]
\centering
\includegraphics[width=0.49\hsize]{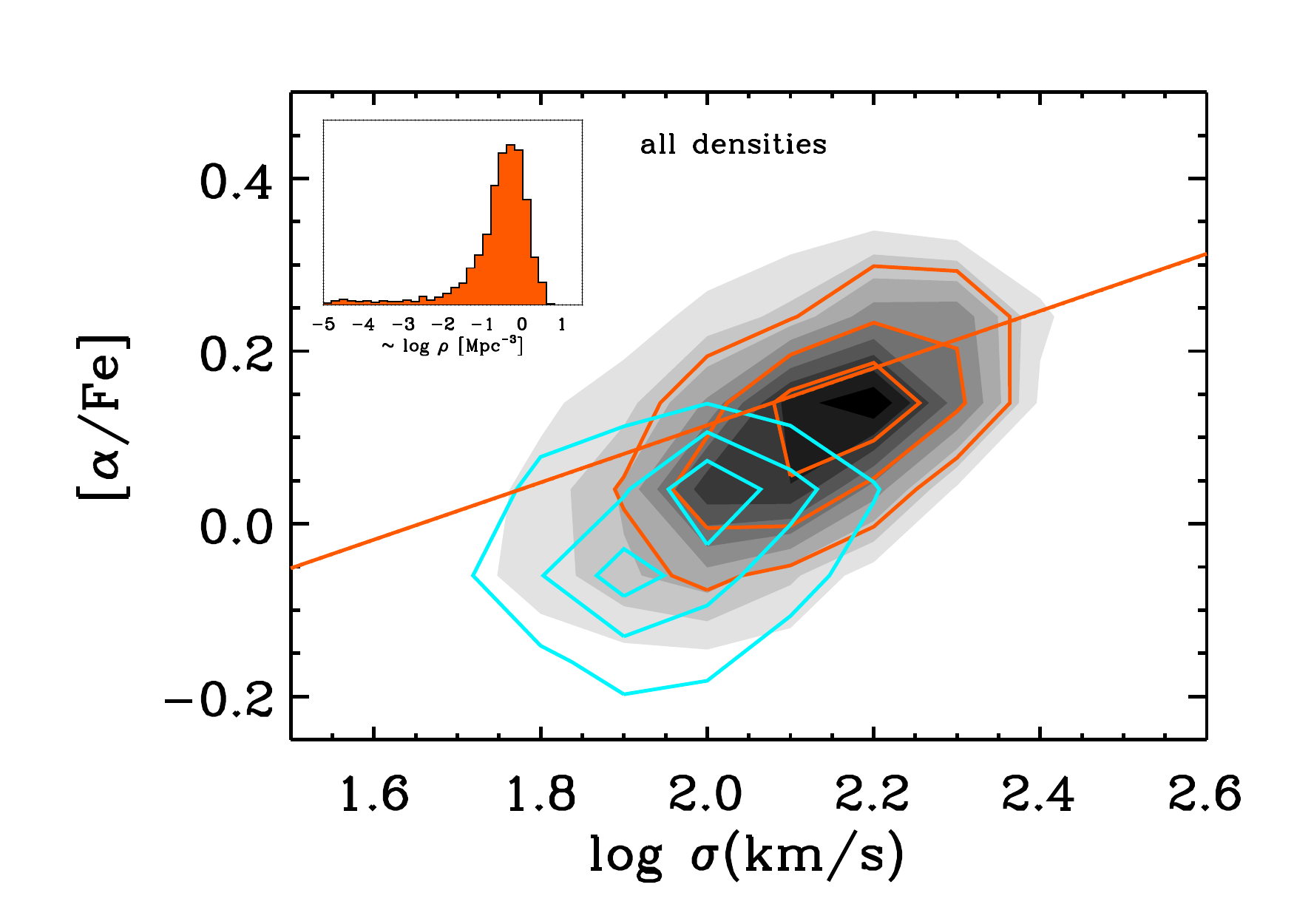} 
\includegraphics[width=0.49\hsize]{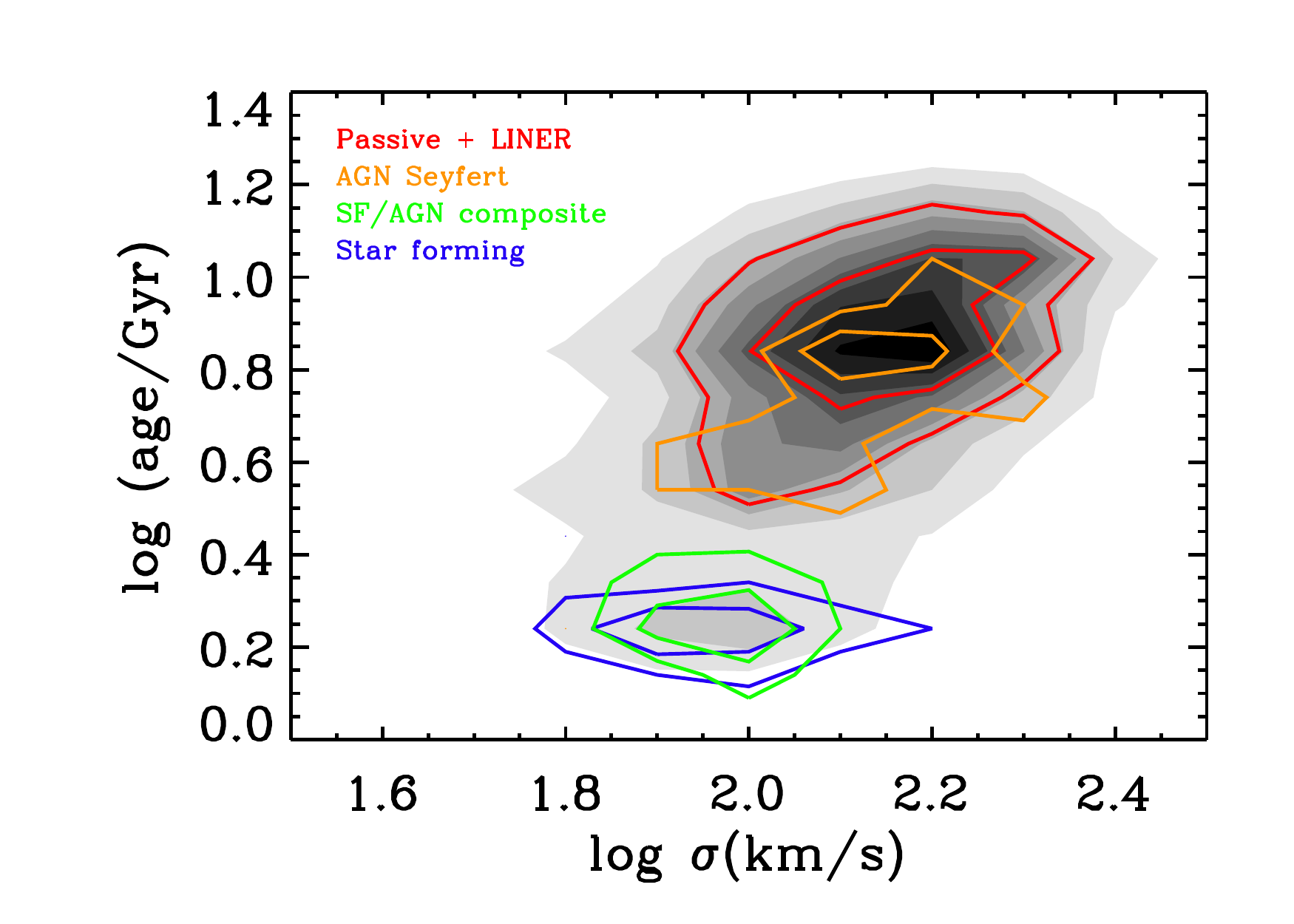} \\
\includegraphics[width=0.52\hsize]{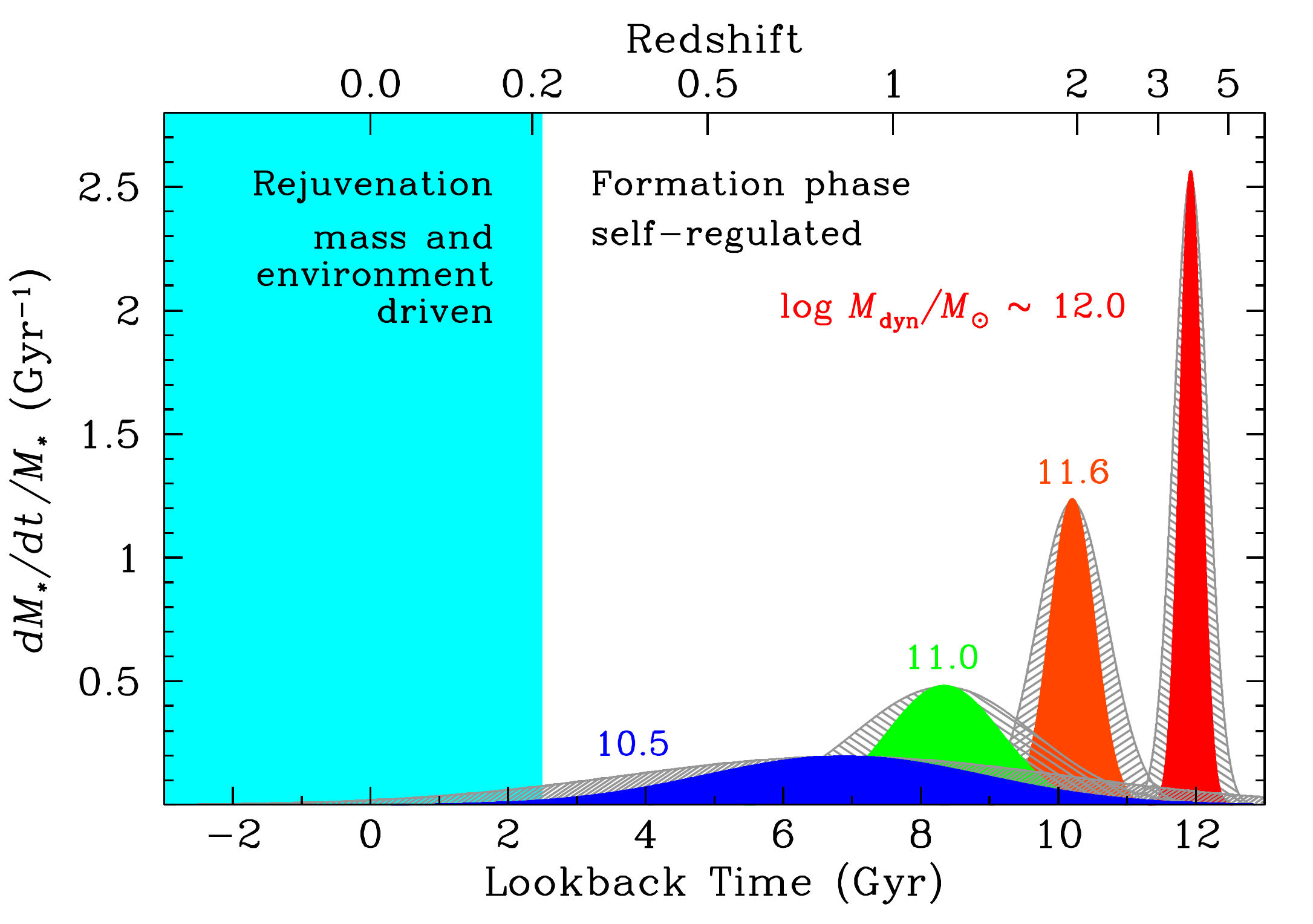} 
\caption{Metallicity ratio and age versus galaxy velocity dispersion
  (i.e. mass) and history of star formation \citep{Thomas+10}}
\label{figdownsize}
\end{figure}

Our understanding of galaxy formation is driven by observations. 
Prior to 2000 or so, it was accepted that hierarchical galaxy
formation predicted that small galaxies form prior to massive
galaxies. 
The first indications that this was in error came from the recognition that
more massive early-type galaxies have redder colors \citep{deVaucouleurs61},
higher metallicities \citep{Faber73} and enhanced 
$[\alpha]/[\rm Fe]$ metallicity ratios \citep{Ziegler+05}, indicative of an older
stellar population with a shorter star formation
time (see Fig.~\ref{figdownsize}). 
This effect is called \emph{downsizing}, as
the most massive galaxies have their stellar populations in place early. 
In effect, we have a cosmic clock: incorporation into stars of debris
from SNe~II ($ \simlt 10^8$ yr) versus SNe~I ($\simgt 10^{9}$ yr) provides a
means of dating the duration of star formation.  
This result was soon
followed by infrared observations that showed that stellar mass assembly
favored more massive systems at earlier epochs \citep{Gonzalez+11}. 

\subsection{Morphological evolution}
We cannot do justice in this review to a largely phenomnological discussion
of morphological evolution of both disk and irregular galaxies. Here,
observations are far ahead of theory.  However, there are strong arguments to
support a continuous sequence between dwarf spheroidal galaxies and 
S0 galaxies
\citep{KB12}. The transformation
applies to the disk components and may involve ram pressure stripping of cold
gas \citep{GG72} as well as galaxy harrassment \citep{MLK98}. This sequence seems to acts
in parallel to the pseudobulges or bulges of S0 galaxies being generated via
stripped/harassed or simply starved disk galaxies \citep{KB12}.

\section{Methods}

\subsection{Observational surveys}

The fundamental driver of progress in astronomy is through
observations. The advent of large galaxy surveys, either wide spectroscopic surveys
probing the nearby Universe (e.g., SDSS) or narrower surveys using
photometric redshifts and often in the infrared domain (e.g., with Spitzer and
Herschel) to probe distant
galaxies in the optical and near-infrared domains, has led to formidable
progress in understanding galaxy formation.
Nevertheless, it is difficult to link the galaxies we see at high redshift
with the ones we see in local Universe, and one is prone to Malmquist bias,
as well as aperture and other selection effects.

\subsection{Semi-Analytical Models}

Several simulation techniques have been developed to be able to link galaxies
from the past to the present, and to obtain a statistical view of the variety
of the evolution histories of galaxies, in terms of star formation, stellar
mass assembly and halo mass assembly.

\begin{figure}[ht]
\centering
\includegraphics[width=7.5cm]{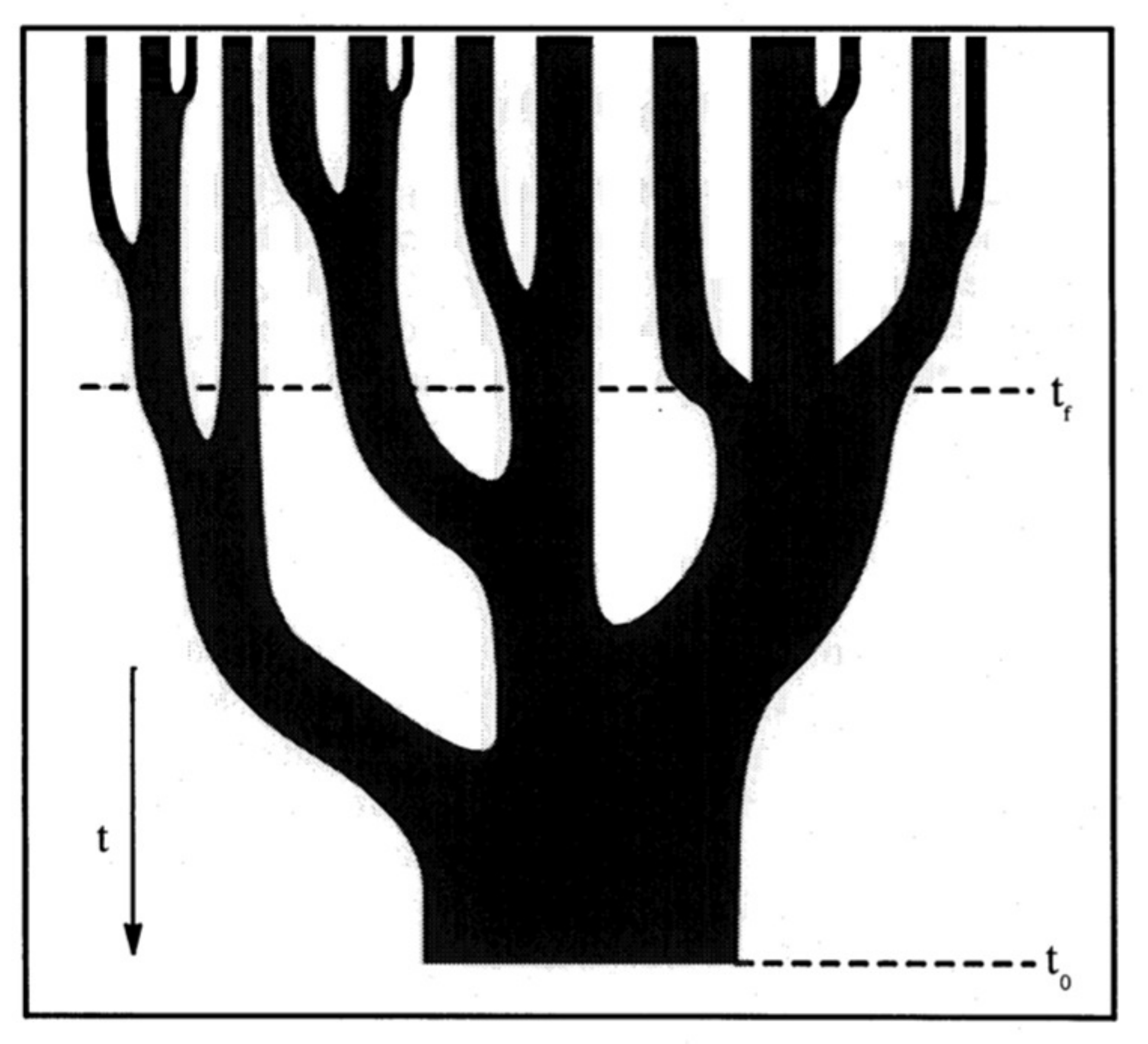}
\caption{Illustration of halo merger tree  \citep{LC93} showing the
  progenitors of a halo selected at time $t_0$}
\label{tree}
\end{figure}

Given current computational constraints,  it is impossible to achieve the
sub-parsec or finer  resolution needed to adequately model star formation and
accretion onto black holes in
a cosmological simulation. Theorists have invented a swindle, wherein the
complex processes of star formation and accretion onto SMBH 
are  hidden inside a black box called
``sub-grid physics'' that can be tagged onto a large-scale simulation. 
In SAMs, galaxies are ``painted'' on halos built from halo merger trees or
detected in cosmological dissipationless (dark matter only) simulations.
The former (see Fig.~\ref{tree}) produce the mass assembly history (MAH) of halos with the
condition that they end up in a halo of mass $M_0$ at epoch $z_0$ (usually
$z_0=0$). The branches are drawn from random samplings given the known
conditional probabilities arising from extensions \citep{Bower91,LC93} and
modifications of the \cite{PS74} formalism. Halo merger trees have the
advantage of being rapid to compute, but lack positional
information. Cosmological simulations, with
up to $10^8$ particles, are becoming increasingly common, but are intensive to
process, in particular to detect halos (\citealp{Knebe+11} and references
therein) and subhalos (\citealp{Onions+12} and references therein) and build halo merger trees.

Once the halo MAH is identified, one follows the branches of the tree from
past to present, to build galaxies. The galaxy formation recipe includes
several ingredients: 
\begin{enumerate}
\item
The gas cooling time must be short for the gas to
dissipatively cool into a disk. 
In particular, gas cannot fall onto low-mass halos because of the cooling
barrier 
and falls less efficiently onto
high-mass halos because of a virial shock, whereas gas can infall along cold
filaments on lower mass halos.
\item Star formation occurs at a rate $\dot m = {\rm cst}\, m_{\rm \rm gas}/t_{\rm
  dyn}$, where $t_{\rm dyn}$ is a measure of the dynamical time of the
  galaxy.
\item Feedback from SNe and from the relativistic jets arising from
  central SMBHs hiding as AGN
  heats up the surrounding gas.
\item While the gas settles into disks, major mergers of galaxies cause disks
  to transform into ellipticals, and after subsequent disk build-up, the
  merger remnant is identified to a bulge inside a spiral galaxy. The bulge
  can also be built-up by repeated minor mergers, as well as starbursts and
  secular evolution of the disk.
\item The SAM keeps track of star formation times and predicts galaxy 
  luminosities in different wavebands using population synthesis codes.
\item When a smaller halo enters a larger one, it becomes a subhalo, its
  galaxy becomes a satellite, and
  usually, the gas that attempts to fall onto the subhalo is now directed
  towards the central galaxy of the halo.
\item Satellite galaxy trajectories are assumed to be  those of the
  subhalos they belong to, and when they are no longer resolved in the
  cosmological simulation, or if one is only using a halo merger tree, the
  galaxies are merged with the central galaxy on a dynamical friction timescale
  (calibrated on simulations, e.g. \citealp{Jiang+08}).

\end{enumerate}

An advantage of SAMs is that it is easy to gauge the importance of various
physical processes by seeing how the outcome is changed when a process is
turned off in the SAM.

Present-day SAMs are increasingly complex, and SAMs can include up to $10^5$
lines of code. The popularity of SAMs has increased with public-domain outputs
\citep{Bower+06,Croton+06,DLB07,Guo+11} and codes \citep{Benson12}.

\subsection{Hydrodynamical simulations}

The weakness of SAMs is that much of the physics is controlled by hand (except for
gravity, when the SAMs are directly applied to cosmological N-body
simulations of the dark matter component). Hydrodynamical simulations provide
the means to treat hydrodynamical processes in a much more self-consistent
manner, and cosmological hydrodynamical simulations have been run for nearly
25 years (starting from \citealp{Evrard88}).

These simulations come in two flavors: cell-based and
particle-based. It was rapidly realized that cell-based methods could not
resolve at the same time the very large cosmological scales and the small
scales within galaxies, and the early progress in the field was 
driven by the Smooth Particle Hydrodynamics (SPH) method
\citep{GM77,Monaghan92,Springel10_ARAA}, in which the 
diffuse gas is treated as a collection of particles, whereas the physical
properties (temperature, metal content, etc.) are smoothed over neighboring
particles using a given SPH-smoothing kernel.
Despite early successes (comparisons of different hydrodynamical codes by
\citealp{Frenk+99}), SPH methods fail to 
resolve shocks as well as Rayleigh-Taylor and Kelvin-Helmholtz
instabilities \citep{Scannapieco+12}. 
This has brought renewed popularity to  cell-based methods,
with the major improvement of resolution within the Adaptive Mesh Refinement
(AMR) scheme \citep{KKK97,OShea+04,Teyssier02}, where cells can be refined into
smaller cells following a condition on density or any other physical
property.
Moreover, schemes with deformable cells (that do not follow the Cartesian
grid) were developed 20 years ago, and are now becoming more widely used
(e.g., AREPO, \citealp{Springel10}).

In these hydrodynamical codes, stars can be formed when the gas is
sufficiently dense, with a convergent flow and a short cooling time. 
Current codes do not have sufficient mass resolution
to resolve individual stars, so the star particles made from the gas are
really collections of stars, with an initial mass function. One can therefore
predict how many core-collapse SNe will explode after the star
particle forms, and the very considerable SN energy is usually
redeposited into the gas by adding velocity kicks to the neighboring gas
particles and possibly also thermal energy. Similarly, AGN can be implemented,
for example by forcing a Magorrian type of relation between the SMBH
mass and the spheroidal mass of the galaxy, and feedback from AGN jets can be
implemented in a similar fashion as is feedback from SNe.
Hydrodynamical codes are therefore not fully self-consistent, as they 
include semi-analytical recipes for the
formation of stars and the feedback from SNe and AGN.

\subsection{New methods}

\subsubsection{Analytical models}

The growing complexity of SAM codes (e.g., the publicly-available GALCTICUS
code of \citealp{Benson12} contains over 120$\,$000 lines of code and
involves over 30 non-cosmological parameters with non-trivial values) has led
some to seek simpler descriptions of galaxy formation.

One level of simplicity is to parameterize the time derivatives of the
different components of galaxies (stars, cold gas, hot gas, dark
matter) as linear combinations of these parameters \citep{NW10}. But one can
go to an even simpler level and characterize the fraction of gas that can cool
\citep{BVM92} or the mass in stars \citep{CMWK11} as a function of halo mass
and epoch. Although such approaches are much too simple to be able to capture
the details of galaxy formation, they are sufficient to study simple
questions. 

Assuming that all the gas in the range of temperatures between $10^4\,\rm K$
and the maximum where the cooling time is shorter than the age of the
Universe (or the dynamical time) effectively cools, and that the gas is
replenished one the timescale where halos grow, \cite{BVM92} showed that
nearly all the baryons should have converted to stars by $z=0$. Since this is
not observed, this simple calculation shows that feedback mechanisms are
required to prevent too high star formation.

\begin{figure}[ht]
\centering
\includegraphics[width=10cm,angle=-90]{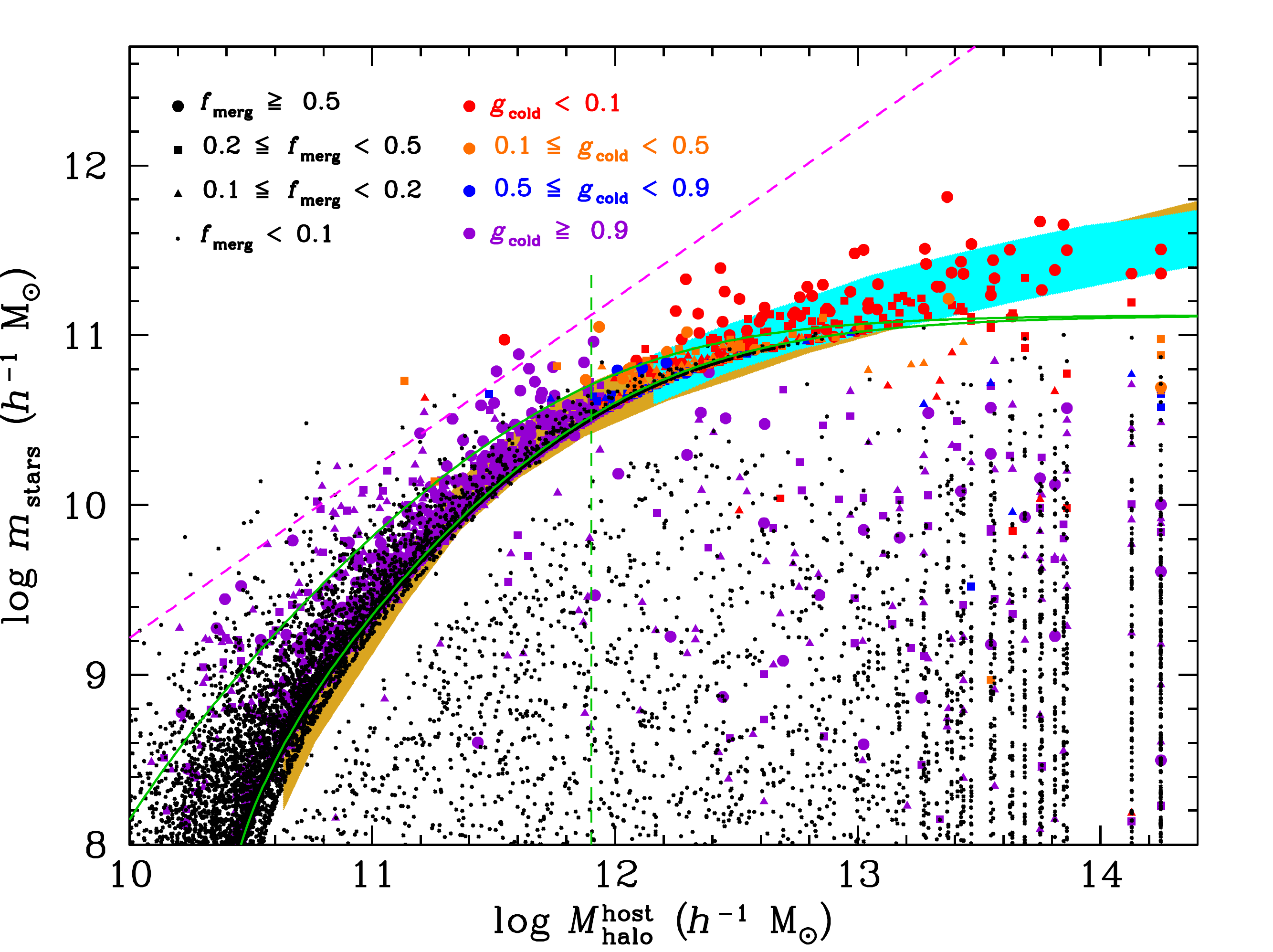} 
\caption{Stellar versus host halo mass from analytical model by \cite{CMWK11} run
  on dark matter simulation. The shaded regions are results obtained from the
  conditional luminosity function (\emph{blue}, \citealp{YMvdB09}) and
  abundance matching (\emph{gold}, \citealp{GWLB10}).
The \emph{large symbols} denote galaxies who have acquired most of their stellar mass
through mergers (rather than smooth gas accretion). The \emph{green curves}
show the galaxy formation model at $z=0$ and $z=3$.
}
\label{mvsMCattaneo}
\end{figure}

\cite{CMWK11} 
apply their simple galaxy formation prescription onto the halos
of a high-resolution cosmological $N$ body simulation and
reproduce the $z=0$ observed stellar mass function with only four parameters,
despite the overly simplistic model. They find a fairly narrow 
stellar versus halo mass relation for the dominant (``central'') galaxies in
halos and a gap between their stellar masses and those of the satellites, in
very good agreement with the relations obtained by \cite{YMvdB09} from the
SDSS using conditional stellar mass functions (Fig.~\ref{mvsMCattaneo}). This gap is remarkable, as it
is less built-in
\citeauthor{CMWK11}'s method than it is in SAMs: for example, halos with two-dominant galaxies
(such as observed in the Coma cluster) are allowed.
Similar ``successful'' analytical models have been proposed by \cite{Peng+10} and \cite{Bouche+10}.

\subsubsection{Halo Occupation Distribution}

A simple way to statistically populate galaxies inside halos, called
\emph{Halo Occupation Distribution} (HOD) is to assume a
functional form for some galaxy statistic in terms of the halo
mass. 
\begin{figure}[ht]
\centering
\includegraphics[width=0.42\hsize]{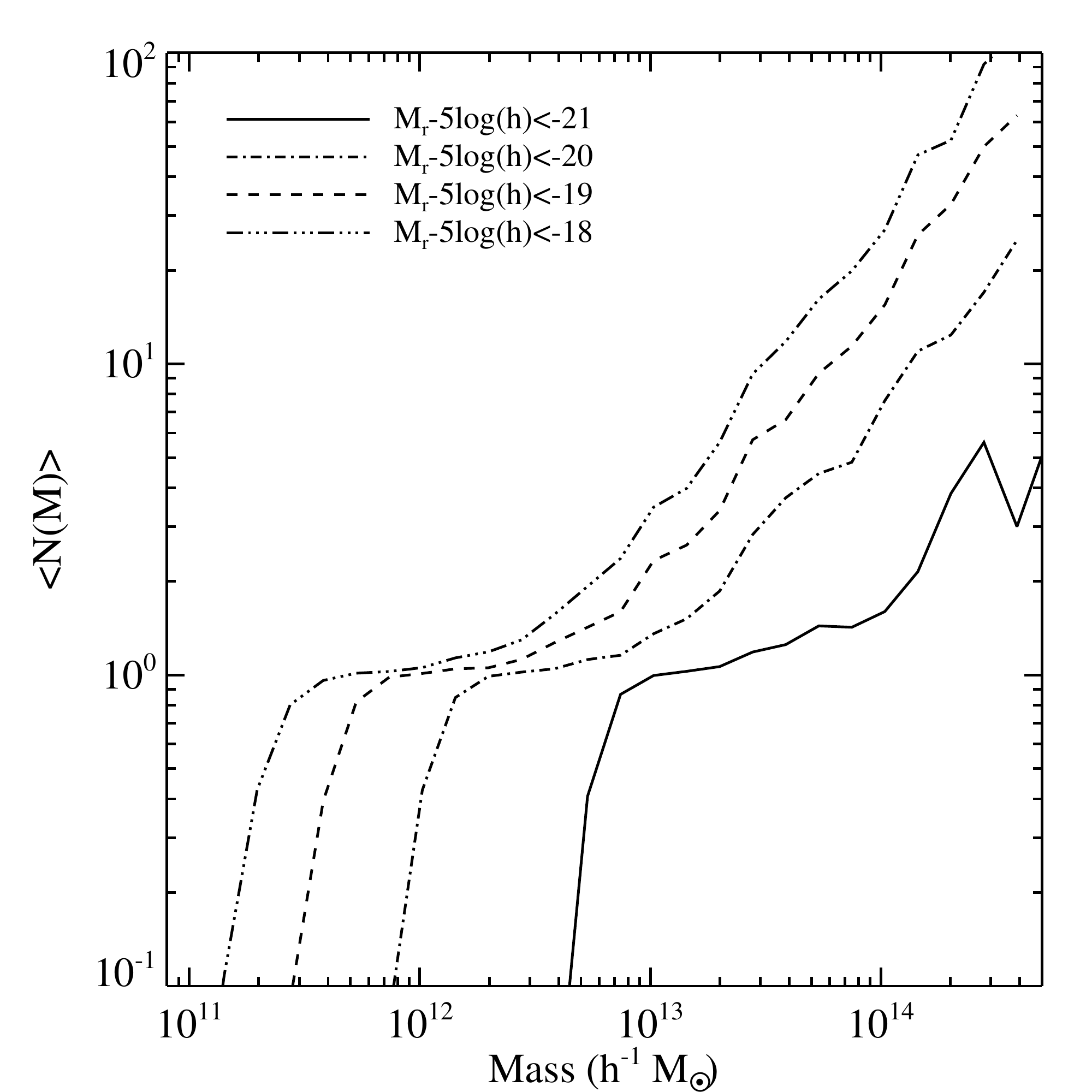}
\includegraphics[width=0.42\hsize]{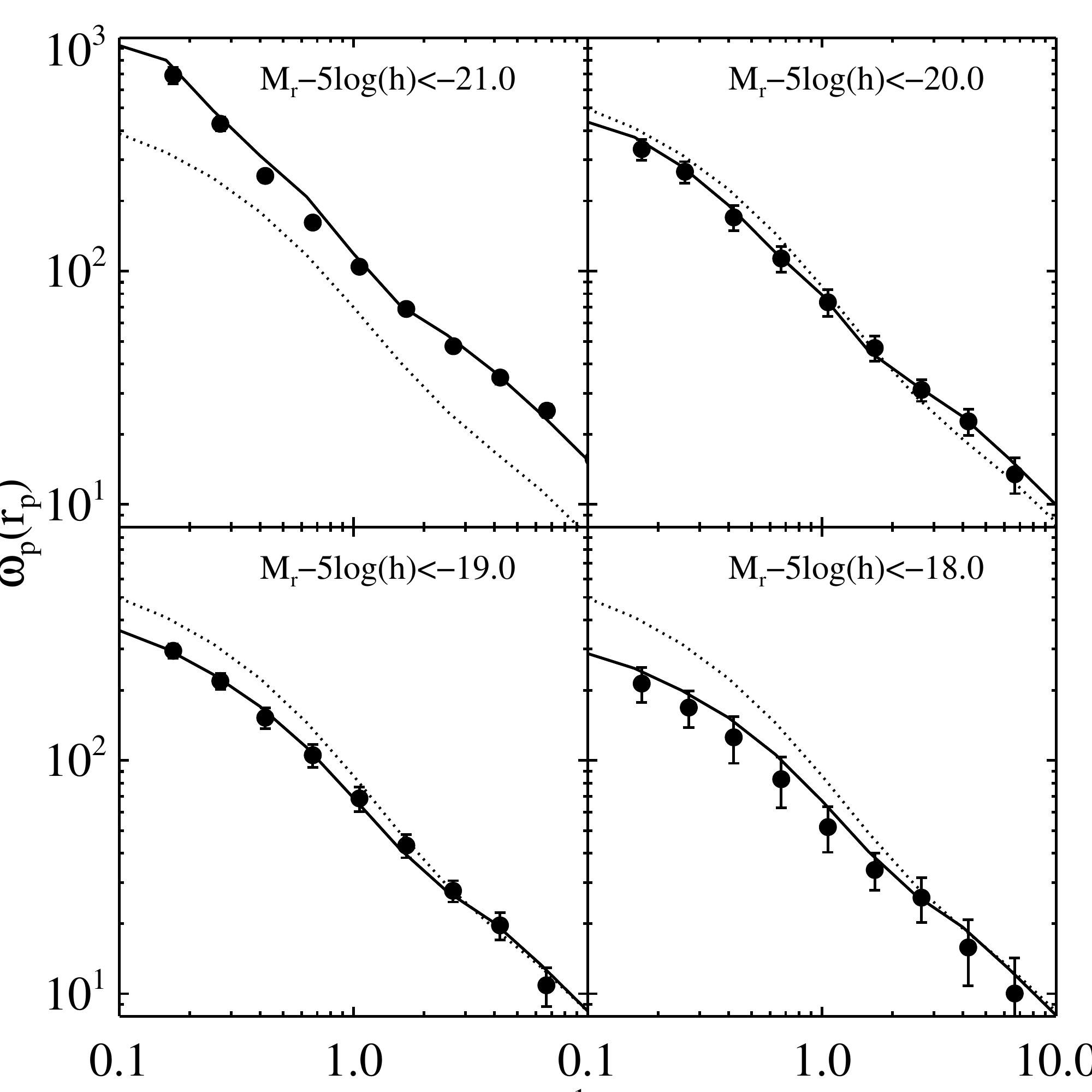}\\
\includegraphics[width=0.6\hsize,angle=-90]{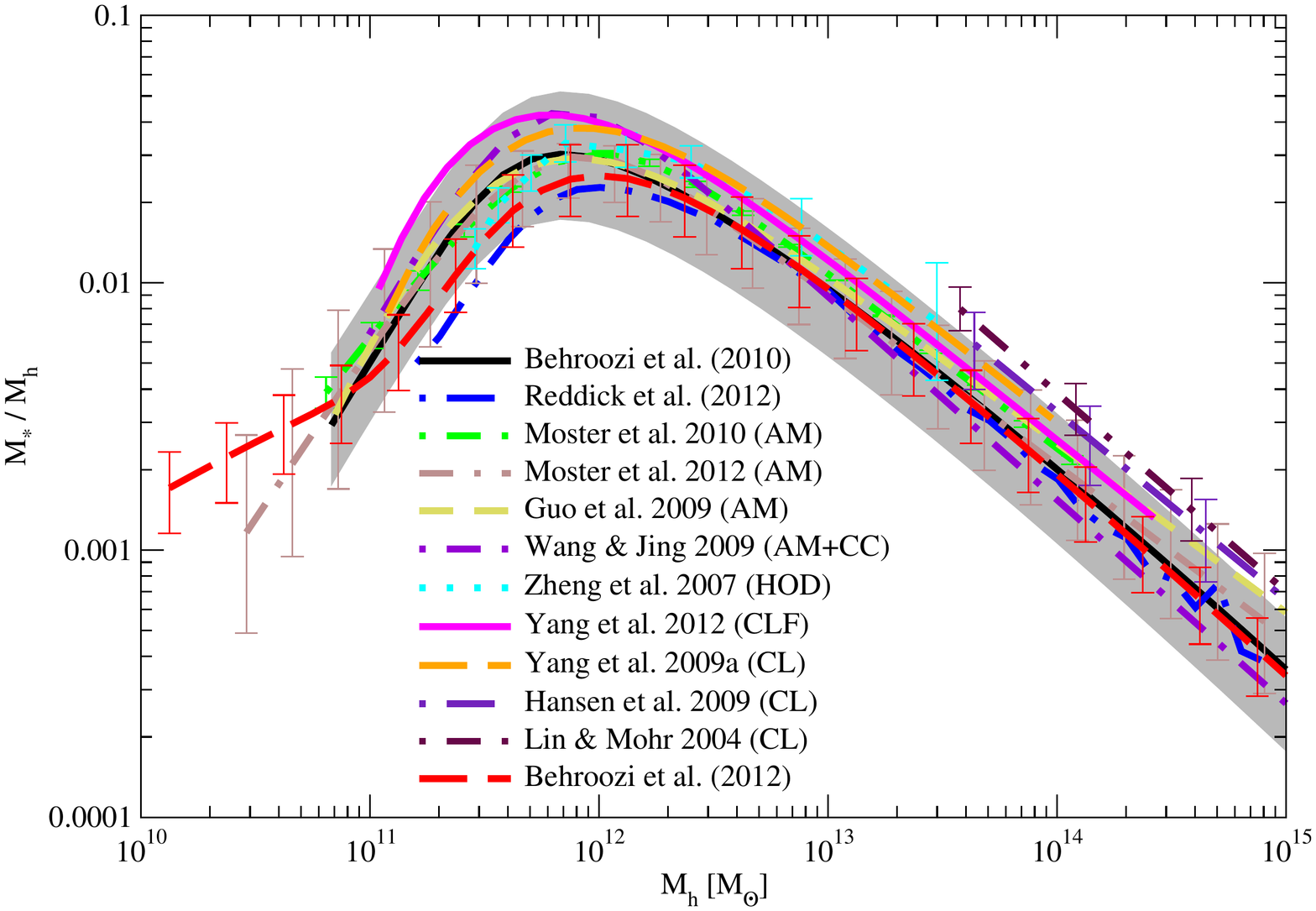} 
\caption{\emph{Top left}: 
Illustration of HOD models of multiplicity functions (per halo) obtained from
abundance matching \citep{CWK06}.
\emph{Top right}:
Abundance matching prediction on the galaxy correlation function compared to
SDSS observations (\emph{symbols}), while the halo-halo correlation function
is shown as \emph{dotted lines} \citep{CWK06}.
\emph{Bottom}:
Halo mass for given stellar mass obtained by abundance matching (AM), HOD,
conditional luminosity function (CLF) and group catalogs (CL) at $z=0.1$ \citep{BWC12}. 
The \emph{shaded region} shows the AM analysis of \cite{BCW10}.
}
\label{AMHOD}
\end{figure}
The galaxy statistic can be the multiplicity function (for galaxies
more massive or more luminous than some threshold, \citealp{BW02}, see upper
left panel of Fig.~\ref{AMHOD}), 
the luminosity or
stellar mass function, generically denoted  CLF for conditional luminosity
function \citep{YMvdB03}.
Although these HOD methods have no underlying
physics, they are a very useful tool to derive galaxy trends with halo mass,
or, in other words to find the effects of the global environment on galaxies.

\subsubsection{Abundance Matching}

An offshoot of HOD models is to link the mean trend of some galaxy property
in terms of the mass of its halo, using so-called Abundance Matching (AM).
The idea is to solve $N(>x) = N(>M_{\rm halo})$, i.e. matching cumulative
distributions of the
observed galaxy property, $x$, with the predicted one for halo masses, determined
either from theory \citep{PS74,SMT01} or from cosmological $N$ body
simulations \citep{WAHT06,Tinker+08,CFCG10,Courtin+11}.

Common uses of AM involve one-to-one
correspondences between 
1) stellar and halo mass for central galaxies in halos, 
2) total stellar mass and halo mass in halos, and 
3) stellar and subhalo mass in galaxies.
\cite{MH02} performed the first such AM analysis to determine $M_{\rm
  halo}/L$ versus $L$; they first had to determine the observed cosmic stellar
mass function, not counting the galaxies within groups, but only the groups
themselves \citep{MHG02}.
\cite{GWLB10} used this third approach (called subhalo abundance matching  or
SHAM, and pioneered independently by \citealp{VO06} and \citealp{CWK06}),
to determine the galaxy
formation efficiency $m_{\rm stars}/M_{\rm halo}$ as a function of $M_{\rm
  halo}$, by matching the observed stellar mass function with the subhalo mass
function that they determined in the Millennium \citep{Springel+05} and
Millennium-II \citep{BoylanKolchin+09} simulations.
Although AM methods are based upon a fine relation between stellar and halo
mass, they can easily be adapted to finite dispersion in this relation \citep{BCW10}.

Not only is AM very useful to determine, \emph{without free parameters}, the
relation of stellar to halo mass (lower panel of Fig.~\ref{AMHOD}), but it
superbly predicts the galaxy correlation function of SDSS galaxies
(\citealp{CWK06}, upper
right panel of Fig.~\ref{AMHOD}). 
The drawback of AM methods is that they do not clarify the underlying physics
of galaxy formation.

\section{Results from numerical simulations}

\subsection{General results from semi-analytical models of galaxy formation}

SAMs have been remarkably successful in constructing mock catalogs of
galaxies at different epochs and are used in motivating and in interpreting
the large surveys of galaxies. 
\begin{figure}[ht]
\centering
\includegraphics[width=0.52\hsize]{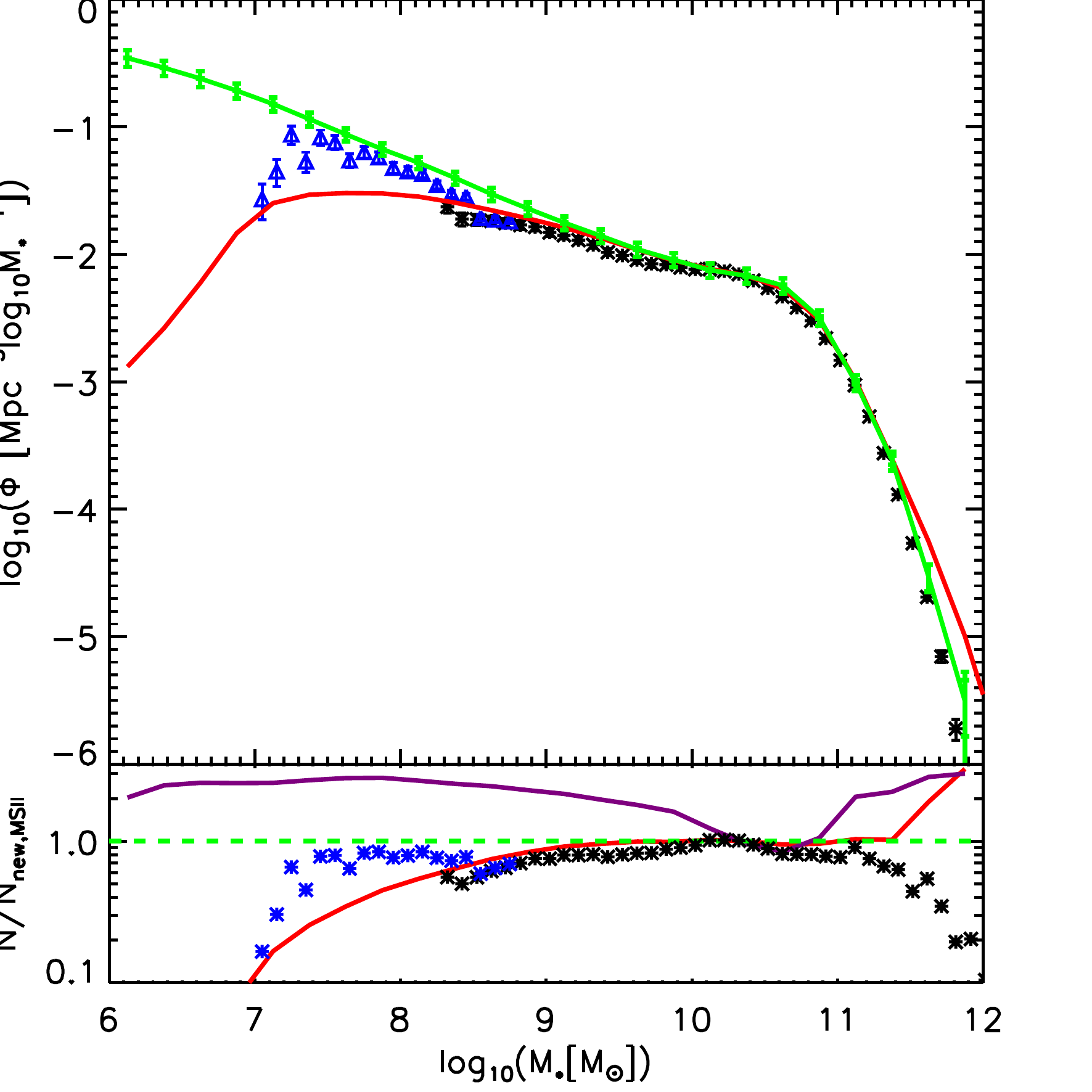}
\includegraphics[width=0.47\hsize]{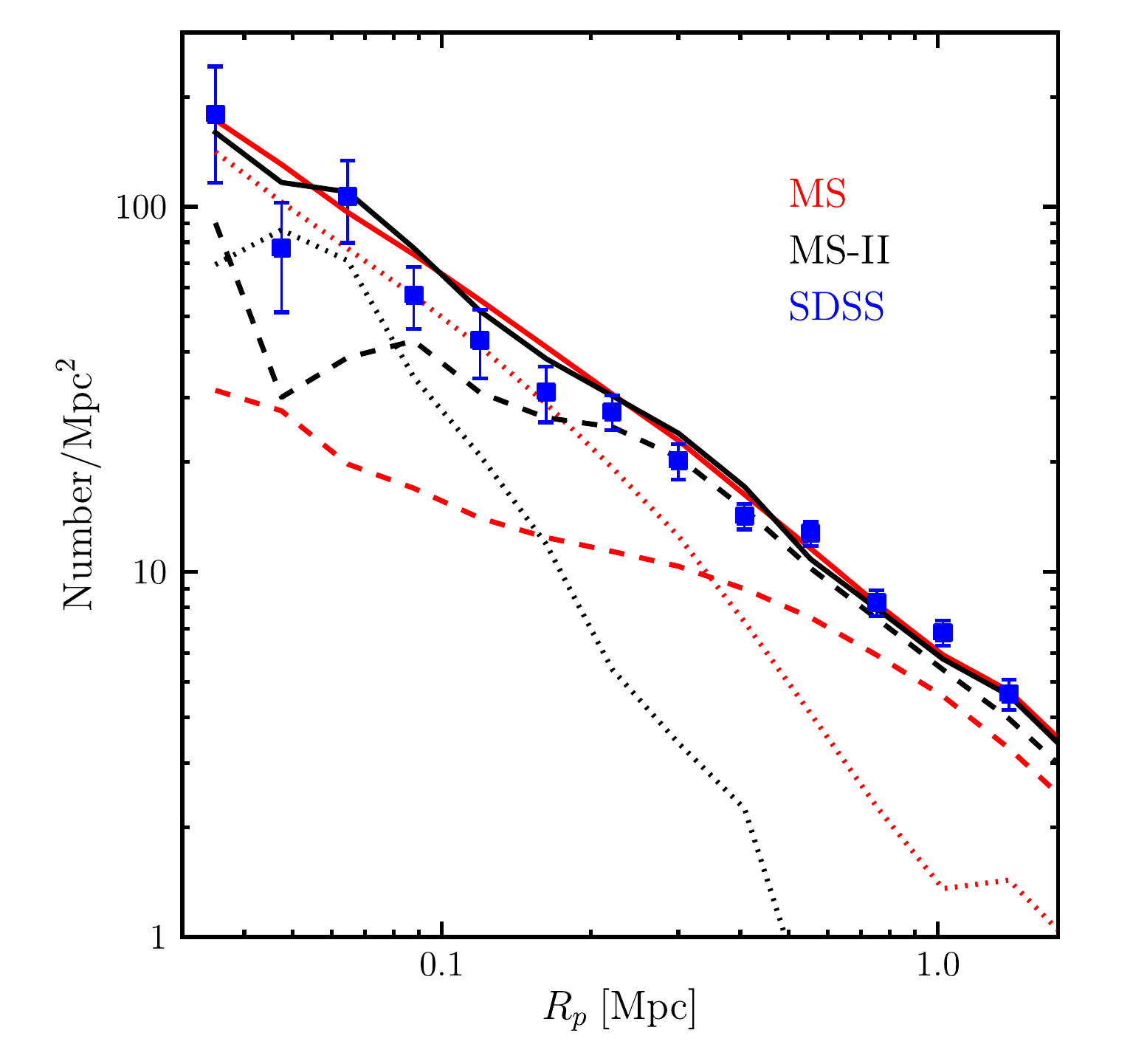}\\
\includegraphics[width=0.5\hsize]{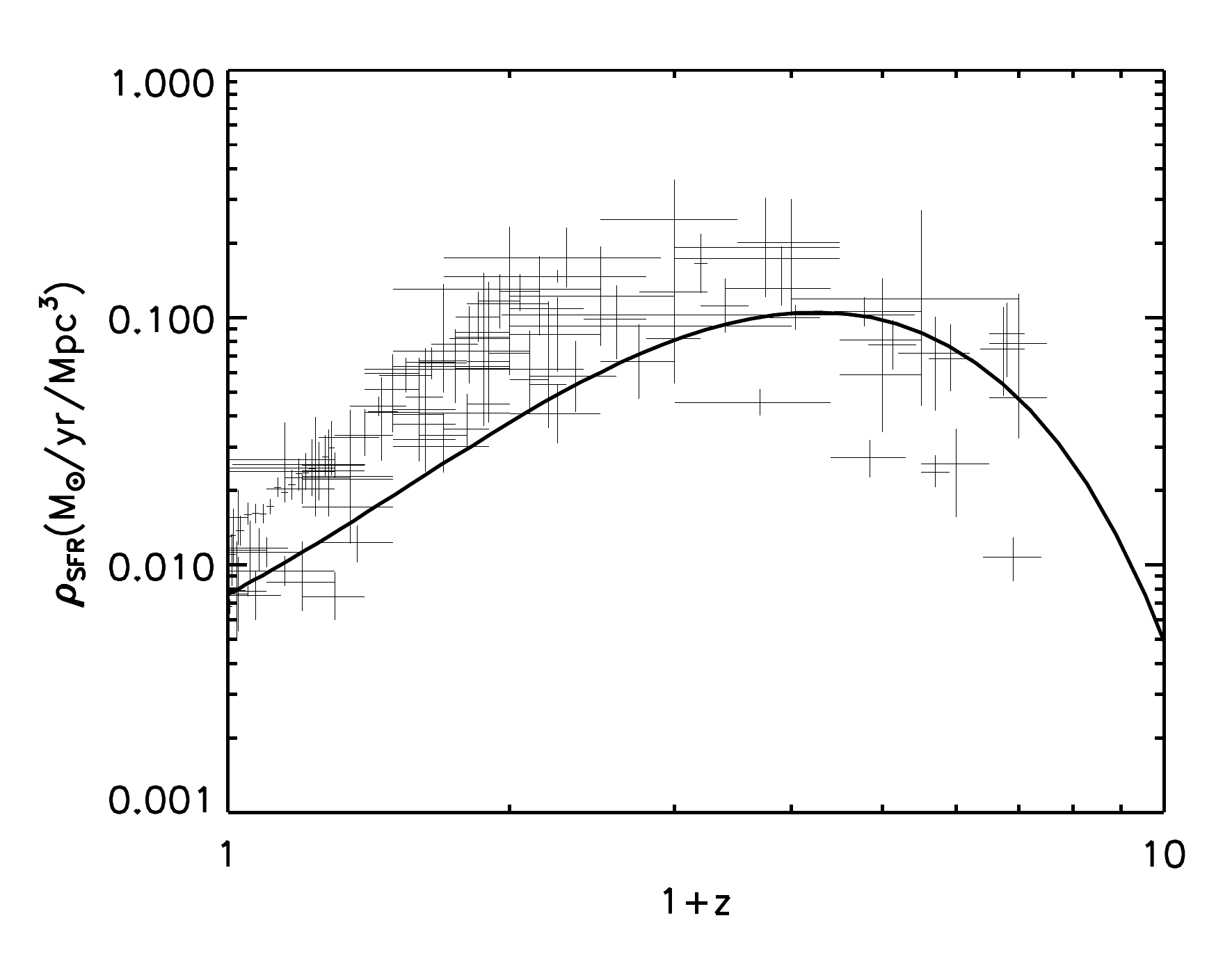}
\includegraphics[width=0.49\hsize]{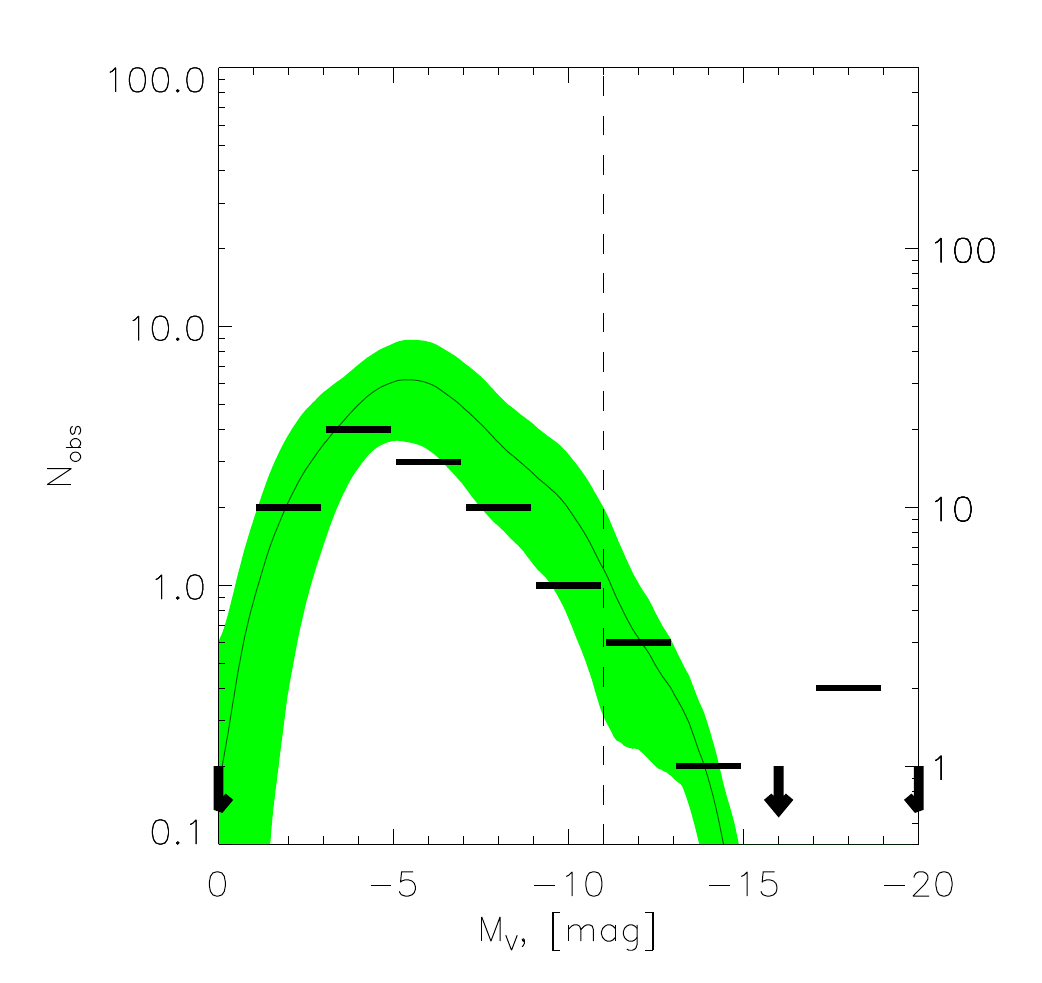}
\caption{Illustrations of predictions of SAMs at $z-0$. 
\emph{Upper left}: Stellar mass functions \citep{Guo+11}: symbols are from SDSS \citep{LW09},
  while curves are from a SAM run on both wide and low-resolution Millennium
  Simulation and on the higher but smaller MS-II simulation. 
\emph{Upper right}: Galaxy correlation functions \citep{Guo+11}.
\emph{Bottom left}: Evolution of the cosmic SFR
\citep{Guo+11}.
\emph{Bottom right}: Very low end of the galaxy luminosity function \citep{Koposov+09}.
}
\label{GuoSAMz0}
\end{figure}
For example, they reproduce very well the
$z=0$ stellar mass function and correlation function  (see Fig.~\ref{GuoSAMz0}).

\begin{figure}[ht]
\centering
\includegraphics[width=\hsize]{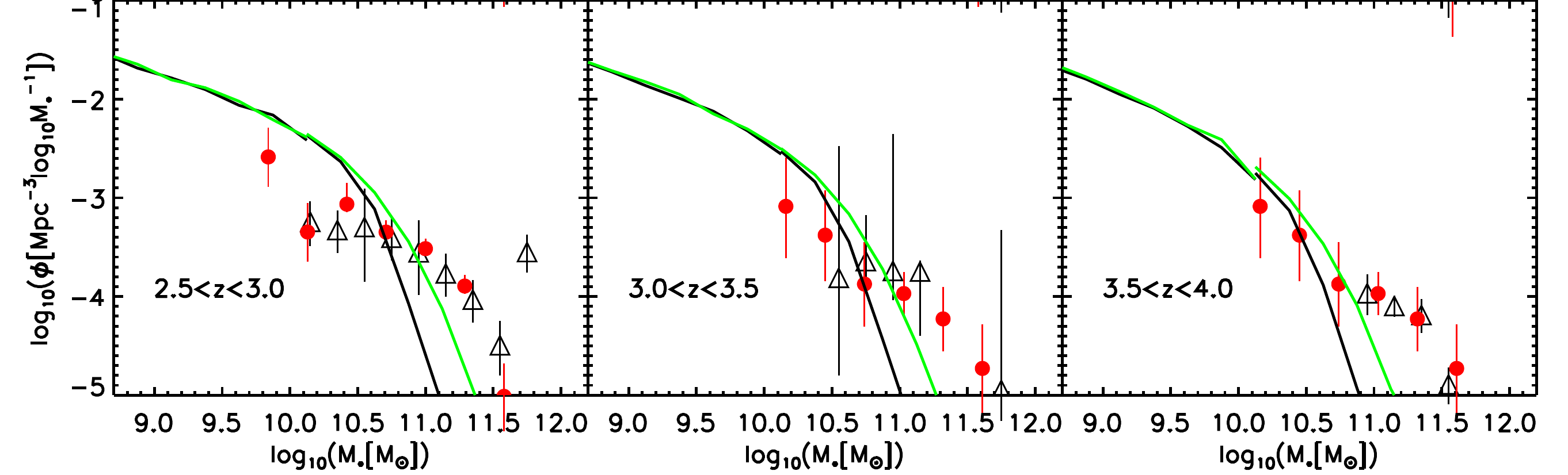}
\caption{Evolution of stellar mass functions predicted by \cite{Guo+11}.
\emph{Open triangles} and \emph{red circles} represent observations by
\cite{PerezGonzalez+08} and \cite{Marchesini+09}, respectively.
\emph{Black} and \emph{green curves} represent the predicted stellar mass functions
of galaxies, respectively  before and after convolving the stellar masses by
0.25 dex measurement errors.}
\label{GuoSAMhiz}
\end{figure}

However, attempts to solve the problems of high redshift galaxies have so far
been woefully inadequate. For example, they cannot reproduce the rapid
decrease in the cosmic SFR since $z=1$ (see Fig.~\ref{GuoSAMhiz}). 
The early SAM feedback models used AGN
quenching, and required excessive dust in early types in the nearby universe
\citep{Bower+06}. Refinements to high redshift attempted to account
simultaneously for galaxy and AGN accounts, and only succeeded by requiring
excessive amounts of dust in order to hide most of the AGNs seen in deep
X-ray surveys \citep{Fanidakis+11}. 
An early indication
that SAMs were entering uncertain territory can be seen in the early
predictions of the cosmic star formation history: as numerical resolution was increased, the predicted SFR
increased without limit \citep{SH03b}. 
This makes one begin to doubt the predictive power
of SAMs.

Clearly, baryon physics is far more complicated than assumed in the early
SAMs of the 1990s. In fact, we still lack an adequate
explanation for the evolution of the stellar mass function. 
Attempts to patch up the problem at low redshift, to avoid an
excess of massive galaxies, exacerbate the inadequacy of the predicted
numbers of massive galaxies at high redshift \citep{Fontanot+09b}. One
attempt to correct the problem at large redshift incorporates for the first
time thermally pulsing AGB (or carbon) stars in the models, and the extra NIR
luminosity reduces the inferred galaxy masses \citep{Henriques+11}.  However
the price is that the lower redshift galaxy count predictions no longer fit
the models.

\subsection{Feedback and dwarfs}

Dwarf spheroidal galaxies are dark matter laboratories, dominated by dark
matter. However the numbers defy interpretation. Feedback is readily adjusted
to reduce the numbers of low mass dwarfs \citep{Koposov+09}, but the most
massive dwarfs predicted by $\Lambda$CDM simulations are not observed \citep{BBK12}.  
This may be a function of the neglect of baryons in the Aquarius simulations: inclusion 
of baryons reduces the central densities of massive dwarfs \citep{Zolotov+12}.
Unorthodox feedback (AGN) may also be a solution \citep{BBK11}.
 Moreover,  most low-mass dwarfs have cores rather than the  cusps  predicted by CDM-only simulations. 
Baryonic feedback may reconcile data on dwarf core profiles with simulations that include star formation and gas cooling \citep{Oh+11, Governato+12}, who find that 
SN-driven outflows help flatten dark matter central density cusps. As mentioned earlier, enhanced early star formation and SN production  creates strong tensions with 
the need for strong late low mass galaxy evolution.
SN feedback at later epochs may turn cusps into cores by sloshing of more recently accreted gas clouds \citep{MCW06},
more recently addressed in \cite{PG12}, who consider bulk gas motions  and require short intense bursts of star formation. There may be evidence for such phenomena in dwarf galaxies \citep{Weisz+12}.


 Multiphase simulations \citep{PSD11} confirm the
effectiveness of SN-driven winds, but find that they do not lead to baryon ejection. In a multi-phase medium
with more realistic filamentary accretion, outflows are only typically $\la 10\%$ of the
gas accretion rate.
 It is not clear whether SN feedback may still
provide enough momentum to yield an acceptable fit to the low mass end
of the galaxy luminosity function for the classical dwarfs.  Ram pressure
stripping \citep{GG72,Mayer+07} remains an alternative or complimentary mechanism,
and morphological transformation of disks into dwarf spheroidals may be
accomplished by repeated rapid encounters, i.e. ``harassment'' \citep{MLK98}
or gravitationally-induced 
resonances \citep{DBCH09}.

SN feedback enables present day disk galaxy properties to be reproduced,
 including the Tully- Fisher relation and sizes, except for massive
 disks. More energetic feedback, from an AGN phase, is envisaged as a
 possible solution \citep{McCarthy+12}. 
Many galaxies, including early types, have extended star formation
histories. Minor mergers provide an adequate gas supply to account for these
\citep{Kaviraj+09}.  However, hydrodynamical studies of the baryonic evolution
and SFR in low mass galaxies disagree about whether or not
one can reproduce their observed properties, including dark matter cores and
baryon fraction. Outflows may reproduce the observed
cores \citep{Governato+12} if the SFE is high at early
epochs, but such models fail to result in the strong evolution observed at
low redshift \citep{Weinmann+12}.

Tidal disruption also plays a role in disrupting satellites whose orbits
intersect the disk or bulge.  Dramatic discoveries due to deep imaging of
nearby galaxies with very small, wide field of view, telescopes confirm the
ubiquity of tidal tails that trace dwarf disruption in the remote past
\citep{MartinezDelgado+10}. Simulations provide a convincing demonstration
that we are seeing tidal disruption in action \citep{Cooper+10}.  An
independent confirmation of disruption in action comes from studies of the
tidal tails around the outermost MW globular star clusters such as
Pal~13. 
Gaps in the tails \citep{Grillmair09} indicate the presence of dark
satellites. Numerical simulations \citep{YJH11} find that high $M/L$
satellites of mass $\sim 10^7\rm\, M_\odot$ are required, again a prediction of
the CDM model.

\begin{figure}[ht]
\centering
\includegraphics[width=8.5cm]{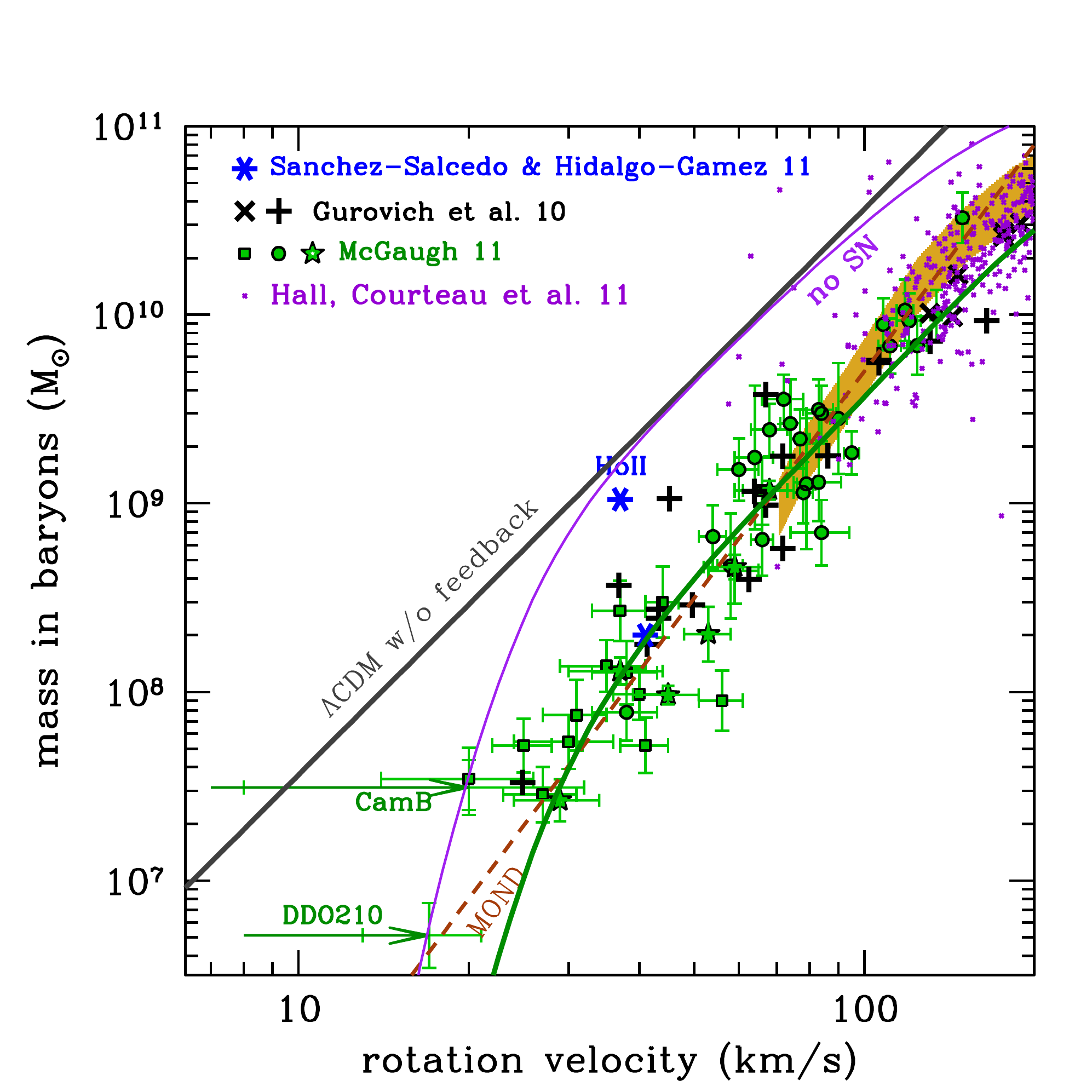} 
\caption{Baryonic Tully-Fisher relation (Mamon \& Silk, in prep.).
\emph{Symbols} are from HI measurements, where the velocity is the flat part of the
rotation curve (\emph{green}, from \citealp{McGaugh12}) or from line-widths
(\emph{black}, \citealp{Gurovich+10}; \emph{magenta}, \citealp{Hall+11}).
The \emph{grey} line is the na\"{\i}ve $\Lambda$CDM (slope 3) prediction with no
feedback, while the \emph{brown dashed line} is the (slope 4) prediction from
MOND.
Note that the inclination of Ho~II is uncertain \citep{Gentile+12}.
}
\label{BTFR}
\end{figure}
At $z=0$, it is possible that SN feedback at intermediate and low masses
combines with entropy feedback from photoionization at low masses to conspire
to give a linear baryonic Tully-Fisher relation (BTFR), as observed (see
Fig.~\ref{BTFR}). This is an important issue as the normalization, slope and
linearity of the BTFR have been used as evidence for MOdified Newtonian
Dynamics (MOND, \citealp{Milgrom83}) and against $\Lambda$CDM. Indeed,
\cite{McGaugh11} has pointed out that the observations of baryonic mass
(stars plus cold gas) as a function of the velocity of the flat part of the
rotation curve is very well matched by the MOND prediction (with no free
parameters). He argues that considerable fine-tuning is
required to bring the na\"{\i}ve $\Lambda$CDM (slope 3) prediction with no
feedback to match the data. Our best-fit model matches the data equally well
(with three free parameters), but the entropy feedback (photoionization)
implies that the relation should curve at low masses, except if one considers
galaxies in which the bulk of the stars formed before the reionization epoch
($z>6$).  \cite{Dutton12} also matched the BTFR data with a SAM.  Moreover,
two of the lowest mass galaxies in the \citeauthor{McGaugh12} sample have
rotation velocities corrected for asymmetric drift (see \citealp{BC04}),
after which the rotation curve of these galaxies is roughly linear with
radius, and therefore depends on the last data point obtained with
radio-observations, contradicting the flat part of the rotation curve sought
by \citeauthor{McGaugh12}.

\begin{figure}[ht]
\centering
\includegraphics[width=0.33\hsize]{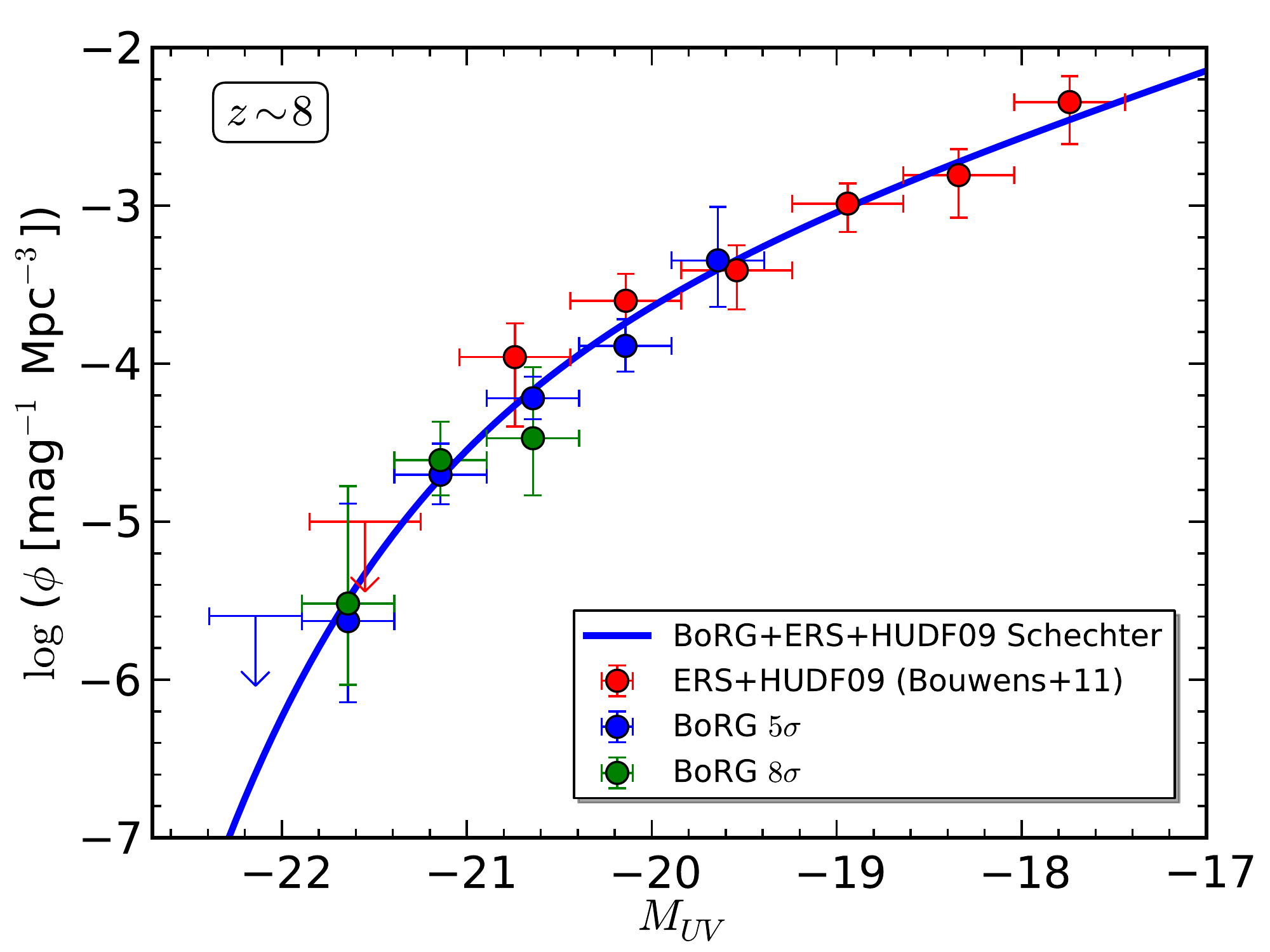} 
\includegraphics[width=0.33\hsize]{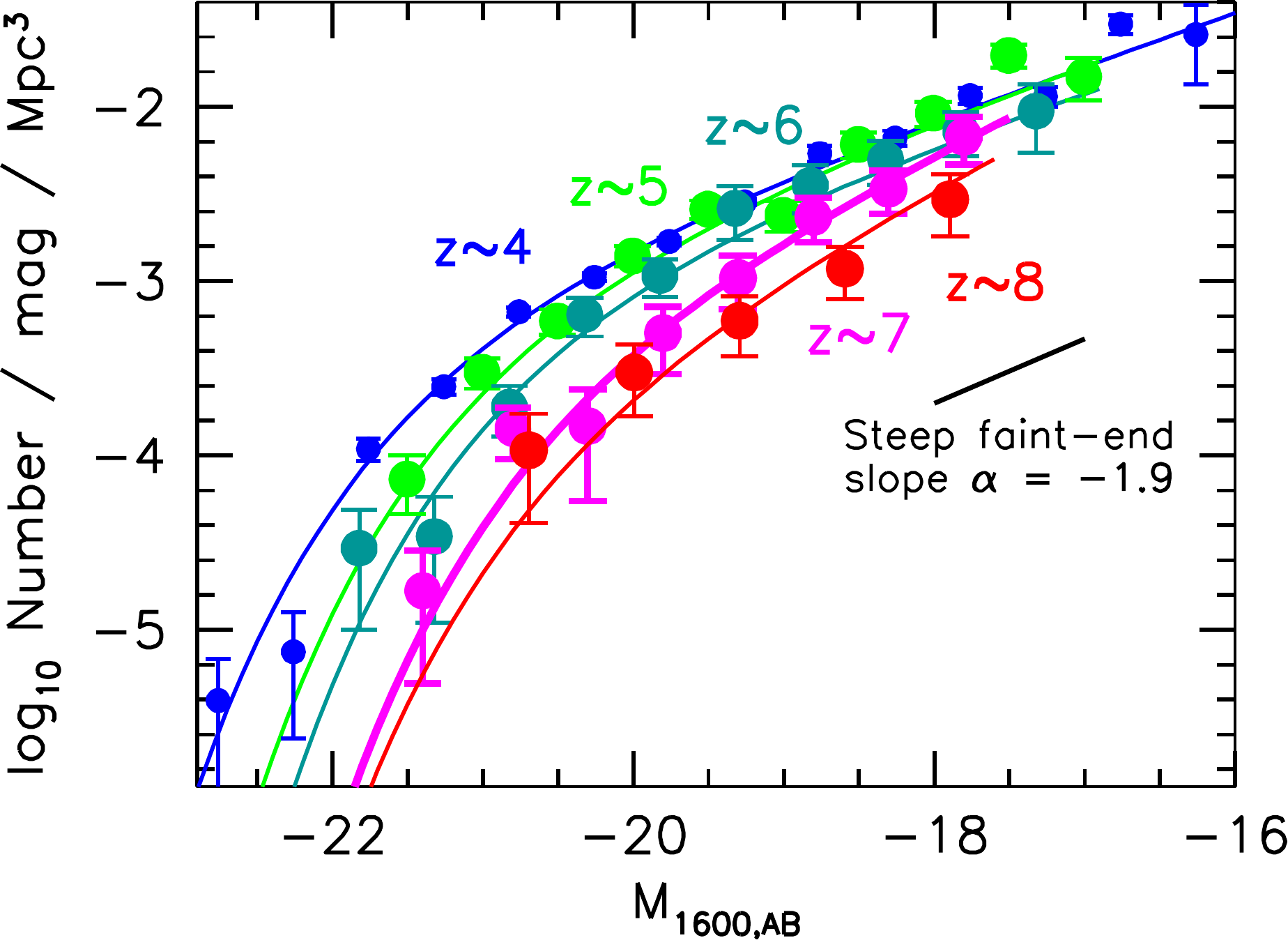} 
\includegraphics[width=0.33\hsize]{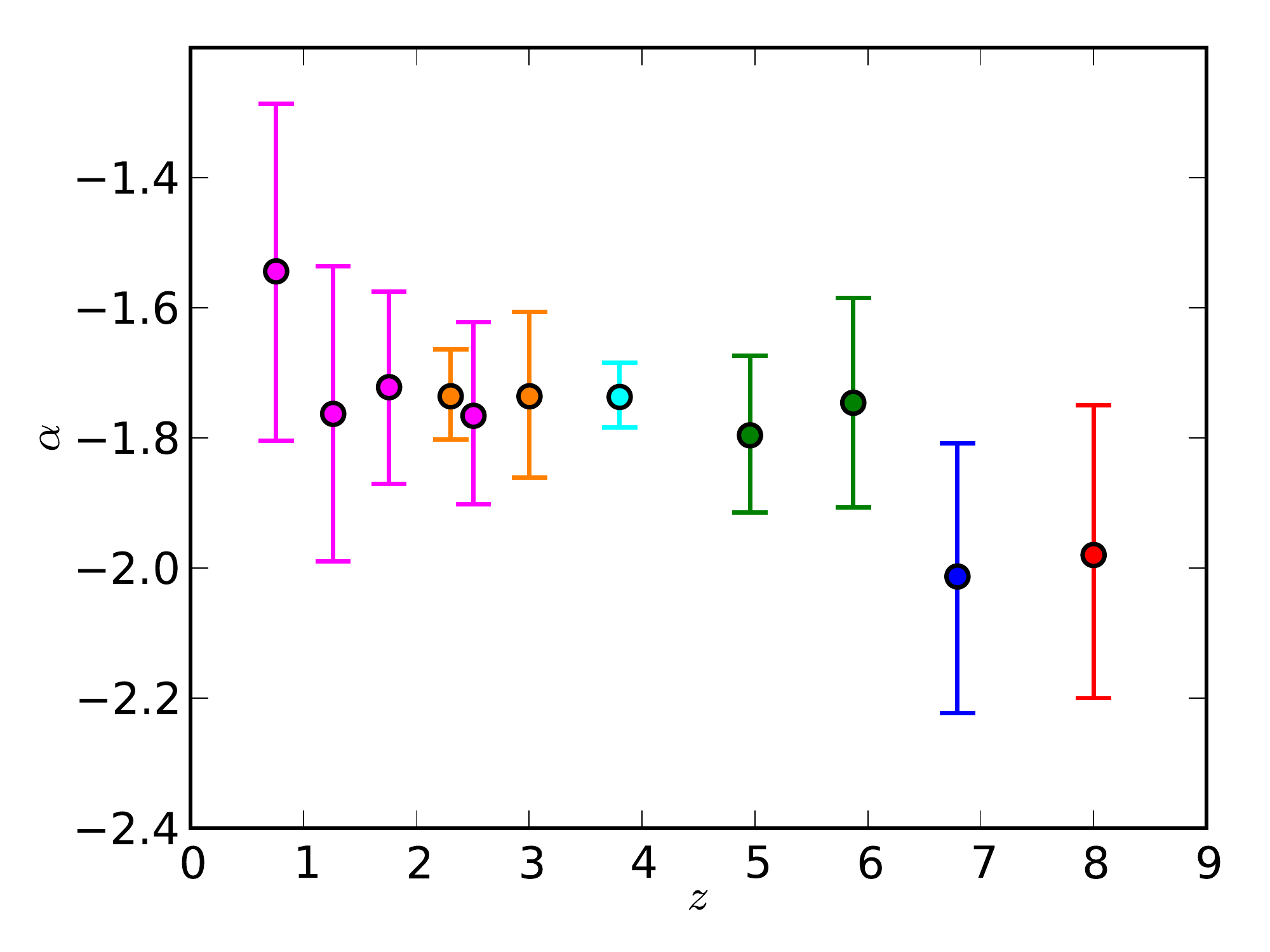} 
\caption{\emph{Left}: Galaxy luminosity function at $z=8$ \citep{Bradley+12}.
\emph{Middle} and \emph{right}: evolution of the galaxy luminosity function
\citep{Bouwens+12} and of its faint-end slope \citep{Bradley+12}.}
\label{lfevol}
\end{figure}
Intermediate-mass dwarfs are present at high redshift and have a steep
luminosity function (\citealp{Bradley+12}, see Fig.~\ref{lfevol}).
They may contribute significantly to the reionization of the Universe.

\subsection{Gas accretion versus mergers}

Star formation seems to be too complex to be simply gravity-induced. Merging
and AGN triggering are culprits for playing possible roles. What seems to be
progressively clear is that there are two distinct modes of star
formation. One mode occurs without any intervention from AGN and is
characteristic of disk galaxies such as the MW, on a time-scale of order at
least several galactic rotation times. Another mode is more intense,
occurring on a relatively rapid time-scale, and involves the intervention of
AGN, at least for quenching and possibly for enhancement or even triggering.

The most important aspect of star formation is the role of the raw material,
cold gas. There are two modes of gas accretion, which may be classified as
cold flows/minor mergers and major mergers/cooling flows. The former provide supplies of cold gas along filaments, the latter 
a source of hot gas which may cool and feed star formation.
\begin{figure}[ht]
\centering
\includegraphics[width=9cm]{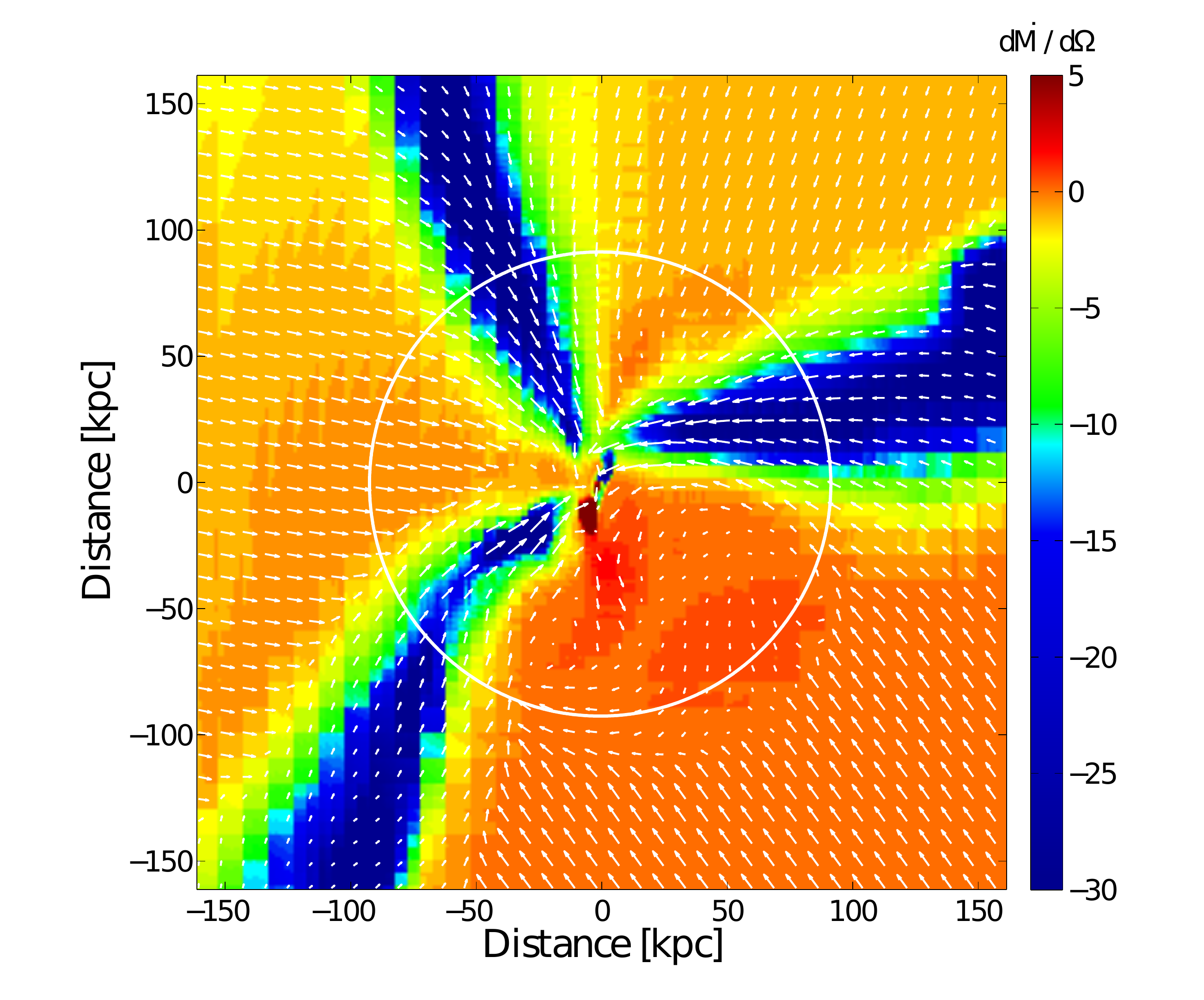} 
\caption{Mass flux map of a $M_v=10^{12}\,\rm  M_\odot$ halo at $z=2.5$ from a
  hydrodynamical simulation \citep{Dekel+09}. The \emph{circle} denotes the
  virial radius.}
\label{fluxmap}
\end{figure} 

The cold flows occur
in filamentary streams that follow the cosmic web of large-scale structure
(see Fig.~\ref{fluxmap}),
and include minor mergers via the dwarf galaxies that similarly trace the web
\citep{Dekel+09}. Theory suggests that, at low redshift, gas accretion by cold
streams is important, and that the cold streams are invariably clumpy and
essentially indistinguishable from minor mergers of gas-rich dwarfs. Major
galaxy mergers account for the observed morphological distortions that are
more common at high $z$,
and generally lead to cloud agglomeration, angular momentum loss and
cooling flows that feed star formation \citep{BPCT11}.

Observationally, one finds that cold flows are rarely if at all
observed. This is presumably because of the small covering factor of the
filaments \citep{Stewart+11_apjl, FK11}. Indirect evidence in favor of cold
accretion comes from studies of star formation in dwarfs. The best example
may be the Carina dwarf where three distinct episodes of star formation are
found \citep{THT09}. However at high redshift, major mergers between galaxies
are common. Indeed, Ultra-Luminous Infrared Galaxies (ULIRGs), whose SFRs are
huge, are invariably undergoing major, often multiple, gas-rich mergers
\citep{BBLC00} 
and dominate the cosmic SFR history at $z\simgt 2,$ whereas
normal star-forming galaxies predominate at low  ($z \simlt 2$) redshift
\citep{LBEOP09}.  
This certainly favors the idea of massive spheroid
formation by major mergers.

Using, their analytical model of galaxy formation on top of
a high-resolution cosmological simulation, \cite{CMWK11}  show that only
in massive galaxies ($m_{\rm stars} >10^{11} h^{-1}\, \rm M_\odot$) do galaxy
mergers contribute to the bulk of the stellar mass growth (see also
\citealp{GW08},  who analyzed a simulation with 11 times worse mass resolution)
and these mergers
are mainly `dry' (gas-poor). As one goes to
lower stellar masses (down to their simulation's effective resolution limit of $10^{10.6}
h^{-1}\, \rm M_\odot$) the role of mergers sharply diminishes, suggesting, by
extrapolation, that mergers are, in general, unimportant for the mass growth of both these
intermediate-mass galaxies and low-mass galaxies, for which
the bulk of the growth
must be by gas accretion. Nevertheless, among those rare intermediate-mass
galaxies built by mergers, the growth in mass is mostly in `wet' (gas-rich) and
minor mergers.
In particular, the non-dominant cluster galaxies, known to be mostly dwarf
ellipticals, are rarely built by mergers.

The sudden dominance of major mergers at high galaxy masses is confirmed by
trends with stellar mass of the colors, color gradients and elongations of
SDSS galaxies (\citealp{Tremonti+04,Bernardi+10}, see also 
\citealp{vanderWel+09,Thomas+10}).
At lower masses (and low redshift), 
minor mergers are required  to account for sizes and masses
\citep{Mclure+12, LopezSanjuan+12}.

Herschel observations of the  Main Sequence of galaxy formation 
(SFR versus stellar mass) suggests that 
starbursts, commonly associated with major mergers,  are displaced to higher
mass and SFR,  but only account for 10\% of the SFR density at $z=2$ 
\citep{Rodighiero+11}.
 However, this conclusion depends critically on the $\sim 100\,\rm Myr$ timescale assumed
 for the starbursts. If the starbursts had shorter duration, say 20 Myr,
 given their effective observation time of $\sim 1\,\rm Gyr$,
 they would account for as much as 50\% of the star formation at $z\sim 2$. It is difficult
 to gauge independent estimates of starburst age, but for example the UV
 continuum flattening observed at high $z$ for luminous star-forming galaxies
 favors a younger starburst age \citep{Gonzalez+12} as would possible SED corrections for nebular emission.

\subsection{Initial stellar mass function}

The IMF of stars forming in galaxies 
is usually treated as universal in galaxy formation modelling. There
has been a recent flurry of papers finding evidence for a systematic
steepening, from the \cite{Chabrier03} to \cite{Salpeter55} IMFs, 
in massive early type galaxies. From
a spectral absorption line analysis,  a correlation of IMF steepening  with
enhanced velocity dispersion, [Mg/Fe] and sodium abundance is reported by
\citep{CvD12}. 
A similar result is reported for stacked massive galaxy spectra 
\citep{Ferreras+12}.
\begin{figure}[ht]
\centering
\includegraphics[width=10.5cm]{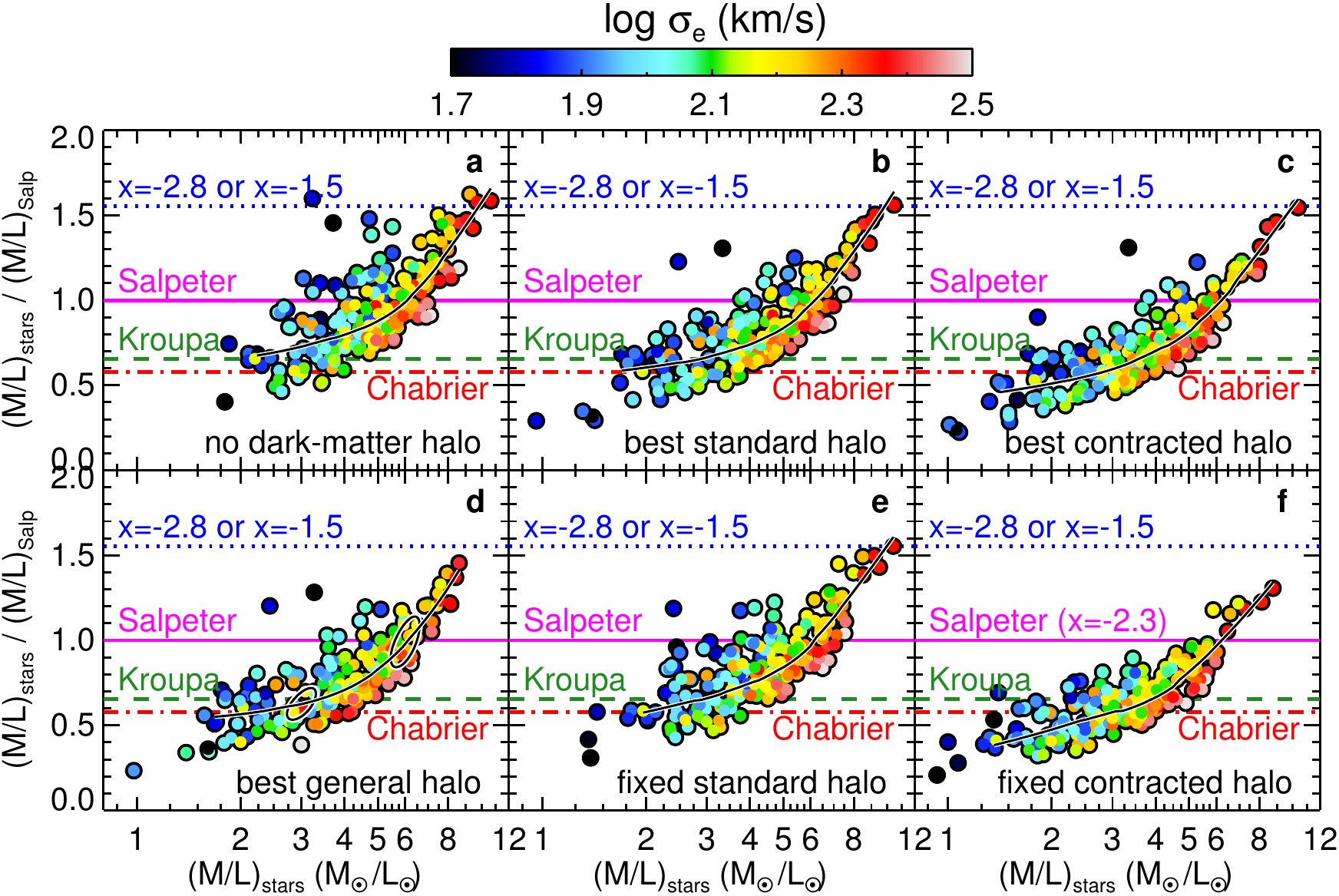} 
\caption{Stellar mass-to-light ratio inferred from kinematical modeling
  (after subtracting off the contribution of the DM component),
  normalized to Salpeter ratio inferred from stellar populations versus
stellar $M/L$ from  kinematical modeling, for six DM models \citep{Cappellari+12}.
}
\label{nonunivIMF}
\end{figure}
The modeling of the internal kinematics of early-type galaxies using
integral field spectroscopy 
provides evidence for steeper IMFs (regardless of many plausible assumptions
on the DM) in increasingly more massive galaxies
(\citealp{Cappellari+12}, see Fig.~\ref{nonunivIMF}).
Lensing plus gas kinematics provides evidence for a Salpeter-like IMF in
several massive ellipticals \citep{Dutton+12}. There may also be a
correlation of a steeper IMF with the densest massive galaxies
\citep{DMS12}.
All of these studies report increasing M/L with increasing spheroid velocity
dispersion and $\alpha/\rm[Fe].$

The possible degeneracy between IMF and DM fraction and shape 
is a concern because the DM profile steepens as a
consequence of adiabatic contraction. 
While, \cite{Cappellari+12} tried a variety of DM  models that do not
significantly 
influence their result (since they only 
probed the region where dark matter accounts for at best 20\% of the mass),
only one study \citep{Sonnenfeld+12} so far has cleanly broken the degeneracy
with the dark matter profile:
By using a double Einstein ring, 
\citeauthor{Sonnenfeld+12} found a strong case for a Salpeter IMF. 
The adiabatic contraction of the DM is within the range found by \cite{GKKN04}.

The implications of a steeper IMF in massive galaxies for galaxy formation
models remain to be explored. The increased efficiency of star formation
required at early epochs will certainly provide further tensions with the
need to leave a substantial gas supply at late epochs for the observed late
evolution observed for low mass galaxies, as discussed below.

\subsection{Feedback and AGN}

Quenching of
star formation has been largely motivated by the apparent success of SMBH
feedback in reproducing the scaling and normalization of the black hole
mass-spheroid velocity dispersion $M_{\rm BH}-\sigma_v$) relation, as first proposed
by \cite{SR98}. 
SAMs  indeed
demonstrate that AGN feedback is able to quench star formation in massive
elliptical galaxies \citep{Croton+06,Bower+06,Cattaneo+06,Somerville+08}. 
One can reproduce the fairly sharp cut-off in the bright end of
the galaxy luminosity function~\citep{Bell+03, PJHC07}. 
These SAMs do not require ``quasar mode'' AGN feedback with Eddington
luminosities.

\begin{figure}[ht]
\centering
\includegraphics[width=0.35\hsize]{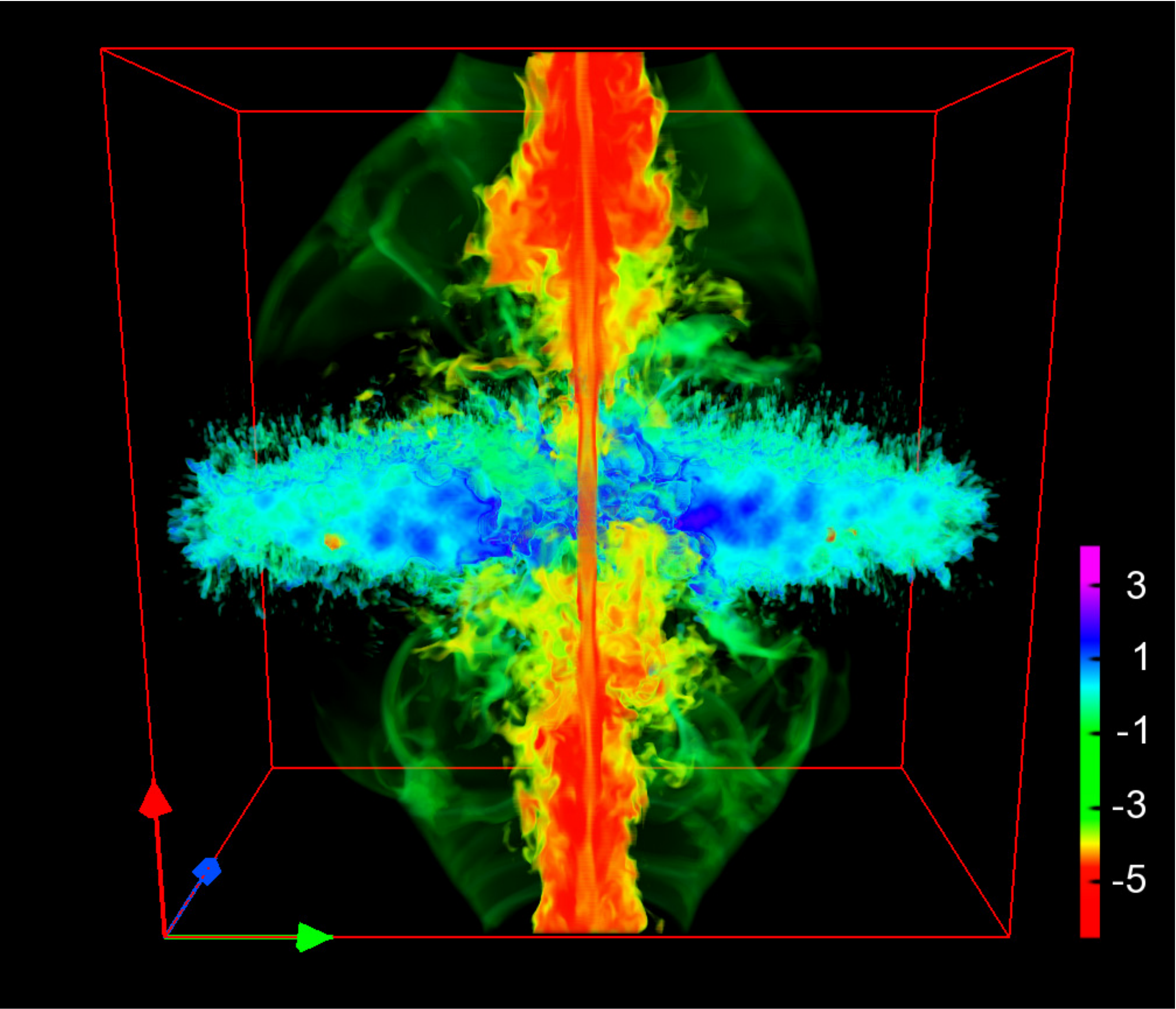}
\includegraphics[width=0.35\hsize]{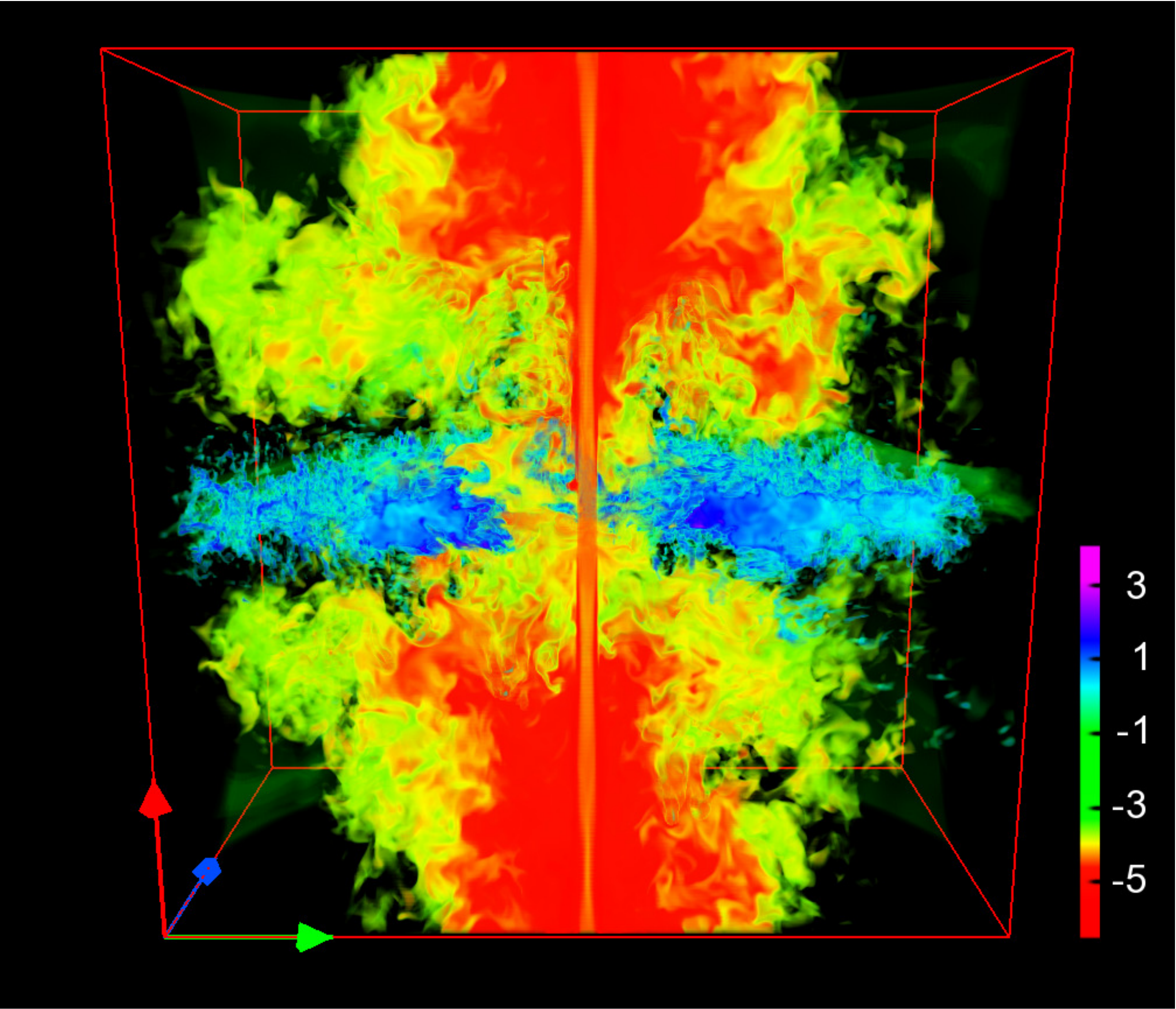} \\
\includegraphics[width=0.35\hsize]{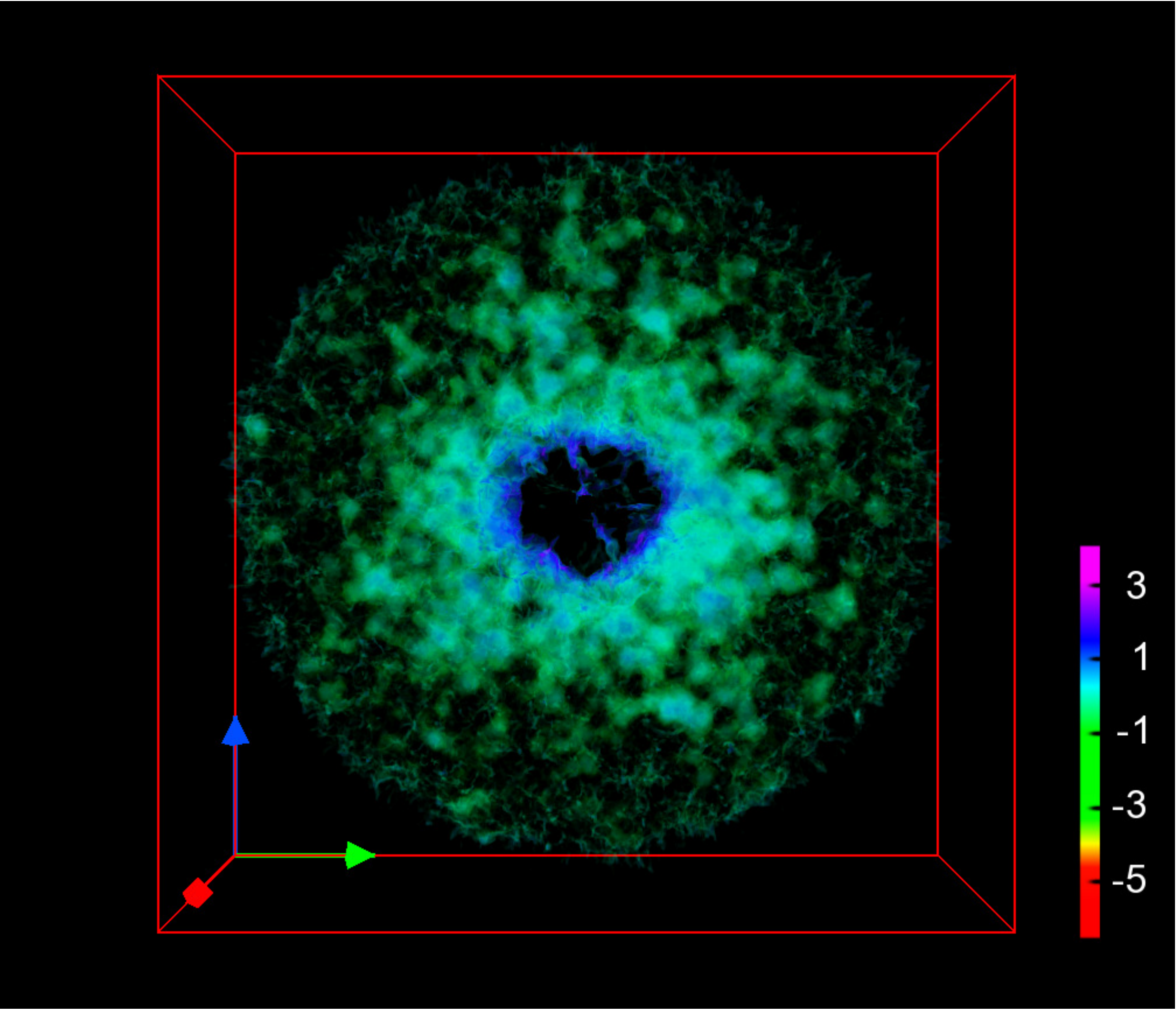}
\includegraphics[width=0.35\hsize]{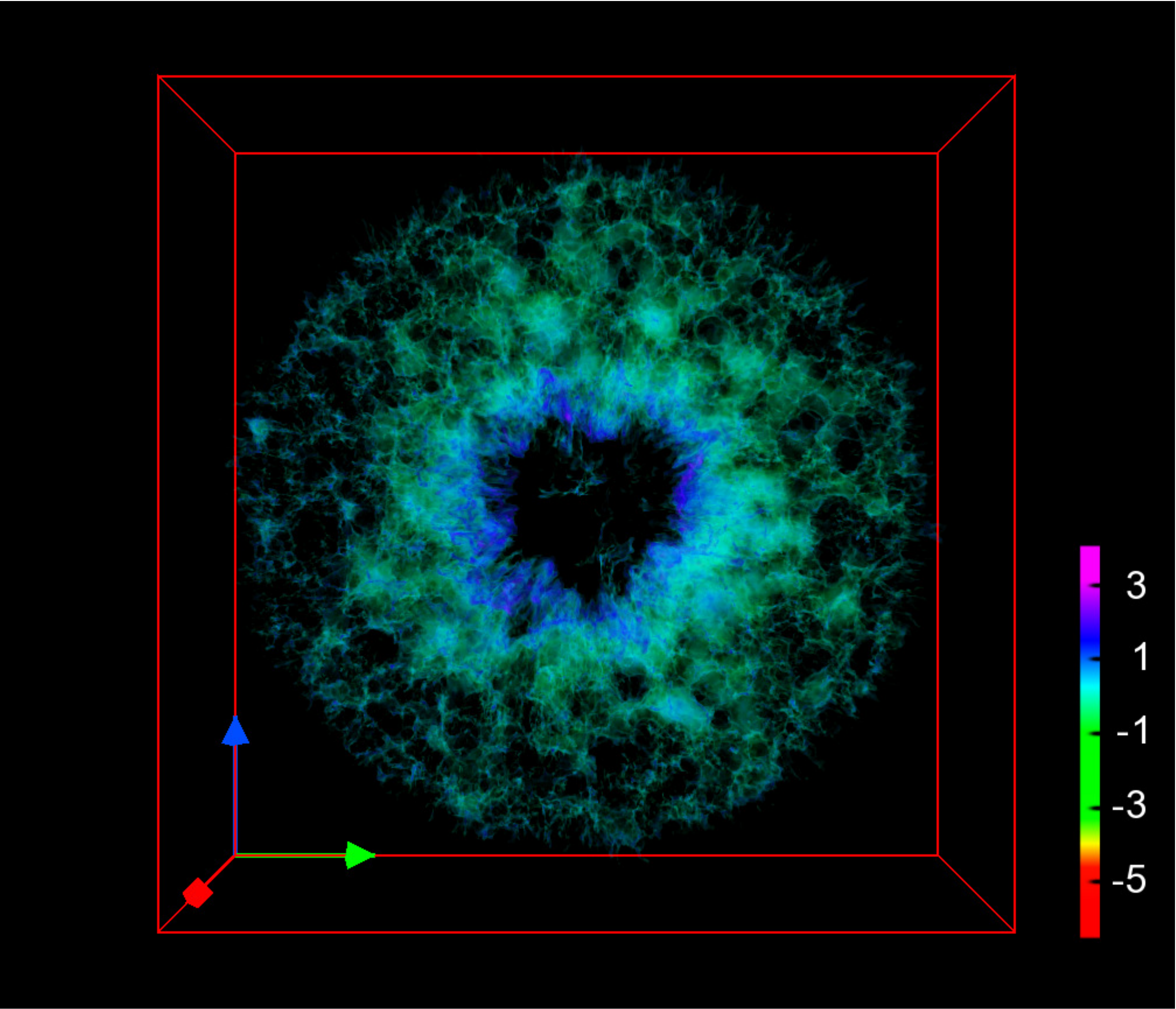} 
\caption{Simulations (in 32 kpc box) of AGN feedback  at 14 (\emph{left}) and
  22 Myr (\emph{right}) after
  the onset of the jet, in edge-on (\emph{top}) and face-on (\emph{bottom})
  views of log density \citep{GKKS11}}
\label{AGNsim}
\end{figure}

 High resolution hydrodynamical cosmological simulations indeed  show that while cold
 streams initially feed the black hole, transferring angular momentum to
 produce central disks \citep{DDST12} that become gravitationally
 unstable and feed the compact bulge through migration of
 clumps (see \citealp{bournaud+11}), the cold flows are eventually interrupted
 by AGN-driven super-winds \citep{Dubois+12_blowout}. 

However the physics of driving SMBH outflows is still not well understood. One issue
is that momentum-driven winds fail to account for the normalization of the
$M_{\rm BH}-\sigma_v$ relation~\citep{SN10, DQM12}, with the shortfall being
about a factor of 10.  This momentum deficit can be supplied by radio
jet-driven outflows \citep{WB11}, which also account for the observed high
velocities of entrained cold gas \citep{WBU12}. Alternative or complementary
possibilities, possibly more relevant to radio-quiet quasars, include
positive feedback from outflow-triggered star formation (\citealp{SilkNorman09,
  GKKS11}, see Fig.~\ref{AGNsim}) and energy-driven outflows
\citep{FQ12}. Nearby AGN show dense molecular rings surrounding circumnuclear
rings of star formation \citep{Sani+12},  reminiscent of the simulated  
triggering of star formation \citep{GKKS11}.

SMBHs are generally found to correlate with bulges rather than  with disks, pseudobulges or dark halos
\citep{Ho07,KBC11,KB11}, although disk galaxies appear to follow a similar $M_{\rm BH}-\sigma_v$
relation, albeit with more scatter \citep{GOAC11}.
This would simplify formation mechanisms, suggesting that bulges and SMBH grow together, perhaps self-regulating each other. 
Massive black hole growth at early epochs seems to be (just) achievable by gas accretion. 
Large cosmological simulations~(\citealp{DiMatteo+12}, see
also~\citealp{Li+07, SSH09, Khandai+12}) have shown that primordial massive
BHs can grow by cold filamentary infall, and acquire masses of up to several
billion solar masses by $z=6$ in the most massive halos ($M_{\rm
  vir}\simeq10^{12-13}\, \rm \,M_\odot$).

Insight into black hole growth is provided by looking for extreme deviations
in the $M_{\rm BH}-\sigma_v$ relation. Massive black holes seem to be in place at
high redshift before spheroids \citep{Wang+11}. This is also the case for a
nearby starbust galaxy containing an AGN but without any matching
spheroid or indeed massive stellar component 
\citep{RD12}. 
On the
other hand, SMGs seem to contain relatively low mass black holes for their
stellar content \citep{Alexander+08}.

\section {Future prospects in observations}
A clue as to the nature of a possible solution may come from the fact that
quasars also reveal luminosity downsizing.  This translates into downsizing
of central SMBH mass. One might be able to connect the two
phenomena if feedback from AGN was initially positive and also a strongly
nonlinear function of SMBH mass. Predictions of positive feedback include
circumnuclear rings on 10--100 pc scales in star-forming AGN. These should be
resolvable with ALMA, via both molecular lines that probe pressurized
molecular gas and FIR fine-structure lines that probe the interplay of
intense FUV radiation fields with photodissociation regions (PDRs and XDRs).

More conventionally, the evidence for AGN quenching of star formation seems
strong. 
Superwinds driven by AGN are capable of depleting the
reservoir of star-forming gas over relatively short time-scales. However,
questions remain as to the relative roles of AGN winds, jet-driven bubbles,
SNe, and radiation pressure especially from OB star clusters. No doubt, JWST,
as well as 30+meter telescopes such as ELT
will complement HST by producing spectacular IR images of star formation, AGN
and outflows  at work. 
Accretion of neutral gas will be studied at
high sensitivity by SKA. Ultimately one needs a spectroscopic survey akin to
SDSS at $z=1-2$ and this will be provided by the Subaru Prime Focus
Spectrograph with optical and NIR capability. The next decade should bring a
vast increase in our phenomenological understanding of the basic processes at
play in galaxy formation and evolution.

\section {Future prospects in astrophysical  theory}

Theory lacks adequate resolution and physics. Of course these issues are
intricately connected. One needs to tackle baryon physics and the associated
possibilities for feedback. Today, state-of-the-art cosmological
simulations of the MW with gas and star formation, such as the ERIS simulation \citep{GCMM11},
provide only $\approx 100\,\rm pc$ resolution. Hence, in  current simulations, the
gas and star formation physics is included in an ad hoc way, because of the
resolution limitation. For example, while stars are known to form in the dense
cores --- of density $\ga 10^5\,\rm cm^{-3}$ --- of Giant Molecular Clouds, the
current hydrodynamical simulations adopt SF thresholds of typically $1\,\rm
cm^{-3}$ and always $\la 10^2\,\rm cm^{-3}$. Sharp increases of the SF
density threshold result in moving the SF regions outside of the nucleus
\citep{TCB10}. 
However,
in reality, it is the unresolved subgrid physics that determines the actual
threshold, if one even exists.  Mastery of the required subparsec-scale
physics will take time, but there is no obvious reason why we cannot achieve
this goal  with orders of
magnitude improvement in computing power.

For the moment, phenomenology drives all modelling. This is true especially
for local star formation. A serious consequence is that physics honed on
local star-forming regions, where one has high resolution probes of
star-forming clouds and of ongoing feedback, may not necessarily apply in the
more extreme conditions of the early universe.

One issue that arises frequently is whether the perceived challenges to
$\Lambda$CDM justify a new theory of gravity. From MOND \citep{Milgrom83} 
onwards, there are
any number of alternative theories that are designed to explain certain
observations.  However, none can explain the ensemble of observations any
better than $\Lambda$CDM, nor do they rely on solid physical grounds. 
But to the extent that any unexplained
anomalies exist, these are invariably at no more than the $2\,\sigma$ level of
significance. It seems that such ``evidence" is not adequate motivation
for abandoning Einstein-Newton gravity.  Indeed, while it is overwhelmingly clear
that there are many potential discrepancies with $\Lambda$CDM, we have
certainly not developed the optimal $\Lambda$CDM theory of galaxy
formation: the current models do not adequately include the baryons nor do we
reliably understand star formation, let alone feedback.
Other MOND-related issues are reviewed  in \cite{FM11}, including challenges raised by the 
apparent emptiness of local voids and satellite phase space correlations.
However, we regard these as more a matter of  absorbing the significance of ever deeper galaxy and 21 cm surveys, on the one hand 
(for example, deep blind HI surveys show that gas-rich galaxies are the least clustered of any galaxy population \citealp{MGHG12}), 
and on the other hand, of 
questioning the details of hitherto inadequately modelled  baryonic  physics,
as developed for example in \cite{Zolotov+12}.
Whether appeal to alternative gravity is justified by inadequate baryonic
physics is a question of judgement at this point. Here is a summary of many
of these failures: we cite some key reasons why $\Lambda$CDM does not yet
provide a robust explanation of the observations: we list below several
examples that represent challenges for theorists.

\begin{enumerate}
\item
Massive bulgeless galaxies with thin disks are reasonably common
\citep{KDBC10}. Simulations invariably make thick disks and bulges. Indeed,
the bulges are typically overly massive relative to the disks for all
galaxies other than S0s.  Massive thin disks are especially hard to simulate
unless very fine-tuned feedback is applied. A consensus is that the feedback
prescriptions are far from unique \citep{Scannapieco+12}.  One appealing
solution involves SN feedback. This drives a galactic fountain that
feeds the bulge. A wind is driven from the bulge where star formation is
largely suppressed for sufficiently high feedback \citep{Brook+12}.
 Another proposal includes radiation pressure from massive stars as well as SNe. The combined feedback helps expand the      
halo expansion, thereby limiting dynamical friction and bulge formation \citep{Maccio+12}.
  
\item
Dark matter cores are generally inferred in dwarf spheroidal galaxies,
whereas $\Lambda$CDM theory predicts a cusp, the NFW profile.  Strong
SN feedback can eject enough baryons from the innermost region to
create a core \citep{Governato+10, PG12}, but this requires high early
SN feedback or a series of implausibly short bursts of star formation.

\item
The excessive predicted numbers of dwarf galaxies are one of the most cited
problems with $\Lambda$CDM. The discrepancy amounts to two orders of
magnitude.  The issue of dwarf visibility is addressed by feedback that
ejects most of the baryons and thereby renders the dwarfs invisible, at least
in the optical bands.  There are three commonly discussed mechanisms for
dwarf galaxy feedback: reionization of the universe at early epochs, SNe, and (ram
pressure and tidal) stripping.  AGN-driven outflows via intermediate mass
black holes provide another alternative to which relatively little attention
has been paid \citep{SN10}.

None of these have so far been demonstrated to provide definitive solutions.
Reionization only works for the lowest mass dwarfs. The ultrafaint dwarfs in
the MW may be fossils of these first galaxies (as checked by detailed models
\citealp{Koposov+09,SF09,BR11b}). 
It is argued that SN feedback solves the problem for the more massive dwarfs \citep{Maccio+10}.
 However, this conclusion is disputed by \cite{BBK11}, who use the Aquarius
 simulations \citep{Springel+08} 
 to predict more massive dwarfs in dark-matter-only simulations than are observed. These authors
 argue that the relatively massive dwarfs should form stars, and we see no
 counterparts of these systems, apart possibly from rare massive dwarfs such
 as the Magellanic Clouds. We have previously remarked that  omission of
 baryonic physics biases the dark matter-only simulations to an overstatement
 of the problem by overpredicting dwarf  central densities
 \citep{Zolotov+12}. 

\item
The SFE in dwarfs is highly debated. Let us put aside
the high SFE at early epochs that is required to obtain
strong feedback in order to generate cores.  For example, it is possible that
intermediate mass black holes could be invoked to solve this problem and
simultaneously generate the required low baryon fraction \citep{PJSP12}.

In order to obtain the required late epoch evolution \citep{Weinmann+12}, one
might appeal to a lower SFE in dwarfs, plausibly
associated with low metallicities and hence low dust and $\rm H_2$ content.
Models based on metallicity-regulated star formation can account for the
numbers and radial distribution of the dwarfs by a decreasing SFE
\citep{Kravtsov10}.  This explanation is
disputed by 
\cite{BBK11}, who infer a range in SFEs for the dwarfs of some two orders of magnitude. 
A similar result appeals to varying the halo mass threshold below which star formation must be suppressed to account
for the dwarf luminosity function, whereas  the stellar masses of many observed dwarfs violate this condition \citep{Ferrero+11}.
Finally,  tidal stripping  may provide a solution \citep{Nickerson+11}, at least for the inner dwarfs.

\item
Another long-standing problem relates to downsizing. Massive galaxies are
in place before lower mass galaxies as measured by stellar mass assembly, and
their star formation time-scales and chemical evolution time-scales at their
formation/assembly epoch are shorter. One popular explanation
\citep{CDFG08} is that
galaxies cannot accrete/retain cold gas in massive halos, either because of
AGN feedback or because of virial shocks that prevent the gas supply of the
disk in cold filaments \citep{BD03}.

\item
It is possible to develop galaxy
formation models with suitable degrees and modes of feedback that address
many of these issues.  However, a major difficulty confronted by all SAMs
is that the evolution of the galaxy luminosity function
contradicts the data, either at high or at low redshift. The SAMs that are
normalized to low redshift and tuned to account for the properties of local
galaxies fail at high redshift by generating too many red galaxies
\citep{Fontanot+09b}. Too few blue galaxies are predicted at $z=0.3.$
This problem has been addressed by including AGB stars in the stellar
populations. This fix results in a more rapid reddening time-scale by
speeding up the evolution of the rest-frame near-infrared galaxy luminosity
function \citep{Henriques+11}. There is a price to be paid however: now there
are excess numbers of blue galaxies predicted at $z=0.5$.

\item 
There is a well-known difficulty in  matching  both the galaxy luminosity function and
  Tully-Fisher scaling relation, even  at $z=0$.
Reconciliation of the Tully-Fisher zero point with the galaxy
luminosity function requires too high an efficiency of star formation
\citep{GWLB10}.
In fact, the problem is even worse: the models of massive spirals  tuned to
fit the Tully-Fisher relation are  too concentrated \citep {McCarthy+12}. 
This is  a reflection of the over-massive bulge problem in disk galaxies that
simply refuses to go away \citep{NS00,ANSE03}. 

\item
The luminosity function problem is most likely related to another
unexplained property of high redshift galaxies. The SSFR evolution at high $z$
is very different from that at low $z$. Essentially, it saturates. One finds
an infrared Main Sequence of galactic SFRs: SFR versus
$M_*$ \citep{Elbaz+11}. Neither the slope nor the scatter are adequately
understood. Starburst galaxies lie above the Main Sequence, but the fraction of cosmic star formation in these systems depends on inadequately justified assumptions about starburst duration. 
For example, nebular emission and dust extinction affect infrerred ages, and one cannot easily understand the blue continuum slopes  oberved at high redshift and lower UV luminosities \citep{Bouwens+12_UV}.

\item The observed rapid growth of early-type galaxy sizes since $z=2$ for fixed stellar
  mass cannot be reproduced in SAMs or analytical models \citep{CNC12}: at
  $z=2$ galaxies are too compact.

\item
Much has been made of nearby rotation curve wiggles that trace similar
dips in the stellar surface density that seemingly reduce the significance of
any dark matter contribution. Maximum disks optimize the contribution of
stars to the rotation curve, and these wiggles are most likely associated
with spiral density waves.  A similar result may be true for low surface
brightness gas-rich dwarf galaxies \citep{SSvAvdH11}.  

\item
High mass-to-light
ratios are sometimes required for maximum disk models of spiral galaxy
rotation curves, but these are easily accommodated if the IMF 
is somewhat bottom-heavy. The case for IMF variations has been made for
several data sets, primarily for early-type galaxies (e.g., see
\citealp{vDC11}). The LSB dwarfs are plausible relics of the building blocks
expected in hierarchical formation theories.

\item
Spiral arms are seen  in the HI distribution in the outer regions of some disks. This tells us that significant 
angular  momentum transfer is helping feed the optical inner disk. The baryon self-gravity is large enough that one does not for example need to appeal to  a flattened halo, which might otherwise be problematic for the DM model
\citep{BA10}.

\item
The slope and normalization of the baryon Tully-Fisher relation do not
agree with the simplest $\Lambda$CDM prediction. The observed slope is
approximately 4, similar to what is found for MOND \citep{Milgrom83},
whereas $\Lambda$CDM (without feedback) gives a slope of 3 \citep{McGaugh11,McGaugh12}, but fails to account for the observed dispersion and curvature.

\item
The baryon fraction in galaxies is some 50\% of the primordial value predicted by light element nucleosynthesis. These baryons are not in hot gaseous halos
\citep{AB10}. Convergence to the universal value on cluster scales is controversial:  convergence to the WMAP value is seen for X-ray clusters above a temperature of 5 keV \citep{DBKR10}, 
but  could be as large as 30\% even for massive clusters \citep{Andreon10, Scannapieco+12}.
If the latter discrepancy were to be confirmed, one would need significant bias of baryons relative to dark matter, presumably due to feedback, on unprecedentedly large scales.

\item The distribution of the MW satellite galaxies in a great circle
  \citep{LyndenBell82} is unexpected in the $\Lambda$CDM context
  \citep{KTB05}. However, infall onto halos is not spherically symmetric
  \citep{APC04}, and subhalos tend to lie in a plane \citep{Libeskind+05}.
The details of the thickness of this plane remained to be settled (e.g.,
\citealp{Kroupa+10} versus \citealp{Libeskind+11}).

\item There is a significant lack of galaxies in comparison with standard
  expectations
in the Local Void close to the
  Local Group \citep{Peebles07,TK09}. But it is not yet clear whether this
  region fairly low galactic latitude region has been surveyed as closely as
  other regions.

\item
Bulk flows are found over 100 Mpc scales that are about two standard deviations
larger than expected in $\Lambda$CDM \citep{FWH10}. The technique primarily
uses Tully-Fisher and Fundamental Plane galaxy calibrators of the distance
scale. An X-ray approach, calibrating via the kinetic \cite{SZ72} effect
(kSZE), claims the existence of a bulk flow out to 800 Mpc
\citep{Kashlinsky+10}. However the discrepancies with $\Lambda$CDM are
controversial because of possible systematics. A recent detection of kSZE
confirms pairwise bulk flows of clusters at $4\,\sigma$ and is consistent
with $\Lambda$CDM \citep{Hand+12}.

\end{enumerate}

Several of these issues may be linked. For example, the analysis of
\cite{Cappellari+12} that the IMF is
non-universal, with shallower (top-heavy) IMFs for galaxies  of lower velocity
dispersion, can be linked with the known relations between velocity dispersion
and metallicity (e.g., \citealp{AHSL09}) to produce a relation between IMF
and metallicity, which goes in the right direction: low-metallicity systems
have top-heavy IMFs. Until now, observers assumed a universal IMF when
deriving stellar masses. They have therefore overestimated the stellar masses of
low-metallicity systems. We would like to think that this overestimation of
$M_*$ might explain at the same time the evolution of the cosmic SSFR and
that of galaxy sizes. Indeed, at high redshift, galaxies are expected to be
more metal-poor, and the overestimate of their typical stellar masses will
lead to an underestimate of their SSFRs, relative to those of lower-redshift
galaxies. Therefore, the cosmic SSFR may not saturate at high redshift, which
will make it  easier to fit to models. At the
same time, if high redshift galaxies have lower stellar masses than inferred
from a universal IMF, then for a given stellar mass, they have larger sizes
than inferred, and the too rapid evolution of galaxy sizes (relative to
models)  might disappear.
We propose that observers replace stellar mass by $K$-band rest-frame luminosity, which,
if properly measured, can serve as a useful proxy for stellar
mass, independently of any assumed IMF.

In summary, it is clear that many problems await refinements in theoretical understanding.
No doubt, these will come about eventually as numerical simulations of galaxy formation are refined to tackle subparsec scales.

We are grateful to A.~Cattaneo, B.~Famaey, A.~Graham, J.~Kormendy, P.~Kroupa, S.~McGaugh,
A.~Pontzen and A.~Tutukov for very useful comments.

\small
\bibliography{master,iau}

\begin{thebibliography}{206}
\expandafter\ifx\csname natexlab\endcsname\relax\def\natexlab#1{#1}\fi

\bibitem[{{Abadi} {et~al.}(2003){Abadi}, {Navarro}, {Steinmetz}, \&
  {Eke}}]{ANSE03}
{Abadi} M.~G., {Navarro} J.~F., {Steinmetz} M., {Eke} V.~R., 2003, \apj, 591,
  499

\bibitem[{{Agertz} {et~al.}(2011){Agertz}, {Teyssier}, \& {Moore}}]{ATM11}
{Agertz} O., {Teyssier} R., {Moore} B., 2011, \mnras, 410, 1391

\bibitem[{{Alexander} {et~al.}(2008){Alexander}, {Brandt}, {Smail}, {Swinbank},
  {Bauer}, {Blain}, {Chapman}, {Coppin}, {Ivison}, \&
  {Men{\'e}ndez-Delmestre}}]{Alexander+08}
{Alexander} D.~M., {Brandt} W.~N., {Smail} I., 
et al.,
  2008, \aj, 135, 1968

\bibitem[{{Allanson} {et~al.}(2009){Allanson}, {Hudson}, {Smith}, \&
  {Lucey}}]{AHSL09}
{Allanson} S.~P., {Hudson} M.~J., {Smith} R.~J., {Lucey} J.~R., 2009, \apj,
  702, 1275

\bibitem[{{Anderson} \& {Bregman}(2010)}]{AB10}
{Anderson} M.~E., {Bregman} J.~N., 2010, \apj, 714, 320

\bibitem[{{Andreon}(2010)}]{Andreon10}
{Andreon} S., 2010, \mnras, 407, 263

\bibitem[{{Aubert} {et~al.}(2004){Aubert}, {Pichon}, \& {Colombi}}]{APC04}
{Aubert} D., {Pichon} C., {Colombi} S., 2004, \mnras, 352, 376

\bibitem[{{Baldry} {et~al.}(2004){Baldry}, {Glazebrook}, {Brinkmann},
  {Ivezi{\'c}}, {Lupton}, {Nichol}, \& {Szalay}}]{Baldry+04}
{Baldry} I.~K., {Glazebrook} K., {Brinkmann} J., 
et al.,
  2004, \apj, 600, 681

\bibitem[{{Begum} \& {Chengalur}(2004)}]{BC04}
{Begum} A., {Chengalur} J.~N., 2004, \aap, 413, 525

\bibitem[{{Behroozi} {et~al.}(2010){Behroozi}, {Conroy}, \& {Wechsler}}]{BCW10}
{Behroozi} P.~S., {Conroy} C., {Wechsler} R.~H., 2010, \apj, 717, 379

\bibitem[{{Behroozi} {et~al.}(2012){Behroozi}, {Wechsler}, \& {Conroy}}]{BWC12}
{Behroozi} P.~S., {Wechsler} R.~H.,  {Conroy} C., 2012, \apj, submitted, arXiv:1207.6105

\bibitem[{{Bell} {et~al.}(2003){Bell}, {Baugh}, {Cole}, {Frenk}, \&
  {Lacey}}]{Bell+03}
{Bell} E.~F., {Baugh} C.~M., {Cole} S., {Frenk} C.~S., {Lacey} C.~G., 2003,
  \mnras, 343, 367

\bibitem[{{Bender} {et~al.}(1992){Bender}, {Burstein}, \& {Faber}}]{BBF92}
{Bender} R., {Burstein} D., {Faber} S.~M., 1992, \apj, 399, 462

\bibitem[{{Benson}(2012)}]{Benson12}
{Benson} A.~J., 2012, \na, 17, 175

\bibitem[{{Berlind} \& {Weinberg}(2002)}]{BW02}
{Berlind} A.~A., {Weinberg} D.~H., 2002, \apj, 575, 587

\bibitem[{{Bernardi} {et~al.}(2010){Bernardi}, {Shankar}, {Hyde}, {Mei},
  {Marulli}, \& {Sheth}}]{Bernardi+10}
{Bernardi} M., {Shankar} F., {Hyde} J.~B., {Mei} S., {Marulli} F., {Sheth}
  R.~K., 2010, \mnras, 404, 2087

\bibitem[{{Bertin} \& {Amorisco}(2010)}]{BA10}
{Bertin} G., {Amorisco} N.~C., 2010, \aap, 512, A17

\bibitem[{{Bigiel} {et~al.}(2011){Bigiel}, {Leroy}, \& {Walter}}]{BLW11}
{Bigiel} F., {Leroy} A., {Walter} F., 2011, in IAU Symposium, Vol. 270,
  Computational Star Formation, {Alves} J., {Elmegreen} B.~G., {Girart} J.~M.,
  {Trimble} V., eds., pp. 327--334

\bibitem[{{Birnboim} \& {Dekel}(2003)}]{BD03}
{Birnboim} Y., {Dekel} A., 2003, \mnras, 345, 349

\bibitem[{{Blanchard} {et~al.}(1992){Blanchard}, {Valls-Gabaud}, \&
  {Mamon}}]{BVM92}
{Blanchard} A., {Valls-Gabaud} D., {Mamon} G.~A., 1992, \aap, 264, 365

\bibitem[{{Boomsma} {et~al.}(2008){Boomsma}, {Oosterloo}, {Fraternali}, {van
  der Hulst}, \& {Sancisi}}]{Boomsma+08}
{Boomsma} R., {Oosterloo} T.~A., {Fraternali} F., {van der Hulst} J.~M.,
  {Sancisi} R., 2008, \aap, 490, 555

\bibitem[{{Borne} {et~al.}(2000){Borne}, {Bushouse}, {Lucas}, \&
  {Colina}}]{BBLC00}
{Borne} K.~D., {Bushouse} H., {Lucas} R.~A., {Colina} L., 2000, \apjl, 529, L77

\bibitem[{{Bouch{\'e}} {et~al.}(2010){Bouch{\'e}}, {Dekel}, {Genzel}, {Genel},
  {Cresci}, {F{\"o}rster Schreiber}, {Shapiro}, {Davies}, \&
  {Tacconi}}]{Bouche+10}
{Bouch{\'e}} N., {Dekel} A., {Genzel} R., 
et al.,
2010, \apj,
  718, 1001

\bibitem[{{Bournaud} {et~al.}(2011{\natexlab{a}}){Bournaud}, {Dekel},
  {Teyssier}, {Cacciato}, {Daddi}, {Juneau}, \& {Shankar}}]{bournaud+11}
{Bournaud} F., {Dekel} A., {Teyssier} R., {Cacciato} M., {Daddi} E., {Juneau}
  S., {Shankar} F., 2011{\natexlab{a}}, \apjl, 741, L33

\bibitem[{{Bournaud} {et~al.}(2011{\natexlab{b}}){Bournaud}, {Powell},
  {Chapon}, \& {Teyssier}}]{BPCT11}
{Bournaud} F., {Powell} L.~C., {Chapon} D., {Teyssier} R., 2011{\natexlab{b}},
  in IAU Symposium, Vol. 271, IAU Symposium, {Brummell} N.~H., {Brun} A.~S.,
  {Miesch} M.~S., {Ponty} Y., eds., pp. 160--169

\bibitem[{{Bouwens} {et~al.}(2011){Bouwens}, {Illingworth}, {Oesch}, {Franx},
  {Labbe}, {Trenti}, {van Dokkum}, {Carollo}, {Gonzalez}, {Smit}, \&
  {Magee}}]{Bouwens+12_UV}
{Bouwens} R.~J., {Illingworth} G.~D., {Oesch} P.~A., 
et al.,
  2011, \apj, in press, arXiv:1109.0994

\bibitem[{{Bouwens} {et~al.}(2012){Bouwens}, {Illingworth}, {Oesch}, {Trenti},
  {Labb{\'e}}, {Franx}, {Stiavelli}, {Carollo}, {van Dokkum}, \&
  {Magee}}]{Bouwens+12}
{Bouwens} R.~J., {Illingworth} G.~D., {Oesch} P.~A., 
et al.,
  2012, \apjl, 752, L5

\bibitem[{{Bovill} \& {Ricotti}(2011)}]{BR11b}
{Bovill} M.~S., {Ricotti} M., 2011, \apj, 741, 18

\bibitem[{{Bower}(1991)}]{Bower91}
{Bower} R.~G., 1991, \mnras, 248, 332

\bibitem[{{Bower} {et~al.}(2006){Bower}, {Benson}, {Malbon}, {Helly}, {Frenk},
  {Baugh}, {Cole}, \& {Lacey}}]{Bower+06}
{Bower} R.~G., {Benson} A.~J., {Malbon} R., 
et al.,
  2006, \mnras, 370, 645

\bibitem[{{Boylan-Kolchin} {et~al.}(2011){Boylan-Kolchin}, {Bullock}, \&
  {Kaplinghat}}]{BBK11}
{Boylan-Kolchin} M., {Bullock} J.~S., {Kaplinghat} M., 2011, \mnras, 415, L40

\bibitem[{{Boylan-Kolchin} {et~al.}(2012){Boylan-Kolchin}, {Bullock}, \&
  {Kaplinghat}}]{BBK12}
---, 2012, \mnras, 422, 1203

\bibitem[{{Boylan-Kolchin} {et~al.}(2009){Boylan-Kolchin}, {Springel}, {White},
  {Jenkins}, \& {Lemson}}]{BoylanKolchin+09}
{Boylan-Kolchin} M., {Springel} V., {White} S.~D.~M., {Jenkins} A., {Lemson}
  G., 2009, \mnras, 398, 1150

\bibitem[{{Bradley} {et~al.}(2012){Bradley}, {Trenti}, {Oesch}, {Stiavelli},
  {Treu}, {Bouwens}, {Shull}, {Holwerda}, \& {Pirzkal}}]{Bradley+12}
{Bradley} L.~D., {Trenti} M., {Oesch} P.~A., 
et al.,
  2012, \apj,
  submitted, arXiv:1204.3641

\bibitem[{{Brook} {et~al.}(2012){Brook}, {Stinson}, {Gibson}, {Ro{\v s}kar},
  {Wadsley}, \& {Quinn}}]{Brook+12}
{Brook} C.~B., {Stinson} G., {Gibson} B.~K., {Ro{\v s}kar} R., {Wadsley} J.,
  {Quinn} T., 2012, \mnras, 419, 771

\bibitem[{{Cappellari} {et~al.}(2012){Cappellari}, {McDermid}, {Alatalo},
  {Blitz}, {Bois}, {Bournaud}, {Bureau}, {Crocker}, {Davies}, {Davis}, {de
  Zeeuw}, {Duc}, {Emsellem}, {Khochfar}, {Krajnovi{\'c}}, {Kuntschner},
  {Lablanche}, {Morganti}, {Naab}, {Oosterloo}, {Sarzi}, {Scott}, {Serra},
  {Weijmans}, \& {Young}}]{Cappellari+12}
{Cappellari} M., {McDermid} R.~M., {Alatalo} K., 
et al.,
  2012, \nat, 484, 485

\bibitem[{{Cattaneo} {et~al.}(2006){Cattaneo}, {Dekel}, {Devriendt},
  {Guiderdoni}, \& {Blaizot}}]{Cattaneo+06}
{Cattaneo} A., {Dekel} A., {Devriendt} J., {Guiderdoni} B., {Blaizot} J., 2006,
  \mnras, 370, 1651

\bibitem[{{Cattaneo} {et~al.}(2008){Cattaneo}, {Dekel}, {Faber}, \&
  {Guiderdoni}}]{CDFG08}
{Cattaneo} A., {Dekel} A., {Faber} S.~M., {Guiderdoni} B., 2008, \mnras, 389,
  567

\bibitem[{{Cattaneo} {et~al.}(2009){Cattaneo}, {Faber}, {Binney}, {Dekel},
  {Kormendy}, {Mushotzky}, {Babul}, {Best}, {Br{\"u}ggen}, {Fabian}, {Frenk},
  {Khalatyan}, {Netzer}, {Mahdavi}, {Silk}, {Steinmetz}, \&
  {Wisotzki}}]{Cattaneo+09}
{Cattaneo} A., {Faber} S.~M., {Binney} J., 
et al.,
  2009, \nat, 460, 213

\bibitem[{{Cattaneo} {et~al.}(2011){Cattaneo}, {Mamon}, {Warnick}, \&
  {Knebe}}]{CMWK11}
{Cattaneo} A., {Mamon} G.~A., {Warnick} K., {Knebe} A., 2011, \aap, 533, A5

\bibitem[{{Ceverino} {et~al.}(2010){Ceverino}, {Dekel}, \& {Bournaud}}]{CDB10}
{Ceverino} D., {Dekel} A., {Bournaud} F., 2010, \mnras, 404, 2151

\bibitem[{{Chabrier}(2003)}]{Chabrier03}
{Chabrier} G., 2003, \pasp, 115, 763

\bibitem[{{Cimatti} {et~al.}(2012){Cimatti}, {Nipoti}, \& {Cassata}}]{CNC12}
{Cimatti} A., {Nipoti} C., {Cassata} P., 2012, \mnras, 422, L62

\bibitem[{{Conroy} \& {van Dokkum}(2012)}]{CvD12}
{Conroy} C., {van Dokkum} P., 2012, \apj, submitted, arXiv:1205.6473

\bibitem[{{Conroy} {et~al.}(2006){Conroy}, {Wechsler}, \& {Kravtsov}}]{CWK06}
{Conroy} C., {Wechsler} R.~H., {Kravtsov} A.~V., 2006, \apj, 647, 201

\bibitem[{{Cooper} {et~al.}(2010){Cooper}, {Cole}, {Frenk}, {White}, {Helly},
  {Benson}, {De Lucia}, {Helmi}, {Jenkins}, {Navarro}, {Springel}, \&
  {Wang}}]{Cooper+10}
{Cooper} A.~P., {Cole} S., {Frenk} C.~S., 
et al.,
  2010, \mnras, 406, 744

\bibitem[{{Courtin} {et~al.}(2011){Courtin}, {Rasera}, {Alimi}, {Corasaniti},
  {Boucher}, \& {F{\"u}zfa}}]{Courtin+11}
{Courtin} J., {Rasera} Y., {Alimi} J.-M., {Corasaniti} P.-S., {Boucher} V.,
  {F{\"u}zfa} A., 2011, \mnras, 410, 1911

\bibitem[{{Crocce} {et~al.}(2010){Crocce}, {Fosalba}, {Castander}, \&
  {Gazta{\~n}aga}}]{CFCG10}
{Crocce} M., {Fosalba} P., {Castander} F.~J., {Gazta{\~n}aga} E., 2010, \mnras,
  403, 1353

\bibitem[{{Croton} {et~al.}(2006){Croton}, {Springel}, {White}, {De Lucia},
  {Frenk}, {Gao}, {Jenkins}, {Kauffmann}, {Navarro}, \& {Yoshida}}]{Croton+06}
{Croton} D.~J., {Springel} V., {White} S.~D.~M., 
et al.,  2006,
  \mnras, 365, 11

\bibitem[{{Dai} {et~al.}(2010){Dai}, {Bregman}, {Kochanek}, \&
  {Rasia}}]{DBKR10}
{Dai} X., {Bregman} J.~N., {Kochanek} C.~S., {Rasia} E., 2010, \apj, 719, 119

\bibitem[{{De Lucia} \& {Blaizot}(2007)}]{DLB07}
{De Lucia} G., {Blaizot} J., 2007, \mnras, 375, 2

\bibitem[{{de Vaucouleurs}(1961)}]{deVaucouleurs61}
{de Vaucouleurs} G., 1961, \apjs, 5, 233

\bibitem[{{Debuhr} {et~al.}(2012){Debuhr}, {Quataert}, \& {Ma}}]{DQM12}
{Debuhr} J., {Quataert} E., {Ma} C.-P., 2012, \mnras, 420, 2221

\bibitem[{{Dekel} {et~al.}(2009){Dekel}, {Birnboim}, {Engel}, {Freundlich},
  {Goerdt}, {Mumcuoglu}, {Neistein}, {Pichon}, {Teyssier}, \&
  {Zinger}}]{Dekel+09}
{Dekel} A., {Birnboim} Y., {Engel} G., 
et al.,  2009,
  \nat, 457, 451

\bibitem[{{Dekel} \& {Silk}(1986)}]{DS86}
{Dekel} A., {Silk} J., 1986, \apj, 303, 39

\bibitem[{{Di Matteo} {et~al.}(2012){Di Matteo}, {Khandai}, {DeGraf}, {Feng},
  {Croft}, {Lopez}, \& {Springel}}]{DiMatteo+12}
{Di Matteo} T., {Khandai} N., {DeGraf} C.,  {Feng} Y., {Croft} R.~A.~C., {Lopez}
 J., {Springel} V., 
  2012, \apjl, 745, L29

\bibitem[{{D'Onghia} {et~al.}(2009){D'Onghia}, {Besla}, {Cox}, \&
  {Hernquist}}]{DBCH09}
{D'Onghia} E., {Besla} G., {Cox} T.~J., {Hernquist} L., 2009, \nat, 460, 605

\bibitem[{{Dubois} {et~al.}(2012{\natexlab{a}}){Dubois}, {Devriendt}, {Slyz},
  \& {Teyssier}}]{DDST12}
{Dubois} Y., {Devriendt} J., {Slyz} A., {Teyssier} R., 2012{\natexlab{a}},
  \mnras, 420, 2662

\bibitem[{{Dubois} {et~al.}(2012{\natexlab{b}}){Dubois}, {Pichon}, {Devriendt},
  {Silk}, {Haehnelt}, {Kimm}, \& {Slyz}}]{Dubois+12_blowout}
{Dubois} Y., {Pichon} C., {Devriendt} J., {Silk} J., {Haehnelt} M., {Kimm} T.,
  {Slyz} A., 2012{\natexlab{b}}, \mnras, submitted, arXiv:1206.5838

\bibitem[{{Dutton}(2012)}]{Dutton12}
{Dutton} A.~A., 2012, \mnras, in press, arXiv:1206.1855

\bibitem[{{Dutton} {et~al.}(2012{\natexlab{a}}){Dutton}, {Mendel}, \&
  {Simard}}]{DMS12}
{Dutton} A.~A., {Mendel} J.~T., {Simard} L., 2012{\natexlab{a}}, \mnras, 422,
  L33

\bibitem[{{Dutton} {et~al.}(2012{\natexlab{b}}){Dutton}, {Treu}, {Brewer},
  {Marshall}, {Auger}, {Barnabe}, {Koo}, {Bolton}, \& {Koopmans}}]{Dutton+12}
{Dutton} A.~A., {Treu} T., {Brewer} B.~J., {Marshall} P.~J., {Auger} M.~W.,
  {Barnabe} M., {Koo} D.~C., {Bolton} A.~S., {Koopmans} L.~V.~E.,
  2012{\natexlab{b}}, \mnras, submitted, arXiv:1206.4310

\bibitem[{{Elbaz} {et~al.}(2011){Elbaz}, {Dickinson}, {Hwang},
  {D{\'{\i}}az-Santos}, {Magdis}, {Magnelli}, {Le Borgne}, {Galliano},
  {Pannella}, {Chanial}, {Armus}, {Charmandaris}, {Daddi}, {Aussel}, {Popesso},
  {Kartaltepe}, {Altieri}, {Valtchanov}, {Coia}, {Dannerbauer}, {Dasyra},
  {Leiton}, {Mazzarella}, {Alexander}, {Buat}, {Burgarella}, {Chary}, {Gilli},
  {Ivison}, {Juneau}, {Le Floc'h}, {Lutz}, {Morrison}, {Mullaney}, {Murphy},
  {Pope}, {Scott}, {Brodwin}, {Calzetti}, {Cesarsky}, {Charlot}, {Dole},
  {Eisenhardt}, {Ferguson}, {F{\"o}rster Schreiber}, {Frayer}, {Giavalisco},
  {Huynh}, {Koekemoer}, {Papovich}, {Reddy}, {Surace}, {Teplitz}, {Yun}, \&
  {Wilson}}]{Elbaz+11}
{Elbaz} D., {Dickinson} M., {Hwang} H.~S., 
et al.,  2011, \aap,
  533, A119

\bibitem[{{Elmegreen}(1997)}]{Elmegreen97}
{Elmegreen} B.~G., 1997, in Revista Mexicana de Astronomia y Astrofisica, vol.
  27, Vol.~6, Revista Mexicana de Astronomia y Astrofisica Conference Series,
  {Franco} J., {Terlevich} R., {Serrano} A., eds., p. 165

\bibitem[{{Evrard}(1988)}]{Evrard88}
{Evrard} A.~E., 1988, \mnras, 235, 911

\bibitem[{{Faber}(1973)}]{Faber73}
{Faber} S.~M., 1973, \apj, 179, 731

\bibitem[{{Faber} \& {Jackson}(1976)}]{FJ76}
{Faber} S.~M., {Jackson} R.~E., 1976, \apj, 204, 668

\bibitem[{{Famaey} \& {McGaugh}(2011)}]{FM11}
{Famaey}, B., {McGaugh}, S., 2011, Living Reviews in Relativity, in press,
arXiv:1112.3960

\bibitem[{{Fanidakis} {et~al.}(2011){Fanidakis}, {Baugh}, {Benson}, {Bower},
  {Cole}, {Done}, \& {Frenk}}]{Fanidakis+11}
{Fanidakis} N., {Baugh} C.~M., {Benson} A.~J., {Bower} R.~G., {Cole} S., {Done}
  C., {Frenk} C.~S., 2011, \mnras, 410, 53

\bibitem[{{Faucher-Gigu{\`e}re} \& {Kere{\v s}}(2011)}]{FK11}
{Faucher-Gigu{\`e}re} C.-A., {Kere{\v s}} D., 2011, \mnras, 412, L118

\bibitem[{{Faucher-Giguere} \& {Quataert}(2012)}]{FQ12}
{Faucher-Giguere} C.-A., {Quataert} E., 2012, \mnras, in press, arXiv:1204.2547

\bibitem[{{Feldman} {et~al.}(2010){Feldman}, {Watkins}, \& {Hudson}}]{FWH10}
{Feldman} H.~A., {Watkins} R., {Hudson} M.~J., 2010, \mnras, 407, 2328

\bibitem[{{Ferrarese} \& {Merritt}(2000)}]{FM00}
{Ferrarese} L., {Merritt} D., 2000, \apjl, 539, L9

\bibitem[{{Ferreras} {et~al.}(2012){Ferreras}, {La Barbera}, {de Carvalho}, {de
  la Rosa}, {Vazdekis}, {Falcon-Barroso}, \& {Ricciardelli}}]{Ferreras+12}
{Ferreras} I., {La Barbera} F., {de Carvalho} R.~R., {de la Rosa} I.~G.,
 {Vazdekis} A., {Falcon-Barroso} J., {Ricciardelli} E., 
  2012, \apjl,
  submitted, arXiv:1206.1594

\bibitem[{{Ferrero} {et~al.}(2011){Ferrero}, {Abadi}, {Navarro}, {Sales}, \&
  {Gurovich}}]{Ferrero+11}
{Ferrero} I., {Abadi} M.~G., {Navarro} J.~F., {Sales} L.~V., {Gurovich} S.,
  2011, \mnras, submitted, arXiv:1111.6609

\bibitem[{{Fontanot} {et~al.}(2009){Fontanot}, {De Lucia}, {Monaco},
  {Somerville}, \& {Santini}}]{Fontanot+09b}
{Fontanot} F., {De Lucia} G., {Monaco} P., {Somerville} R.~S., {Santini} P.,
  2009, \mnras, 397, 1776

\bibitem[{{Frenk} {et~al.}(1999){Frenk}, {White}, {Bode}, {Bond}, {Bryan},
  {Cen}, {Couchman}, {Evrard}, {Gnedin}, {Jenkins}, {Khokhlov}, {Klypin},
  {Navarro}, {Norman}, {Ostriker}, {Owen}, {Pearce}, {Pen}, {Steinmetz},
  {Thomas}, {Villumsen}, {Wadsley}, {Warren}, {Xu}, \& {Yepes}}]{Frenk+99}
{Frenk} C.~S., {White} S.~D.~M., {Bode} P., 
et al.,  1999, \apj, 525, 554

\bibitem[{{Gaibler} {et~al.}(2011){Gaibler}, {Khochfar}, {Krause}, \&
  {Silk}}]{GKKS11}
{Gaibler} V., {Khochfar} S., {Krause} M., {Silk} J., 2011, \mnras, in press,
  arXiv:1111.4478

\bibitem[{{Gentile} {et~al.}(2012){Gentile}, {Angus}, {Famaey}, {Oh}, \& {de
  Blok}}]{Gentile+12}
{Gentile} G., {Angus} G.~W., {Famaey} B., {Oh} S.-H., {de Blok} W.~J.~G., 2012,
  \aap, 543, A47

\bibitem[{{Genzel} {et~al.}(2010){Genzel}, {Tacconi}, {Gracia-Carpio},
  {Sternberg}, {Cooper}, {Shapiro}, {Bolatto}, {Bouch{\'e}}, {Bournaud},
  {Burkert}, {Combes}, {Comerford}, {Cox}, {Davis}, {Schreiber},
  {Garcia-Burillo}, {Lutz}, {Naab}, {Neri}, {Omont}, {Shapley}, \&
  {Weiner}}]{Genzel+10}
{Genzel} R., {Tacconi} L.~J., {Gracia-Carpio} J., 
et al.,  2010, \mnras, 407, 2091

\bibitem[{{Gingold} \& {Monaghan}(1977)}]{GM77}
{Gingold} R.~A., {Monaghan} J.~J., 1977, \mnras, 181, 375

\bibitem[{{Gnat} \& {Sternberg}(2007)}]{GS07}
{Gnat} O., {Sternberg} A., 2007, \apjs, 168, 213

\bibitem[{{Gnedin}(2000)}]{Gnedin00}
{Gnedin} N.~Y., 2000, \apj, 542, 535

\bibitem[{{Gnedin} {et~al.}(2004){Gnedin}, {Kravtsov}, {Klypin}, \&
  {Nagai}}]{GKKN04}
{Gnedin} O.~Y., {Kravtsov} A.~V., {Klypin} A.~A., {Nagai} D., 2004, \apj, 616,
  16

\bibitem[{{Gonz{\'a}lez} {et~al.}(2011b){Gonzalez}, {Bouwens}, {Labbe},
  {Illingworth}, {Oesch}, {Franx}, \& {Magee}}]{Gonzalez+12}
{Gonz{\'a}lez} V., {Bouwens} R., {Labb{\'e}} I., {Illingworth} G., {Oesch} P., {Franx}
  M., {Magee} D., 2011b, \apj, submitted, arXiv:1110.6441

\bibitem[{{Gonz{\'a}lez} {et~al.}(2011a){Gonz{\'a}lez}, {Labb{\'e}}, {Bouwens},
  {Illingworth}, {Franx}, \& {Kriek}}]{Gonzalez+11}
{Gonz{\'a}lez} V., {Labb{\'e}} I., {Bouwens} R.~J., {Illingworth} G., {Franx}
  M., {Kriek} M., 2011a, \apjl, 735, L34

\bibitem[{{Governato} {et~al.}(2010){Governato}, {Brook}, {Mayer}, {Brooks},
  {Rhee}, {Wadsley}, {Jonsson}, {Willman}, {Stinson}, {Quinn}, \&
  {Madau}}]{Governato+10}
{Governato} F., {Brook} C., {Mayer} L., 
et al.,  2010, \nat,
  463, 203

\bibitem[{{Governato} {et~al.}(2012){Governato}, {Zolotov}, {Pontzen},
  {Christensen}, {Oh}, {Brooks}, {Quinn}, {Shen}, \& {Wadsley}}]{Governato+12}
{Governato} F., {Zolotov} A., {Pontzen} A., 
et al.,  2012, \mnras, 422, 1231

\bibitem[{{Graham} {et~al.}(2011){Graham}, {Onken}, {Athanassoula}, \&
  {Combes}}]{GOAC11}
{Graham} A.~W., {Onken} C.~A., {Athanassoula} E., {Combes} F., 2011, \mnras,
  412, 2211

\bibitem[{{Grillmair}(2009)}]{Grillmair09}
{Grillmair} C.~J., 2009, \apj, 693, 1118

\bibitem[{{Guedes} {et~al.}(2011){Guedes}, {Callegari}, {Madau}, \&
  {Mayer}}]{GCMM11}
{Guedes} J., {Callegari} S., {Madau} P., {Mayer} L., 2011, \apj, 742, 76

\bibitem[{{Gunn} \& {Gott}(1972)}]{GG72}
{Gunn} J.~E., {Gott} J.~R., 1972, \apj, 176, 1

\bibitem[{{Guo} {et~al.}(2011){Guo}, {White}, {Boylan-Kolchin}, {De Lucia},
  {Kauffmann}, {Lemson}, {Li}, {Springel}, \& {Weinmann}}]{Guo+11}
{Guo} Q., {White} S., {Boylan-Kolchin} M., {De Lucia} G., {Kauffmann} G.,
  {Lemson} G., {Li} C., {Springel} V., {Weinmann} S., 2011, \mnras, 413, 101

\bibitem[{{Guo} {et~al.}(2010){Guo}, {White}, {Li}, \&
  {Boylan-Kolchin}}]{GWLB10}
{Guo} Q., {White} S., {Li} C., {Boylan-Kolchin} M., 2010, \mnras, 404, 1111

\bibitem[{{Guo} \& {White}(2008)}]{GW08}
{Guo} Q., {White} S.~D.~M., 2008, \mnras, 384, 2

\bibitem[{{Gurovich} {et~al.}(2010){Gurovich}, {Freeman}, {Jerjen},
  {Staveley-Smith}, \& {Puerari}}]{Gurovich+10}
{Gurovich} S., {Freeman} K., {Jerjen} H., {Staveley-Smith} L., {Puerari} I.,
  2010, \aj, 140, 663

\bibitem[{{Hall} {et~al.}(2011){Hall}, {Courteau}, {Dutton}, {McDonald}, \&
  {Zhu}}]{Hall+11}
{Hall} M., {Courteau} S., {Dutton} A.~A., {McDonald} M., {Zhu} Y., 2011,
  \mnras, submitted, arXiv:1111.5009

\bibitem[{{Hand} {et~al.}(2012){Hand}, {Addison}, {Aubourg}, {Battaglia},
  {Battistelli}, {Bizyaev}, {Bond}, {Brewington}, {Brinkmann}, {Brown}, {Das},
  {Dawson}, {Devlin}, {Dunkley}, {Dunner}, {Eisenstein}, {Fowler}, {Gralla},
  {Hajian}, {Halpern}, {Hilton}, {Hincks}, {Hlozek}, {Hughes}, {Infante},
  {Irwin}, {Kosowsky}, {Lin}, {Malanushenko}, {Malanushenko}, {Marriage},
  {Marsden}, {Menanteau}, {Moodley}, {Niemack}, {Nolta}, {Oravetz}, {Page},
  {Palanque-Delabrouille}, {Pan}, {Reese}, {Schlegel}, {Schneider}, {Sehgal},
  {Shelden}, {Sievers}, {Sifon}, {Simmons}, {Snedden}, {Spergel}, {Staggs},
  {Swetz}, {Switzer}, {Trac}, {Weaver}, {Wollack}, {Yeche}, \&
  {Zunckel}}]{Hand+12}
{Hand} N., {Addison} G.~E., {Aubourg} E., 
et al.,  2012, \prl, 109, 041101

\bibitem[{{Heald} {et~al.}(2011){Heald}, {J{\'o}zsa}, {Serra}, {Zschaechner},
  {Rand}, {Fraternali}, {Oosterloo}, {Walterbos}, {J{\"u}tte}, \&
  {Gentile}}]{Heald+11a}
{Heald} G., {J{\'o}zsa} G., {Serra} P., {Zschaechner} L., {Rand} R.,
  {Fraternali} F., {Oosterloo} T., {Walterbos} R., {J{\"u}tte} E., {Gentile}
  G., 2011, \aap, 526, A118

\bibitem[{{Henriques} {et~al.}(2011){Henriques}, {Maraston}, {Monaco},
  {Fontanot}, {Menci}, {De Lucia}, \& {Tonini}}]{Henriques+11}
{Henriques} B., {Maraston} C., {Monaco} P., {Fontanot} F., {Menci} N., {De
  Lucia} G., {Tonini} C., 2011, \mnras, 415, 3571

\bibitem[{{Ho}(2007)}]{Ho07}
{Ho} L.~C., 2007, \apj, 668, 94

\bibitem[{{Jiang} {et~al.}(2008){Jiang}, {Jing}, {Faltenbacher}, {Lin}, \&
  {Li}}]{Jiang+08}
{Jiang} C.~Y., {Jing} Y.~P., {Faltenbacher} A., {Lin} W.~P., {Li} C., 2008,
  \apj, 675, 1095

\bibitem[{{Kashlinsky} {et~al.}(2010){Kashlinsky}, {Atrio-Barandela},
  {Ebeling}, {Edge}, \& {Kocevski}}]{Kashlinsky+10}
{Kashlinsky} A., {Atrio-Barandela} F., {Ebeling} H., {Edge} A., {Kocevski} D.,
  2010, \apjl, 712, L81

\bibitem[{{Kaviraj} {et~al.}(2009){Kaviraj}, {Peirani}, {Khochfar}, {Silk}, \&
  {Kay}}]{Kaviraj+09}
{Kaviraj} S., {Peirani} S., {Khochfar} S., {Silk} J., {Kay} S., 2009, \mnras,
  394, 1713

\bibitem[{{Kennicutt} {et~al.}(2007){Kennicutt}, {Calzetti}, {Walter}, {Helou},
  {Hollenbach}, {Armus}, {Bendo}, {Dale}, {Draine}, {Engelbracht}, {Gordon},
  {Prescott}, {Regan}, {Thornley}, {Bot}, {Brinks}, {de Blok}, {de Mello},
  {Meyer}, {Moustakas}, {Murphy}, {Sheth}, \& {Smith}}]{Kennicutt+07}
{Kennicutt} Jr. R.~C., {Calzetti} D., {Walter} F., 
et al.,  2007, \apj, 671, 333

\bibitem[{{Khandai} {et~al.}(2012){Khandai}, {Feng}, {DeGraf}, {Di Matteo}, \&
  {Croft}}]{Khandai+12}
{Khandai} N., {Feng} Y., {DeGraf} C., {Di Matteo} T., {Croft} R.~A.~C., 2012,
  \mnras, 423, 2397

\bibitem[{{Khochfar} \& {Silk}(2011)}]{KS11}
{Khochfar} S., {Silk} J., 2011, \mnras, 410, L42

\bibitem[{{Knebe} {et~al.}(2011){Knebe}, {Knollmann}, {Muldrew}, {Pearce},
  {Aragon-Calvo}, {Ascasibar}, {Behroozi}, {Ceverino}, {Colombi}, {Diemand},
  {Dolag}, {Falck}, {Fasel}, {Gardner}, {Gottl{\"o}ber}, {Hsu}, {Iannuzzi},
  {Klypin}, {Luki{\'c}}, {Maciejewski}, {McBride}, {Neyrinck}, {Planelles},
  {Potter}, {Quilis}, {Rasera}, {Read}, {Ricker}, {Roy}, {Springel}, {Stadel},
  {Stinson}, {Sutter}, {Turchaninov}, {Tweed}, {Yepes}, \& {Zemp}}]{Knebe+11}
{Knebe} A., {Knollmann} S.~R., {Muldrew} S.~I., 
et al.,  2011, \mnras, 415, 2293

\bibitem[{{Koposov} {et~al.}(2009){Koposov}, {Yoo}, {Rix}, {Weinberg},
  {Macci{\`o}}, \& {Escud{\'e}}}]{Koposov+09}
{Koposov} S.~E., {Yoo} J., {Rix} H.-W., {Weinberg} D.~H., {Macci{\`o}} A.~V.,
  {Escud{\'e}} J.~M., 2009, \apj, 696, 2179

\bibitem[{{Kormendy} \& {Bender}(2011)}]{KB11}
{Kormendy} J., {Bender} R., 2011, \nat, 469, 377

\bibitem[{{Kormendy} \& {Bender}(2012)}]{KB12}
--- , 2012, \apjs, 198, 2

\bibitem[{{Kormendy} {et~al.}(2011){Kormendy}, {Bender}, \& {Cornell}}]{KBC11}
{Kormendy} J., {Bender} R., {Cornell} M.~E., 2011, \nat, 469, 374

\bibitem[{{Kormendy} {et~al.}(2010){Kormendy}, {Drory}, {Bender}, \&
  {Cornell}}]{KDBC10}
{Kormendy} J., {Drory} N., {Bender} R., {Cornell} M.~E., 2010, \apj, 723, 54

\bibitem[{{Kravtsov}(2010)}]{Kravtsov10}
{Kravtsov} A., 2010, Advances in Astronomy, 2010, 281913

\bibitem[{{Kravtsov} {et~al.}(1997){Kravtsov}, {Klypin}, \& {Khokhlov}}]{KKK97}
{Kravtsov} A.~V., {Klypin} A.~A., {Khokhlov} A.~M., 1997, \apjs, 111, 73

\bibitem[{{Kroupa} {et~al.}(2010){Kroupa}, {Famaey}, {de Boer},
  {Dabringhausen}, {Pawlowski}, {Boily}, {Jerjen}, {Forbes}, {Hensler}, \&
  {Metz}}]{Kroupa+10}
{Kroupa} P., {Famaey} B., {de Boer} K.~S., 
et~al.,
  2010, \aap, 523, A32

\bibitem[{{Kroupa} {et~al.}(2005){Kroupa}, {Theis}, \& {Boily}}]{KTB05}
{Kroupa} P., {Theis} C., {Boily} C.~M., 2005, \aap, 431, 517

\bibitem[{{Krumholz} \& {Dekel}(2012)}]{KD12}
{Krumholz} M.~R., {Dekel} A., 2012, \apj, 753, 16

\bibitem[{{Krumholz} {et~al.}(2012){Krumholz}, {Dekel}, \& {McKee}}]{KDM12}
{Krumholz} M.~R., {Dekel} A., {McKee} C.~F., 2012, \apj, 745, 69

\bibitem[{{Lacey} \& {Cole}(1993)}]{LC93}
{Lacey} C., {Cole} S., 1993, \mnras, 262, 627

\bibitem[{{Le Borgne} {et~al.}(2009){Le Borgne}, {Elbaz}, {Ocvirk}, \&
  {Pichon}}]{LBEOP09}
{Le Borgne} D., {Elbaz} D., {Ocvirk} P., {Pichon} C., 2009, \aap, 504, 727

\bibitem[{{Li} \& {White}(2009)}]{LW09}
{Li} C., {White} S.~D.~M., 2009, \mnras, 398, 2177

\bibitem[{{Li} {et~al.}(2007){Li}, {Hernquist}, {Robertson}, {Cox}, {Hopkins},
  {Springel}, {Gao}, {Di Matteo}, {Zentner}, {Jenkins}, \& {Yoshida}}]{Li+07}
{Li} Y., {Hernquist} L., {Robertson} B., {Cox} T.~J., {Hopkins} P.~F.,
  {Springel} V., {Gao} L., {Di Matteo} T., {Zentner} A.~R., {Jenkins} A.,
  {Yoshida} N., 2007, \apj, 665, 187

\bibitem[{{Libeskind} {et~al.}(2005){Libeskind}, {Frenk}, {Cole}, {Helly},
  {Jenkins}, {Navarro}, \& {Power}}]{Libeskind+05}
{Libeskind} N.~I., {Frenk} C.~S., {Cole} S., {Helly} J.~C., {Jenkins} A.,
  {Navarro} J.~F., {Power} C., 2005, \mnras, 363, 146

\bibitem[{{Libeskind} {et~al.}(2011){Libeskind}, {Knebe}, {Hoffman},
  {Gottl{\"o}ber}, {Yepes}, \& {Steinmetz}}]{Libeskind+11}
{Libeskind} N.~I., {Knebe} A., {Hoffman} Y., {Gottl{\"o}ber} S., {Yepes} G.,
  {Steinmetz} M., 2011, \mnras, 411, 1525

\bibitem[{{L{\'o}pez-Sanjuan} {et~al.}(2012){L{\'o}pez-Sanjuan}, {Le
  F{\`e}vre}, {Ilbert}, {Tasca}, {Bridge}, {Cucciati}, {Kampczyk}, {Pozzetti},
  {Xu}, {Carollo}, {Contini}, {Kneib}, {Lilly}, {Mainieri}, {Renzini},
  {Sanders}, {Scodeggio}, {Scoville}, {Taniguchi}, {Zamorani}, {Aussel},
  {Bardelli}, {Bolzonella}, {Bongiorno}, {Capak}, {Caputi}, {de la Torre}, {de
  Ravel}, {Franzetti}, {Garilli}, {Iovino}, {Knobel}, {Kova{\v c}},
  {Lamareille}, {Le Borgne}, {Le Brun}, {Le Floc'h}, {Maier}, {McCracken},
  {Mignoli}, {Pell{\'o}}, {Peng}, {P{\'e}rez-Montero}, {Presotto},
  {Ricciardelli}, {Salvato}, {Silverman}, {Tanaka}, {Tresse}, {Vergani},
  {Zucca}, {Barnes}, {Bordoloi}, {Cappi}, {Cimatti}, {Coppa}, {Koekoemoer},
  {Liu}, {Moresco}, {Nair}, {Oesch}, {Schawinski}, \&
  {Welikala}}]{LopezSanjuan+12}
{L{\'o}pez-Sanjuan} C., {Le F{\`e}vre} O., {Ilbert} O., 
et al.,  2012, \aap, submitted,
  arXiv:1202.4674

\bibitem[{{Lynden-Bell}(1982)}]{LyndenBell82}
{Lynden-Bell} D., 1982, The Observatory, 102, 202

\bibitem[{{Macci{\`o}} {et~al.}(2010){Macci{\`o}}, {Kang}, {Fontanot},
  {Somerville}, {Koposov}, \& {Monaco}}]{Maccio+10}
{Macci{\`o}} A.~V., {Kang} X., {Fontanot} F., {Somerville} R.~S., {Koposov} S.,
  {Monaco} P., 2010, \mnras, 402, 1995

\bibitem[{{Macci{\`o}} {et~al.}(2012){Macci{\`o}}, {Stinson}, {Brook},
  {Wadsley}, {Couchman}, {Shen}, {Gibson}, \& {Quinn}}]{Maccio+12}
{Macci{\`o}} A.~V., {Stinson} G., {Brook} C.~B., {Wadsley} J., {Couchman}
  H.~M.~P., {Shen} S., {Gibson} B.~K., {Quinn} T., 2012, \apjl, 744, L9

\bibitem[{{Magorrian} {et~al.}(1998){Magorrian}, {Tremaine}, {Richstone},
  {Bender}, {Bower}, {Dressler}, {Faber}, {Gebhardt}, {Green}, {Grillmair},
  {Kormendy}, \& {Lauer}}]{Magorrian+98}
{Magorrian} J., {Tremaine} S., {Richstone} D., 
et al.,  1998, \aj, 115, 2285

\bibitem[{{Mamon} {et~al.}(2012){Mamon}, {Tweed}, {Thuan}, \&
  {Cattaneo}}]{MTTC12}
{Mamon} G.~A., {Tweed} D., {Thuan} T.~X., {Cattaneo} A., 2012, in Dwarf
  Galaxies: Keys to Galaxy Formation and Evolution, {Papaderos} P., {Recchi}
  S., {Hensler} G., eds., Springer Verlag, Berlin, Heidelberg, pp. 39--46,
  arXiv:1103.5349

\bibitem[{{Marchesini} {et~al.}(2009){Marchesini}, {van Dokkum}, {F{\"o}rster
  Schreiber}, {Franx}, {Labb{\'e}}, \& {Wuyts}}]{Marchesini+09}
{Marchesini} D., {van Dokkum} P.~G., {F{\"o}rster Schreiber} N.~M., {Franx} M.,
  {Labb{\'e}} I., {Wuyts} S., 2009, \apj, 701, 1765

\bibitem[{{Marinoni} \& {Hudson}(2002)}]{MH02}
{Marinoni} C., {Hudson} M.~J., 2002, \apj, 569, 101

\bibitem[{{Marinoni} {et~al.}(2002){Marinoni}, {Hudson}, \& {Giuricin}}]{MHG02}
{Marinoni} C., {Hudson} M.~J., {Giuricin} G., 2002, \apj, 569, 91

\bibitem[{{Martin} {et~al.}(2012)}]{MGHG12}
{Martin}, A.~M., {Giovanelli}, R., {Haynes}, M.~P., {Guzzo}, L.,
2012, \apj, 750, 38

\bibitem[{{Mart{\'{\i}}nez-Delgado} {et~al.}(2010){Mart{\'{\i}}nez-Delgado},
  {Gabany}, {Crawford}, {Zibetti}, {Majewski}, {Rix}, {Fliri},
  {Carballo-Bello}, {Bardalez-Gagliuffi}, {Pe{\~n}arrubia}, {Chonis}, {Madore},
  {Trujillo}, {Schirmer}, \& {McDavid}}]{MartinezDelgado+10}
{Mart{\'{\i}}nez-Delgado} D., {Gabany} R.~J., {Crawford} K., 
et al.,  2010, \aj, 140, 962

\bibitem[{{Mashchenko} {et~al.}(2006){Mashchenko}, {Couchman}, \&
  {Wadsley}}]{MCW06}
{Mashchenko} S., {Couchman} H.~M.~P., {Wadsley} J., 2006, \nat, 442, 539

\bibitem[{{Mayer} {et~al.}(2007){Mayer}, {Kazantzidis}, {Madau}, {Colpi},
  {Quinn}, \& {Wadsley}}]{Mayer+07}
{Mayer} L., {Kazantzidis} S., {Madau} P., {Colpi} M., {Quinn} T., {Wadsley} J.,
  2007, Science, 316, 1874


\bibitem[{{McCarthy} {et~al.}(2012){McCarthy}, {Schaye}, {Font},
  {Theuns}, {Frenk}, {Crain}, \& {Dalla Vecchia}}]{McCarthy+12}
{McCarthy} I.~G., {Schaye} J., {Font} A.~S., {Theuns} T., {Frenk} C.~S.,
{Crain} R.~A., {Dalla Vecchia} C., 2012, \mnras, submitted, arXiv:1204.5195

\bibitem[{{McConnell} {et~al.}(2011){McConnell}, {Ma}, {Gebhardt}, {Wright},
  {Murphy}, {Lauer}, {Graham}, \& {Richstone}}]{McConnell+11}
{McConnell} N.~J., {Ma} C.-P., {Gebhardt} K., {Wright} S.~A., {Murphy} J.~D.,
  {Lauer} T.~R., {Graham} J.~R., {Richstone} D.~O., 2011, \nat, 480, 215

\bibitem[{{McGaugh}(2011)}]{McGaugh11}
{McGaugh} S.~S., 2011, Physical Review Letters, 106, 121303

\bibitem[{{McGaugh}(2012)}]{McGaugh12}
---, 2012, \aj, 143, 40

\bibitem[{{McLure} {et~al.}(2012){McLure}, {Pearce}, {Dunlop}, {Cirasuolo},
  {Curtis-Lake}, {Bruce}, {Caputi}, {Almaini}, {Bonfield}, {Bradshaw},
  {Buitrago}, {Chuter}, {Foucaud}, {Hartley}, \& {Jarvis}}]{Mclure+12}
{McLure} R.~J., {Pearce} H.~J., {Dunlop} J.~S., 
et al.,  2012, \mnras, submitted, arXiv:1205.4058

\bibitem[{{Milgrom}(1983)}]{Milgrom83}
{Milgrom} M., 1983, \apj, 270, 365

\bibitem[{{Monaghan}(1992)}]{Monaghan92}
{Monaghan} J.~J., 1992, \araa, 30, 543

\bibitem[{{Moore} {et~al.}(1998){Moore}, {Lake}, \& {Katz}}]{MLK98}
{Moore} B., {Lake} G., {Katz} N., 1998, \apj, 495, 139

\bibitem[{{Navarro} \& {Steinmetz}(2000)}]{NS00}
{Navarro} J.~F., {Steinmetz} M., 2000, \apj, 538, 477

\bibitem[{{Neistein} \& {Weinmann}(2010)}]{NW10}
{Neistein} E., {Weinmann} S.~M., 2010, \mnras, 405, 2717

\bibitem[{{Nickerson} {et~al.}(2011){Nickerson}, {Stinson}, {Couchman},
  {Bailin}, \& {Wadsley}}]{Nickerson+11}
{Nickerson} S., {Stinson} G., {Couchman} H.~M.~P., {Bailin} J., {Wadsley} J.,
  2011, \mnras, 415, 257

\bibitem[{{Oh} {et~al.}(2011){Oh}, {Brook}, {Governato}, {Brinks}, {Mayer}, {de
  Blok}, {Brooks}, \& {Walter}}]{Oh+11}
{Oh} S.-H., {Brook} C., {Governato} F., {Brinks} E., {Mayer} L., {de Blok}
  W.~J.~G., {Brooks} A., {Walter} F., 2011, \aj, 142, 24

\bibitem[{{Okamoto}(2012)}]{Okamoto12}
{Okamoto} T., 2012, \mnras, submitted, arXiv:1203.5372

\bibitem[{{Okamoto} {et~al.}(2008){Okamoto}, {Gao}, \& {Theuns}}]{OGT08}
{Okamoto} T., {Gao} L., {Theuns} T., 2008, \mnras, 390, 920

\bibitem[{{Onions} {et~al.}(2012){Onions}, {Knebe}, {Pearce}, {Muldrew}, {Lux},
  {Knollmann}, {Ascasibar}, {Behroozi}, {Elahi}, {Han}, {Maciejewski},
  {Merch{\'a}n}, {Neyrinck}, {Ruiz}, {Sgr{\'o}}, {Springel}, \&
  {Tweed}}]{Onions+12}
{Onions} J., {Knebe} A., {Pearce} F.~R., 
et al.,  2012, \mnras, 423, 1200

\bibitem[{{O'Shea} {et~al.}(2005){O'Shea}, {Bryan}, {Bordner}, {Norman},
  {Abel}, {Harkness}, \& {Kritsuk}}]{OShea+04}
{O'Shea} B.~W., {Bryan} G., {Bordner} J., {Norman} M.~L., {Abel} T., {Harkness}
  R., {Kritsuk} A., 2005, in Lecture Notes in Computational Science and
  Engineering, Vol.~41, Adaptive Mesh Refinement --- Theory and Applications,
  {Plewa} T., {Linde} T., {Weirs} V.~G., eds., Springer, Berlin, Heidelberg,
  pp. 341--350, arXiv:astro-ph/0403044

\bibitem[{{Panter} {et~al.}(2007){Panter}, {Jimenez}, {Heavens}, \&
  {Charlot}}]{PJHC07}
{Panter} B., {Jimenez} R., {Heavens} A.~F., {Charlot} S., 2007, \mnras, 378,
  1550

\bibitem[{{Peebles}(2007)}]{Peebles07}
{Peebles} P.~J.~E., 2007, Nuovo Cimento B Serie, 122, 1035

\bibitem[{{Peirani} {et~al.}(2012){Peirani}, {Jung}, {Silk}, \&
  {Pichon}}]{PJSP12}
{Peirani} S., {Jung} I., {Silk} J., {Pichon} C., 2012, \mnras, submitted,
  arXiv:1205.4694

\bibitem[{{Peng} {et~al.}(2010){Peng}, {Lilly}, {Kova{\v c}}, {Bolzonella},
  {Pozzetti}, {Renzini}, {Zamorani}, {Ilbert}, {Knobel}, {Iovino}, {Maier},
  {Cucciati}, {Tasca}, {Carollo}, {Silverman}, {Kampczyk}, {de Ravel},
  {Sanders}, {Scoville}, {Contini}, {Mainieri}, {Scodeggio}, {Kneib}, {Le
  F{\`e}vre}, {Bardelli}, {Bongiorno}, {Caputi}, {Coppa}, {de la Torre},
  {Franzetti}, {Garilli}, {Lamareille}, {Le Borgne}, {Le Brun}, {Mignoli},
  {Perez Montero}, {Pello}, {Ricciardelli}, {Tanaka}, {Tresse}, {Vergani},
  {Welikala}, {Zucca}, {Oesch}, {Abbas}, {Barnes}, {Bordoloi}, {Bottini},
  {Cappi}, {Cassata}, {Cimatti}, {Fumana}, {Hasinger}, {Koekemoer},
  {Leauthaud}, {Maccagni}, {Marinoni}, {McCracken}, {Memeo}, {Meneux}, {Nair},
  {Porciani}, {Presotto}, \& {Scaramella}}]{Peng+10}
{Peng} Y.-j., {Lilly} S.~J., {Kova{\v c}} K., 
et al.,  2010, \apj, 721, 193

\bibitem[{{P{\'e}rez-Gonz{\'a}lez} {et~al.}(2008){P{\'e}rez-Gonz{\'a}lez},
  {Rieke}, {Villar}, {Barro}, {Blaylock}, {Egami}, {Gallego}, {Gil de Paz},
  {Pascual}, {Zamorano}, \& {Donley}}]{PerezGonzalez+08}
{P{\'e}rez-Gonz{\'a}lez} P.~G., {Rieke} G.~H., {Villar} V., 
et al.,  2008, \apj, 675, 234

\bibitem[{{Pontzen} \& {Governato}(2012)}]{PG12}
{Pontzen} A., {Governato} F., 2012, \mnras, 421, 3464

\bibitem[{{Powell} {et~al.}(2011){Powell}, {Slyz}, \& {Devriendt}}]{PSD11}
{Powell} L.~C., {Slyz} A., {Devriendt} J., 2011, \mnras, 414, 3671

\bibitem[{{Press} \& {Schechter}(1974)}]{PS74}
{Press} W.~H., {Schechter} P., 1974, \apj, 187, 425

\bibitem[{{Rees}(1986)}]{Rees86}
{Rees} M.~J., 1986, \mnras, 218, 25P

\bibitem[{{Reines} \& {Deller}(2012)}]{RD12}
{Reines} A.~E., {Deller} A.~T., 2012, \apjl, 750, L24

\bibitem[{{Rodighiero} {et~al.}(2011){Rodighiero}, {Daddi}, {Baronchelli},
  {Cimatti}, {Renzini}, {Aussel}, {Popesso}, {Lutz}, {Andreani}, {Berta},
  {Cava}, {Elbaz}, {Feltre}, {Fontana}, {F{\"o}rster Schreiber},
  {Franceschini}, {Genzel}, {Grazian}, {Gruppioni}, {Ilbert}, {Le Floch},
  {Magdis}, {Magliocchetti}, {Magnelli}, {Maiolino}, {McCracken}, {Nordon},
  {Poglitsch}, {Santini}, {Pozzi}, {Riguccini}, {Tacconi}, {Wuyts}, \&
  {Zamorani}}]{Rodighiero+11}
{Rodighiero} G., {Daddi} E., {Baronchelli} I., 
et al.,  2011, \apjl, 739,
  L40

\bibitem[{{Salpeter}(1955)}]{Salpeter55}
{Salpeter} E.~E., 1955, \apj, 121, 161

\bibitem[{{Salvadori} \& {Ferrara}(2009)}]{SF09}
{Salvadori} S., {Ferrara} A., 2009, \mnras, 395, L6

\bibitem[{{Sani} {et~al.}(2012){Sani}, {Davies}, {Sternberg}, {Gracia-Carpio},
  {Hicks}, {Krips}, {Tacconi}, {Genzel}, {Vollmer}, {Schinnerer},
  {Garcia-Burillo}, {Usero}, \& {Orban de Xivry}}]{Sani+12}
{Sani} E., {Davies} R.~I., {Sternberg} A., 
et al.,  2012,
  \mnras, in press, arXiv:1205.4242

\bibitem[{{Scannapieco} {et~al.}(2012){Scannapieco}, {Wadepuhl}, {Parry},
  {Navarro}, {Jenkins}, {Springel}, {Teyssier}, {Carlson}, {Couchman}, {Crain},
  {Vecchia}, {Frenk}, {Kobayashi}, {Monaco}, {Murante}, {Okamoto}, {Quinn},
  {Schaye}, {Stinson}, {Theuns}, {Wadsley}, {White}, \&
  {Woods}}]{Scannapieco+12}
{Scannapieco} C., {Wadepuhl} M., {Parry} O.~H., 
et al.,  2012, \mnras, 423, 1726

\bibitem[{{Schawinski}(2012)}]{Schawinski12}
{Schawinski} K., 2012, in Frank N. Bash Symposium: New Horizons in Astronomy,
  arXiv:1206.2661

\bibitem[{{Schawinski} {et~al.}(2007){Schawinski}, {Thomas}, {Sarzi},
  {Maraston}, {Kaviraj}, {Joo}, {Yi}, \& {Silk}}]{Schawinski+07}
{Schawinski} K., {Thomas} D., {Sarzi} M., {Maraston} C., {Kaviraj} S., {Joo}
  S.-J., {Yi} S.~K., {Silk} J., 2007, \mnras, 382, 1415

\bibitem[{{Schechter}(1976)}]{Schechter76}
{Schechter} P., 1976, \apj, 203, 297

\bibitem[{{Sheth} {et~al.}(2001){Sheth}, {Mo}, \& {Tormen}}]{SMT01}
{Sheth} R.~K., {Mo} H.~J., {Tormen} G., 2001, \mnras, 323, 1

\bibitem[{{Sijacki} {et~al.}(2009){Sijacki}, {Springel}, \& {Haehnelt}}]{SSH09}
{Sijacki} D., {Springel} V., {Haehnelt} M.~G., 2009, \mnras, 400, 100

\bibitem[{{Silk}(1977)}]{Silk77}
{Silk} J., 1977, \apj, 211, 638

\bibitem[{{Silk}(1997)}]{Silk97}
---, 1997, \apj, 481, 703

\bibitem[{{Silk} \& {Norman}(2009)}]{SilkNorman09}
{Silk} J., {Norman} C., 2009, \apj, 700, 262

\bibitem[{{Silk} \& {Nusser}(2010)}]{SN10}
{Silk} J., {Nusser} A., 2010, \apj, 725, 556

\bibitem[{{Silk} \& {Rees}(1998)}]{SR98}
{Silk} J., {Rees} M.~J., 1998, \aap, 331, L1

\bibitem[{{Somerville} {et~al.}(2008){Somerville}, {Hopkins}, {Cox},
  {Robertson}, \& {Hernquist}}]{Somerville+08}
{Somerville} R.~S., {Hopkins} P.~F., {Cox} T.~J., {Robertson} B.~E.,
  {Hernquist} L., 2008, \mnras, 391, 481

\bibitem[{{Sonnenfeld} {et~al.}(2012){Sonnenfeld}, {Treu}, {Gavazzi},
  {Marshall}, {Auger}, {Suyu}, {Koopmans}, \& {Bolton}}]{Sonnenfeld+12}
{Sonnenfeld} A., {Treu} T., {Gavazzi} R., {Marshall} P.~J., {Auger} M.~W.,
  {Suyu} S.~H., {Koopmans} L.~V.~E., {Bolton} A.~S., 2012, \apj, 752, 163

\bibitem[{{Springel}(2010{\natexlab{a}})}]{Springel10}
{Springel} V., 2010{\natexlab{a}}, \mnras, 401, 791

\bibitem[{{Springel}(2010{\natexlab{b}})}]{Springel10_ARAA}
---, 2010{\natexlab{b}}, \araa, 48, 391

\bibitem[{{Springel} \& {Hernquist}(2003)}]{SH03b}
{Springel} V., {Hernquist} L., 2003, \mnras, 339, 312

\bibitem[{{Springel} {et~al.}(2008){Springel}, {Wang}, {Vogelsberger},
  {Ludlow}, {Jenkins}, {Helmi}, {Navarro}, {Frenk}, \& {White}}]{Springel+08}
{Springel} V., {Wang} J., {Vogelsberger} M., 
et al.,  2008, \mnras, 391, 1685

\bibitem[{{Springel} {et~al.}(2005){Springel}, {White}, {Jenkins}, {Frenk},
  {Yoshida}, {Gao}, {Navarro}, {Thacker}, {Croton}, {Helly}, {Peacock}, {Cole},
  {Thomas}, {Couchman}, {Evrard}, {Colberg}, \& {Pearce}}]{Springel+05}
{Springel} V., {White} S.~D.~M., {Jenkins} A., 
et al.,  2005, \nat, 435, 629

\bibitem[{{Stewart} {et~al.}(2011){Stewart}, {Kaufmann}, {Bullock}, {Barton},
  {Maller}, {Diemand}, \& {Wadsley}}]{Stewart+11_apjl}
{Stewart} K.~R., {Kaufmann} T., {Bullock} J.~S., {Barton} E.~J., {Maller}
  A.~H., {Diemand} J., {Wadsley} J., 2011, \apjl, 735, L1

\bibitem[{{Sunyaev} \& {Zeldovich}(1972)}]{SZ72}
{Sunyaev} R.~A., {Zeldovich} Y.~B., 1972, Comments on Astrophysics and Space
  Physics, 4, 173

\bibitem[{{Swaters} {et~al.}(2011){Swaters}, {Sancisi}, {van Albada}, \& {van
  der Hulst}}]{SSvAvdH11}
{Swaters} R.~A., {Sancisi} R., {van Albada} T.~S., {van der Hulst} J.~M., 2011,
  \apj, 729, 118

\bibitem[{{Teyssier}(2002)}]{Teyssier02}
{Teyssier} R., 2002, \aap, 385, 337

\bibitem[{{Teyssier} {et~al.}(2010){Teyssier}, {Chapon}, \& {Bournaud}}]{TCB10}
{Teyssier} R., {Chapon} D., {Bournaud} F., 2010, \apjl, 720, L149

\bibitem[{{Thomas} {et~al.}(2010){Thomas}, {Maraston}, {Schawinski}, {Sarzi},
  \& {Silk}}]{Thomas+10}
{Thomas} D., {Maraston} C., {Schawinski} K., {Sarzi} M., {Silk} J., 2010,
  \mnras, 404, 1775

\bibitem[{{Tikhonov} \& {Klypin}(2009)}]{TK09}
{Tikhonov} A.~V., {Klypin} A., 2009, \mnras, 395, 1915

\bibitem[{{Tinker} {et~al.}(2008){Tinker}, {Kravtsov}, {Klypin}, {Abazajian},
  {Warren}, {Yepes}, {Gottl{\"o}ber}, \& {Holz}}]{Tinker+08}
{Tinker} J., {Kravtsov} A.~V., {Klypin} A., {Abazajian} K., {Warren} M.,
  {Yepes} G., {Gottl{\"o}ber} S., {Holz} D.~E., 2008, \apj, 688, 709

\bibitem[{{Tollerud} {et~al.}(2011){Tollerud}, {Bullock}, {Graves}, \&
  {Wolf}}]{TBGW11}
{Tollerud} E.~J., {Bullock} J.~S., {Graves} G.~J., {Wolf} J., 2011, \apj, 726,
  108

\bibitem[{{Tolstoy} {et~al.}(2009){Tolstoy}, {Hill}, \& {Tosi}}]{THT09}
{Tolstoy} E., {Hill} V., {Tosi} M., 2009, \araa, 47, 371

\bibitem[{{Tremonti} {et~al.}(2004){Tremonti}, {Heckman}, {Kauffmann},
    {Brinchmann}, {Charlot}, {White}, {Seibert}, {Peng}, {Schlegel},
    {Uomoto}, {Fukugita}, \& {Brinkmann}}]{Tremonti+04}
{Tremonti}, C.~A., {Heckman}, T.~M., {Kauffmann}, G.,
et al., 2004, \apj, 613, 898

\bibitem[{{Vale} \& {Ostriker}(2006)}]{VO06}
{Vale} A., {Ostriker} J.~P., 2006, \mnras, 371, 1173

\bibitem[{{van der Wel} {et~al.}(2009){van der Wel}, {Rix}, {Holden}, {Bell},
  \& {Robaina}}]{vanderWel+09}
{van der Wel} A., {Rix} H.-W., {Holden} B.~P., {Bell} E.~F., {Robaina} A.~R.,
  2009, \apjl, 706, L120

\bibitem[{{van Dokkum} \& {Conroy}(2011)}]{vDC11}
{van Dokkum} P.~G., {Conroy} C., 2011, \apjl, 735, L13

\bibitem[{{Wagner} \& {Bicknell}(2011)}]{WB11}
{Wagner} A.~Y., {Bicknell} G.~V., 2011, \apj, 728, 29

\bibitem[{{Wagner} {et~al.}(2012){Wagner}, {Bicknell}, \& {Umemura}}]{WBU12}
{Wagner} A.~Y., {Bicknell} G.~V., {Umemura} M., 2012, \apj, submitted,
  arXiv:1205.0542

\bibitem[{{Wang} {et~al.}(2011){Wang}, {Wagg}, {Carilli}, {Walter}, {Riechers},
  {Willott}, {Bertoldi}, {Omont}, {Beelen}, {Cox}, {Strauss}, {Bergeron},
  {Forveille}, {Menten}, \& {Fan}}]{Wang+11}
{Wang} R., {Wagg} J., {Carilli} C.~L., 
et al.,  2011, \apjl, 739,
  L34

\bibitem[{{Warren} {et~al.}(2006){Warren}, {Abazajian}, {Holz}, \&
  {Teodoro}}]{WAHT06}
{Warren} M.~S., {Abazajian} K., {Holz} D.~E., {Teodoro} L., 2006, \apj, 646,
  881

\bibitem[{{Weinmann} {et~al.}(2011){Weinmann}, {Neistein}, \& {Dekel}}]{WND11}
{Weinmann} S.~M., {Neistein} E., {Dekel} A., 2011, \mnras, 417, 2737

\bibitem[{{Weinmann} {et~al.}(2012){Weinmann}, {Pasquali}, {Oppenheimer},
  {Finlator}, {Mendel}, {Crain}, \& {Maccio}}]{Weinmann+12}
{Weinmann} S.~M., {Pasquali} A., {Oppenheimer} B.~D., {Finlator} K., {Mendel}
  J.~T., {Crain} R.~A., {Maccio} A.~V., 2012, \mnras, submitted,
  arXiv:1204.4184

\bibitem[{{Weisz} {et~al.}(2012){Weisz}, {Johnson}, {Johnson}, {Skillman},
  {Lee}, {Kennicutt}, {Calzetti}, {van Zee}, {Bothwell}, {Dalcanton}, {Dale},
  {Williams}}]{Weisz+12}
{Weisz}, D.~R., {Johnson}, B.~D., {Johnson}, L.~C.
et~al., 2012, \apj, 744, 44

\bibitem[{{Yang} {et~al.}(2003){Yang}, {Mo}, \& {van den Bosch}}]{YMvdB03}
{Yang} X., {Mo} H.~J., {van den Bosch} F.~C., 2003, \mnras, 339, 1057

\bibitem[{{Yang} {et~al.}(2009){Yang}, {Mo}, \& {van den Bosch}}]{YMvdB09}
---, 2009, \apj, 695, 900

\bibitem[{{Yoon} {et~al.}(2011){Yoon}, {Johnston}, \& {Hogg}}]{YJH11}
{Yoon} J.~H., {Johnston} K.~V., {Hogg} D.~W., 2011, \apj, 731, 58

\bibitem[{{Ziegler} {et~al.}(2005){Ziegler}, {Thomas}, {B{\"o}hm}, {Bender},
  {Fritz}, \& {Maraston}}]{Ziegler+05}
{Ziegler} B.~L., {Thomas} D., {B{\"o}hm} A., {Bender} R., {Fritz} A.,
  {Maraston} C., 2005, \aap, 433, 519

\bibitem[{{Zolotov} {et~al.}(2012){Zolotov}, {Brooks}, {Willman}, {Governato},
  {Pontzen}, {Christensen}, {Dekel}, {Quinn}, {Shen}, \&
  {Wadsley}}]{Zolotov+12}
{Zolotov} A., {Brooks} A.~M., {Willman} B., 
et al.,  2012,
  \apj, submitted, arXiv:1207.0007

\end{thebibliography}
\end{document}